\documentclass[journal,twoside]{IEEEtaes} 
\usepackage{cite}

\usepackage{hyperref}
\hypersetup{hidelinks}

\usepackage{textcomp}
\def\BibTeX{{\rm B\kern-.05em{\sc i\kern-.025em b}\kern-.08em
    T\kern-.1667em\lower.7ex\hbox{E}\kern-.125emX}}
\usepackage{booktabs}

\usepackage{tipa}

\newcommand{\turnt}{\text{\rotatebox[origin=c]{180}{$\mathbf{t}$}}}
\newcommand{\nturnt}{\text{\rotatebox[origin=c]{180}{${t}$}}}

\usepackage{graphicx} 
\usepackage{xcolor}
\usepackage{comment}

\usepackage[cmex10]{amsmath}
\usepackage{amssymb}
\usepackage{amsfonts}
\usepackage{textcomp}
\usepackage{bm}
\usepackage{cancel}
\usepackage{upgreek}
\usepackage{algorithm}
\usepackage{algorithmicx}
\usepackage{algpseudocode}
\usepackage{mathrsfs}
\usepackage{mathtools}
\usepackage{scalerel}
\usepackage{accents} 

\usepackage{pgfplots}

\usepackage{support-caption}
\usepackage{subcaption}

\usepackage{tabularx}
\usepackage{makecell}

\providecommand{\dashline}{\raisebox{0.1em}{\scalebox{0.75}{\rotatebox[origin=c]{90}{$\dabar@\dabar@\dabar@$}}}}

\DeclarePairedDelimiter{\norm}{\lVert}{\rVert}

\providecommand{\grade}[1]{{\langle #1 \rangle}}

\makeatother
\newcommand{\dd}{\mathop{}\!\mathrm{d}}

\let\oldsqrt\sqrt
\def\sqrt{\mathpalette\DHLhksqrt}
\def\DHLhksqrt#1#2{%
\setbox0=\hbox{$#1\oldsqrt{#2\,}$}\dimen0=\ht0
\advance\dimen0-0.2\ht0
\setbox2=\hbox{\vrule height\ht0 depth -\dimen0}%
{\box0\lower0.4pt\box2}}

\providecommand*{\eu}{\ensuremath{\mathrm{e}}}

\usepackage{xcolor}
\definecolor{MyGrey5}{RGB}{150,150,150}
\definecolor{MyGrey}{RGB}{234,234,234}
\definecolor{MyBlue}{RGB}{0,0,255}
\definecolor{DarkSlateGray}{rgb}{0.1836,  0.3086  ,  0.3086}
\definecolor{Chocolate}{rgb}{0.8203    ,  0.4102  ,  0.1172}
\definecolor{FireBrick}{rgb}{ 0.6953   ,  0.1328  ,  0.1328}
\definecolor{MediumBlue}{rgb}{ 0       ,  0       ,  0.8008}
\definecolor{MyGreen}{rgb}{ 0          ,  0.5000  ,  0     }
\definecolor{MyRed}{rgb}{ 1        ,  0.0000  ,  0     }
\definecolor{MyPurple}{rgb}{0.5     ,    0  ,0.5}
\definecolor{WatermarkGrey}{rgb}{0.6,0.6,0.6}
\definecolor{WatermarkGrey2}{rgb}{0.8,0.8,0.8}
\definecolor{MyBlack}{rgb}{ 0        ,  0  ,  0    }
\definecolor{myWhite}{rgb}{1,1,1}
\definecolor{mySnow}{rgb}{1,0.97917,0.97917}
\definecolor{myHoneydew}{rgb}{0.9375,1,0.9375}
\definecolor{myMintCream}{rgb}{0.95833,1,0.97917}
\definecolor{myAzure}{rgb}{0.9375,1,1}
\definecolor{myAliceBlue}{rgb}{0.9375,0.97083,1}
\definecolor{myGhostWhite}{rgb}{0.97083,0.97083,1}
\definecolor{myWhiteSmoke}{rgb}{0.95833,0.95833,0.95833}
\definecolor{mySeashell}{rgb}{1,0.95833,0.92969}
\definecolor{myBeige}{rgb}{0.95833,0.95833,0.85938}
\definecolor{myOldLace}{rgb}{0.99167,0.95833,0.89844}
\definecolor{myFloralWhite}{rgb}{1,0.97917,0.9375}
\definecolor{myIvory}{rgb}{1,1,0.9375}
\definecolor{myAntiqueWhite}{rgb}{0.97917,0.91797,0.83984}
\definecolor{myLinen}{rgb}{0.97917,0.9375,0.89844}
\definecolor{myLavenderBlush}{rgb}{1,0.9375,0.95833}
\definecolor{myMistyRose}{rgb}{1,0.89063,0.87891}
\definecolor{myGray}{rgb}{0.5,0.5,0.5}
\definecolor{myGainsboro}{rgb}{0.85938,0.85938,0.85938}
\definecolor{myLightGray}{rgb}{0.82422,0.82422,0.82422}
\definecolor{mySilver}{rgb}{0.75,0.75,0.75}
\definecolor{myDarkGray}{rgb}{0.66016,0.66016,0.66016}
\definecolor{myDimGray}{rgb}{0.41016,0.41016,0.41016}
\definecolor{myLightSlateGray}{rgb}{0.46484,0.53125,0.59766}
\definecolor{mySlateGray}{rgb}{0.4375,0.5,0.5625}
\definecolor{myDarkSlateGray}{rgb}{0.18359,0.30859,0.30859}
\definecolor{myBlack}{rgb}{0,0,0}
\definecolor{myRed}{rgb}{1,0,0}
\definecolor{myLightSalmon}{rgb}{1,0.625,0.47656}
\definecolor{mySalmon}{rgb}{0.97917,0.5,0.44531}
\definecolor{myDarkSalmon}{rgb}{0.91016,0.58594,0.47656}
\definecolor{myLightCoral}{rgb}{0.9375,0.5,0.5}
\definecolor{myIndianRed}{rgb}{0.80078,0.35938,0.35938}
\definecolor{myCrimson}{rgb}{0.85938,0.078125,0.23438}
\definecolor{myFireBrick}{rgb}{0.69531,0.13281,0.13281}
\definecolor{myDarkRed}{rgb}{0.54297,0,0}
\definecolor{myPink}{rgb}{1,0.75,0.79297}
\definecolor{myLightPink}{rgb}{1,0.71094,0.75391}
\definecolor{myHotPink}{rgb}{1,0.41016,0.70313}
\definecolor{myDeepPink}{rgb}{1,0.078125,0.57422}
\definecolor{myPaleVioletRed}{rgb}{0.85547,0.4375,0.57422}
\definecolor{myMediumVioletRed}{rgb}{0.77734,0.082031,0.51953}
\definecolor{myOrange}{rgb}{1,0.64453,0}
\definecolor{myDarkOrange}{rgb}{1,0.54688,0}
\definecolor{myCoral}{rgb}{1,0.49609,0.3125}
\definecolor{myTomato}{rgb}{1,0.38672,0.27734}
\definecolor{myOrangeRed}{rgb}{1,0.26953,0}
\definecolor{myYellow}{rgb}{1,1,0}
\definecolor{myLightYellow}{rgb}{1,1,0.875}
\definecolor{myLemonChiffon}{rgb}{1,0.97917,0.80078}
\definecolor{myLightGoldenrodYellow}{rgb}{0.97917,0.97917,0.82031}
\definecolor{myPapayaWhip}{rgb}{1,0.93359,0.83203}
\definecolor{myMoccasin}{rgb}{1,0.89063,0.70703}
\definecolor{myPeachPuff}{rgb}{1,0.85156,0.72266}
\definecolor{myPaleGoldenrod}{rgb}{0.92969,0.90625,0.66406}
\definecolor{myKhaki}{rgb}{0.9375,0.89844,0.54688}
\definecolor{myDarkKhaki}{rgb}{0.73828,0.71484,0.41797}
\definecolor{myGold}{rgb}{1,0.83984,0}
\definecolor{myBrown}{rgb}{0.64453,0.16406,0.16406}
\definecolor{myCornsilk}{rgb}{1,0.97083,0.85938}
\definecolor{myBlanchedAlmond}{rgb}{1,0.91797,0.80078}
\definecolor{myBisque}{rgb}{1,0.89063,0.76563}
\definecolor{myNavajoWhite}{rgb}{1,0.86719,0.67578}
\definecolor{myWheat}{rgb}{0.95833,0.86719,0.69922}
\definecolor{myBurlyWood}{rgb}{0.86719,0.71875,0.52734}
\definecolor{myTan}{rgb}{0.82031,0.70313,0.54688}
\definecolor{myRosyBrown}{rgb}{0.73438,0.55859,0.55859}
\definecolor{mySandyBrown}{rgb}{0.95417,0.64063,0.375}
\definecolor{myGoldenrod}{rgb}{0.85156,0.64453,0.125}
\definecolor{myDarkGoldenrod}{rgb}{0.71875,0.52344,0.042969}
\definecolor{myPeru}{rgb}{0.80078,0.51953,0.24609}
\definecolor{myChocolate}{rgb}{0.82031,0.41016,0.11719}
\definecolor{mySaddleBrown}{rgb}{0.54297,0.26953,0.074219}
\definecolor{mySienna}{rgb}{0.625,0.32031,0.17578}
\definecolor{myMaroon}{rgb}{0.5,0,0}
\definecolor{myGreen}{rgb}{0,0.5,0}
\definecolor{myPaleGreen}{rgb}{0.59375,0.98333,0.59375}
\definecolor{myLightGreen}{rgb}{0.5625,0.92969,0.5625}
\definecolor{myYellowGreen}{rgb}{0.60156,0.80078,0.19531}
\definecolor{myGreenYellow}{rgb}{0.67578,1,0.18359}
\definecolor{myChartreuse}{rgb}{0.49609,1,0}
\definecolor{myLawnGreen}{rgb}{0.48438,0.9875,0}
\definecolor{myLime}{rgb}{0,1,0}
\definecolor{myLimeGreen}{rgb}{0.19531,0.80078,0.19531}
\definecolor{myMediumSpringGreen}{rgb}{0,0.97917,0.60156}
\definecolor{mySpringGreen}{rgb}{0,1,0.49609}
\definecolor{myMediumAquamarine}{rgb}{0.39844,0.80078,0.66406}
\definecolor{myAquamarine}{rgb}{0.49609,1,0.82813}
\definecolor{myLightSeaGreen}{rgb}{0.125,0.69531,0.66406}
\definecolor{myMediumSeaGreen}{rgb}{0.23438,0.69922,0.44141}
\definecolor{mySeaGreen}{rgb}{0.17969,0.54297,0.33984}
\definecolor{myDarkSeaGreen}{rgb}{0.55859,0.73438,0.55859}
\definecolor{myForestGreen}{rgb}{0.13281,0.54297,0.13281}
\definecolor{myDarkGreen}{rgb}{0,0.39063,0}
\definecolor{myOliveDrab}{rgb}{0.41797,0.55469,0.13672}
\definecolor{myOlive}{rgb}{0.5,0.5,0}
\definecolor{myDarkOliveGreen}{rgb}{0.33203,0.41797,0.18359}
\definecolor{myTeal}{rgb}{0,0.5,0.5}
\definecolor{myBlue}{rgb}{0,0,1}
\definecolor{myLightBlue}{rgb}{0.67578,0.84375,0.89844}
\definecolor{myPowderBlue}{rgb}{0.6875,0.875,0.89844}
\definecolor{myPaleTurquoise}{rgb}{0.68359,0.92969,0.92969}
\definecolor{myTurquoise}{rgb}{0.25,0.875,0.8125}
\definecolor{myMediumTurquoise}{rgb}{0.28125,0.81641,0.79688}
\definecolor{myDarkTurquoise}{rgb}{0,0.80469,0.81641}
\definecolor{myLightCyan}{rgb}{0.875,1,1}
\definecolor{myCyan}{rgb}{0,1,1}
\definecolor{myAqua}{rgb}{0,1,1}
\definecolor{myDarkCyan}{rgb}{0,0.54297,0.54297}
\definecolor{myCadetBlue}{rgb}{0.37109,0.61719,0.625}
\definecolor{myLightSteelBlue}{rgb}{0.6875,0.76563,0.86719}
\definecolor{mySteelBlue}{rgb}{0.27344,0.50781,0.70313}
\definecolor{myLightSkyBlue}{rgb}{0.52734,0.80469,0.97917}
\definecolor{mySkyBlue}{rgb}{0.52734,0.80469,0.91797}
\definecolor{myDeepSkyBlue}{rgb}{0,0.74609,1}
\definecolor{myDodgerBlue}{rgb}{0.11719,0.5625,1}
\definecolor{myCornflowerBlue}{rgb}{0.39063,0.58203,0.92578}
\definecolor{myRoyalBlue}{rgb}{0.25391,0.41016,0.87891}
\definecolor{myMediumBlue}{rgb}{0,0,0.80078}
\definecolor{myDarkBlue}{rgb}{0,0,0.54297}
\definecolor{myNavy}{rgb}{0,0,0.5}
\definecolor{myMidnightBlue}{rgb}{0.097656,0.097656,0.4375}
\definecolor{myPurple}{rgb}{0.5,0,0.5}
\definecolor{myLavender}{rgb}{0.89844,0.89844,0.97917}
\definecolor{myThistle}{rgb}{0.84375,0.74609,0.84375}
\definecolor{myPlum}{rgb}{0.86328,0.625,0.86328}
\definecolor{myViolet}{rgb}{0.92969,0.50781,0.92969}
\definecolor{myOrchid}{rgb}{0.85156,0.4375,0.83594}
\definecolor{myFuchsia}{rgb}{1,0,1}
\definecolor{myMagenta}{rgb}{1,0,1}
\definecolor{myMediumOrchid}{rgb}{0.72656,0.33203,0.82422}
\definecolor{myMediumPurple}{rgb}{0.57422,0.4375,0.85547}
\definecolor{myAmethyst}{rgb}{0.59766,0.39844,0.79688}
\definecolor{myBlueViolet}{rgb}{0.53906,0.16797,0.88281}
\definecolor{myDarkViolet}{rgb}{0.57813,0,0.82422}
\definecolor{myDarkOrchid}{rgb}{0.59766,0.19531,0.79688}
\definecolor{myDarkMagenta}{rgb}{0.54297,0,0.54297}
\definecolor{mySlateBlue}{rgb}{0.41406,0.35156,0.80078}
\definecolor{myDarkSlateBlue}{rgb}{0.28125,0.23828,0.54297}
\definecolor{myMediumSlateBlue}{rgb}{0.48047,0.40625,0.92969}
\definecolor{myIndigo}{rgb}{0.29297,0,0.50781}
\definecolor{myGrey}{rgb}{0.5,0.5,0.5}
\definecolor{myLightGrey}{rgb}{0.82422,0.82422,0.82422}
\definecolor{myDarkGrey}{rgb}{0.66016,0.66016,0.66016}
\definecolor{myDimGrey}{rgb}{0.41016,0.41016,0.41016}
\definecolor{myLightSlateGrey}{rgb}{0.46484,0.53125,0.59766}
\definecolor{mySlateGrey}{rgb}{0.4375,0.5,0.5625}
\definecolor{myDarkSlateGrey}{rgb}{0.18359,0.30859,0.30859}
\definecolor{MyC1}{rgb}{0.2,0.2,0.2}
\definecolor{MyC2}{rgb}{0.4,0.4,0.4}
\definecolor{MyC3}{rgb}{0.6,0.6,0.6}
\definecolor{MyC4}{rgb}{0.5,0,0}
\definecolor{myCmap1}{rgb}{0.5151,0.0482,0.6697}
\definecolor{myCmap2}{rgb}{0.51986,0.15038,0.78122}
\definecolor{myCmap3}{rgb}{0.50624,0.24502,0.88013}
\definecolor{myCmap4}{rgb}{0.46845,0.32734,0.95715}
\definecolor{myCmap5}{rgb}{0.42262,0.40041,0.99163}
\definecolor{myCmap6}{rgb}{0.39171,0.47446,0.97876}
\definecolor{myCmap7}{rgb}{0.34868,0.54328,0.9273}
\definecolor{myCmap8}{rgb}{0.30536,0.60642,0.86689}
\definecolor{myCmap9}{rgb}{0.25736,0.66546,0.79834}
\definecolor{myCmap10}{rgb}{0.2215,0.71818,0.71517}
\definecolor{myCmap11}{rgb}{0.24346,0.75604,0.63608}
\definecolor{myCmap12}{rgb}{0.27135,0.79747,0.54425}
\definecolor{myCmap13}{rgb}{0.30037,0.83048,0.45329}
\definecolor{myCmap14}{rgb}{0.32929,0.86008,0.3625}
\definecolor{myCmap15}{rgb}{0.36336,0.89117,0.29194}
\definecolor{myCmap16}{rgb}{0.4385,0.91084,0.2995}
\definecolor{myCmap17}{rgb}{0.55047,0.92321,0.31979}
\definecolor{myCmap18}{rgb}{0.64985,0.9255,0.3352}
\definecolor{myCmap19}{rgb}{0.72705,0.9255,0.34385}
\definecolor{myCmap20}{rgb}{0.8,0.9255,0.3529}

\usepackage{tikz}
\usetikzlibrary{arrows.meta}
\usetikzlibrary{calc}
\usetikzlibrary{arrows.meta,bending}
\usepackage{tikz-3dplot}
\usetikzlibrary{3d,backgrounds,intersections}
\pgfplotsset{compat=1.18}

\newcommand{\roll}{\upphi}
\newcommand{\pitch}{\uptheta}
\newcommand{\yaw}{\uppsi}

\providecommand*{\ju}{\ensuremath{\mathrm{j}}}

\providecommand*{\eu}{\ensuremath{\e}}

\newcommand{\e}{\mathbf{e}}

\newcommand{\bv}{\mathbf{b}}
\newcommand{\bft}{\mathbf{t}}

\newcommand{\bfxi}{\bm{\xi}}
\newcommand{\xiDot}{\dot{\bfxi}}
\newcommand{\xiDDot}{\ddot{\bfxi}}

\newcommand{\thrust}{f}

\newcommand{\wind}{\bm{w}}
\newcommand\Tau{\scalerel*{\uptau}{T}}

\newcommand\smallT{\scalerel*{T}{\tau}}

\newcommand{\bfOmega}{\bm{\Omega}}
\newcommand{\bfOmegaDot}{\dot{\bfOmega}}

\newcommand{\Julia}{{\protect\sc Julia }}

\newcommand{\xiCol}{\underline{\bfxi}}
\newcommand{\xiDotCol}{\underline{\dot{\bfxi}}}

\newcommand{\transpose}{\top}

\newtheorem{assumption}{Assumption}
\newtheorem{theorem}{Theorem}
\newtheorem{proposition}{Proposition}

\newtheorem{definition}{Definition}

\usetikzlibrary{dsp,chains}
\tikzset{myptr/.style={decoration={markings,mark=at position 1 with {\arrow[scale=0.5,>=stealth,ultra thick]{>}}},postaction={decorate}}}
\usetikzlibrary{shapes.geometric, arrows}
\tikzstyle{gain} = [isosceles triangle, inner sep=1.25pt, draw=black, fill=white]

\newcommand\solidSrule[1][1cm]{\rule[0.5ex]{#1}{.75pt}}

\newcommand\solidLrule[1][1cm]{\rule[0.5ex]{#1}{2.25pt}}

\newcommand\dashedrule{\mbox{\solidSrule[.75mm]\hspace{0.25mm}\solidSrule[0.75mm]\hspace{0.25mm}\solidSrule[0.75mm]}}

\tikzset{test/.style n args={3}{
    postaction={
    decorate,
    decoration={
    markings,
    mark=between positions 0 and \pgfdecoratedpathlength step 0.5pt with {
    \pgfmathsetmacro\myval{multiply(
        divide(
        \pgfkeysvalueof{/pgf/decoration/mark info/distance from start}, \pgfdecoratedpathlength
        ),
        100
    )};
    \pgfsetfillcolor{#3!\myval!#2};
    \pgfpathcircle{\pgfpointorigin}{#1};
    \pgfusepath{fill};}
}}}}

\newcommand{\ignacio}[1]{{\color{teal}\footnote{\color{teal}{\textbf{Ignacio:} #1}}}}

\newcommand{\sandoval}[1]{{\color{myPurple}\footnote{\color{myPurple}{\textbf{Steven:} #1}}}}

\newcommand{\hernan}[1]{{\color{myDarkTurquoise}\footnote{\color{myDarkTurquoise}{\textbf{Hernan:} #1}}}}

\usetikzlibrary{shadows}

\makeatletter
\def\convertto#1#2{\strip@pt\dimexpr #2*65536/\number\dimexpr 1#1}
\makeatother

\usepackage{placeins}

\begin{document}
\title{Globally Asymptotically Stable Trajectory Tracking of Underactuated UAVs using Geometric Algebra}
%
\author{Ignacio Rubio Scola}
\affil{Department of Industrial Products Engineering INTI, CONICET and UNR, Rosario, Argentina}
\author{Omar Alejandro Garcia Alcantara}
\affil{Klipsch School of Electrical and Computer Engineering, New Mexico State University (NMSU), Las Cruces NM 88003 USA}
\author{Steven Sandoval}
\affil{Klipsch School of Electrical and Computer Engineering, New Mexico State University (NMSU), Las Cruces NM 88003 USA}
\author{Eduardo Steed Espinoza Quesada} 
\member{Senior, IEEE} 
\affil{Center for Research and Advanced Studies on the National Polytechnic Institute, Mexico City 07360 Mexico}
\author{Hernan Haimovich}
\member{Senior, IEEE} 
\affil{Centro Internacional Franco-Argentino de Ciencias de la Información y Sistemas (CIFASIS), UNR and CONICET, Rosario, Argentina}
\author{Luis Rodolfo Garcia Carrillo}
\member{Senior, IEEE}
\affil{Klipsch School of Electrical and Computer Engineering, New Mexico State University (NMSU), Las Cruces NM 88003 USA}
\corresp{{\itshape Corresponding author: L. R. Garcia Carrillo}}
\authoraddress{I. Rubio Scola is with Department of Industrial Products Engineering INTI, CONICET and UNR, Rosario, Argentina. e-mail: irubio@inti.gob.ar. E. S. Espinoza Quesada is with Secihti and with the Center for Research and Advanced Studies on the National Polytechnic Institute, Mexico City 07360 Mexico. e-mail: eduardo.espinoza@cinvestav.mx. H. Haimovich is with the Centro Internacional Franco-Argentino de Ciencias de la Informaci\'on y Sistemas (CIFASIS), UNR and CONICET, Rosario, Argentina. e-mail: haimovich@cifasis-conicet.gov.ar. O. A. Garcia Alcantara, S. Sandoval, and L. R. Garcia Carrillo are with Klipsch School of Electrical and Computer Engineering, New Mexico State University (NMSU), Las Cruces NM 88003 USA. e-mail: \{omargalc,spsandov,luisillo\}@nmsu.edu}
\receiveddate{\textcolor{myFireBrick}{\textbf{Draft:} \today.}}
\markboth{AUTHOR ET AL.}{SHORT ARTICLE TITLE}
\maketitle
%
\begin{center}
\textit{\small This work has been submitted to the IEEE for possible publication.  
Copyright may be transferred without notice, after which this version  
may no longer be accessible.}
\end{center}
\begin{abstract}
This paper employs Geometric Algebra (GA) tools to model the dynamics of objects in 3-dimensional space, serving as a proof of concept to facilitate control design for trajectory tracking in underactuated systems. For control purposes, the model is structured as a cascade system, where a rotational subsystem drives a translational one. The rotational subsystem is linear, while the translational subsystem follows a linear-plus-perturbation form, thereby reducing the complexity of control design. A control strategy requiring only simple operations, no memory, and no iterative search loops is presented to illustrate the main features of the GA model. 
 By employing GA to model both translations and rotations, a singularity-free and geometrically intuitive representation can be achieved through the use of the geometric product. Closed-loop stability is rigorously established using input-to-state stability methods. Numerical simulations of a quad tilt-rotorcraft performing trajectory tracking in a windy environment validate the controller’s stability and performance.
\color{black}

\end{abstract}


\section{INTRODUCTION}

G{\scshape eometric} Algebra (GA) and the quaternion algebra are 
powerful mathematical frameworks used in fields like physics, computer graphics, robotics, and aerospace to represent rotations, transformations, and dynamics \cite{doran2003geometric,corrochano2001geometric,Bau24a,Bau24b}. 
Quaternions are specifically efficient for 3-dimensional ($3$D) rotation representation, avoiding gimbal lock (a limitation of Euler angles), and offering a compact representation. For this reason, quaternions are widely applied in areas such as robotics, computer graphics, avionics, and space systems \cite{spring1986euler,conway2003quaternions,altmann2005rotations,hanson2005visualizing,vince2011quaternions,voight2021quaternion,bayro2021survey
}.  
While quaternions excel at efficiently handling $3$D rotations,  they are limited in addressing other types of transformations or higher-dimensional dynamics \cite{spring1986euler}.

\subsection{Mathematical Frameworks for Dynamic Modeling}

Translational dynamics are easily represented by vectors and combinations of translations are then expressed as sums of vectors. Rotational dynamics, on the other hand, are far more complicated to deal with due to their inherently nonlinear behavior. Rigid-body motions combine translations and rotations, therefore, they naturally inherit the nonlinear characteristics of rotations~\cite{Bau24a}. 
The effective description of rotation has led to the development of numerous parameterization techniques presenting various properties and advantages. The Cartesian rotation vector, the Euler-Rodrigues parameters, or the Wiener-Milenkovic parameters all are examples of vectorial parameterizations \cite{bauchau2011flexible}. These are all characterized by a minimal set of three parameters that behave as the Cartesian components of a vector. A concise analysis of different parameterizations of rotations is presented in~\cite{stuelpnagel1964parametrization}, showing that four parameter representations, such as the quaternion representation, are singularity free, in contrast with minimal set parameterizations, which always involve singularities. 

A survey of the literature reveals that different parameterization techniques come with their own unique advantages and limitations. For both theoretical and numerical applications, the choice of parameterization is often driven by convention and personal preference rather than a rigorous cost-benefit analysis. Given the numerous approaches to the problem and its mathematical complexity, it is unsurprising that the multibody dynamics and robotics communities have yet to reach a consensus on the optimal treatment of kinematics for complex systems.
Vectors, quaternions, and matrices are often jointly employed in problems that require both rotation and translation, even though these mathematical systems were not designed from their inception to be used in conjunction with each other. The mixing of mathematical frameworks often complicates the modeling by requiring the definition of additional products and/or mappings. A clear example of such mixture of mathematical frameworks is presented in \cite{Nguyen2020}, where, for modeling the translation dynamics of an underactuated system, a $4\times 4$ rotation matrix constructed from quaternion coefficients is employed to map vectors from the body-fixed coordinate frame to the inertial reference frame. 
Additionally, a 4-dimensional column vector representation of the angular rates extended with a zero first entry is obtained by reformulating the quaternion Kronecker product as the multiplication of a $4\times 4$ matrix with a column vector. Moreover, a cross product operation from linear algebra is utilized to model the rotational subsystem, a product that is also inherited in the model-based computed torque attitude controller.

In contrast, GA utilizes a graded linear space of multivectors (sums of $k$-vectors) which allows for a richer space of mathematical objects that all share the same geometric product. This general framework is coordinate-free and not only generalizes vector algebra but also naturally subsumes systems such as complex numbers, quaternions, Plücker coordinates, spinors, Lie groups and Lie algebra representations---making GA particularly well-suited for modeling dynamic systems \cite{hildenbrand2013foundations, Hestenes2017Genesis}. Eliminating coordinate-based matrix constructs not only simplifies the mathematical formulation but also unveils the underlying geometric structures that are often obscured in conventional matrix-based representations. This leads to a unified, coordinate-free framework for representing and manipulating geometric objects and transformations in a more transparent and comprehensive manner~\cite{hestenes2003reforming, Hestenes2017Genesis, hestenes2012new, hestenes2012clifford, macdonald2017survey, bayro2010geometric, bayro2018geometric, bayro2020geometric, baylis2004electrodynamics, doran2003geometric, taylord2021geoemtric, hitzer2013quaternion}. Table~\ref{tab:modeling_comparison} provides a comparative overview of several mathematical frameworks that are widely employed for representing rotations (Modified Rodrigues Parameters (MRP)~\cite{Crassidis1996}, Euler Angles~\cite{hover2016}, Direction Cosine Matrix (DCM)~\cite{musa2018}, Quaternions~\cite{fresk2013full}) and rotations/translations (DCM + Vectors~\cite{DCM2015}, Quaternions + Vectors~\cite{Nguyen2020}, Geometric Algebra~\cite{Bau24a}). From this comparison, it becomes clear that GA offers a unique, unified, and elegant alternative. By relying on the geometric product, GA enables a singularity-free and coordinate-free formulation in which both rotations and translations can be represented and manipulated within the same algebraic structure. This unification not only reduces the need for additional mappings or ad hoc operations but also enhances geometric intuition, making the modeling process more transparent and robust.

Although 3D GA multivectors span an 8‑dimensional linear space, this structure allows efficient optimization strategies that achieve high runtime performance and robustness~\cite{hildenbrand2013foundations, worsdorfer2009optimizations}.
In this work, we adopt an efficient and lightweight computational implementation that separates even- and odd-grade multivectors, following an approach similar to that described in~\cite{SimpleGA}. This separation significantly reduces computational complexity by exploiting the algebraic structure of geometric algebra, allowing operations to be performed on smaller, well-structured subspaces rather than on the full 8-dimensional space. As a result, this implementation achieves improved runtime efficiency and numerical robustness while preserving the full expressive power of the geometric product. 
%
This approach has demonstrated a reduction of approximately $\sim$33\% in the computation time required to rotate a $3$D vector, as opposed to using a $3 \times 3$ rotation matrix~\cite{GALevelUp, SimpleGALecture}.


        





The modern resurgence of the development and application of GA is due to Hestenes \cite{hestenesSTA, hestenes2003reforming, sobczyk2012new}. GA was extended to Geometric Calculus which incorporates operations like differentiation and integration into the GA framework, allowing for the study of continuous change and motion in GA~\cite{hestenes2012clifford}. Formulation of the application of GA to classical physics, quantum theory, and electromagnetism can be found in \cite{hestenes2012new, doran2003geometric}. Building on these advancements, this work investigates the use of GA to design a globally asymptotically stable trajectory tracking controller for underactuated unmanned aircraft systems (UASs).

\begin{table*}[ht]
    \centering
    \caption{Various mathematical frameworks used for the representation of rotations and translations.}
    \renewcommand{\arraystretch}{1.3}
    \begin{tabular}{
        >{\raggedright\arraybackslash}p{3.0cm}
        >{\centering\arraybackslash}p{1.6cm}
        >{\centering\arraybackslash}p{2.6cm}
        >{\raggedright\arraybackslash}p{6.0cm}
        >{\centering\arraybackslash}p{2.5cm}
    }
        \toprule
        \textbf{Mathematical Framework} & \textbf{Singularity-Free} & \textbf{Geometric Intuition} & \textbf{Required Operations} & \textbf{Representation} \\
        \midrule
        Modified Rodrigues Parameters & No & No & Cross product, matrix-vector product, mapping from quaternion to a vector of MRP & $\mathrm{SO}(3)$ \\
        Euler Angles & No & Yes & Matrix-vector product, matrix product & $\mathrm{SO}(3)$ \\
        Direction Cosine Matrix & Yes & Yes & Matrix-vector product, matrix product & $\mathrm{SO}(3)$\\
        Quaternions & Yes & No & Kronecker product & $\mathrm{SO}(3)$ \\
        DCM + Vectors & Yes & Yes & Matrix-vector product, matrix product & $\mathrm{SE}(3)$ \\
        Quaternions + Vectors & Yes & No & Kronecker product, cross product, matrix-vector product, mapping from quaternion to a vector of quaternion coefficients, rotation matrices using quaternion coefficients & $\mathrm{SE}(3)$ \\
        Geometric Algebra & Yes & Yes & Geometric product & $\mathrm{SE}(3)$ \\
        \bottomrule
    \end{tabular}
    \label{tab:modeling_comparison}
\end{table*}


\color{black}
\subsection{Related Work} \label{sec:related_work}

The authors in~\cite{lee2010geometric} introduced a globally defined dynamic model for a quadrotor unmanned aerial vehicle (UAV) and designed a geometric tracking controller directly on the special Euclidean group $\mathrm{SE}(3)$. This intrinsic, coordinate-free approach avoids the singularities of Euler angles and the ambiguities of quaternions in attitude representation, which may exhibit unwinding behavior. This method enables the underactuated UAV to stabilize six degrees of freedom—both translational and rotational—using four thrust inputs while asymptotically tracking four outputs: position and heading direction. 
The controller is shown to be effective for executing complex, acrobatic maneuvers, such as recovering from an upside-down orientation. %
Similarly, a geometric adaptive robust hierarchical control strategy for an underactuated quadrotor, effectively addressing the negative effects of nonlinearities, uncertainties, and coupling within $\mathrm{SO}(3)$ is proposed in~\cite{liang2021geometric}. Geometric control is employed to manage nonlinearities, where the attitude error is directly defined on the tangent space of the rotation group, leading to improved attitude tracking performance.

A quaternion-based sliding variable that defines exponentially convergent error dynamics for desired attitude trajectories is introduced in~\cite{lopez2021sliding}. This sliding variable operates within the non-Euclidean space of quaternions and explicitly accounts for the double-covering property, enabling global attitude tracking in feedback control. Additionally, it prevents the unwinding phenomenon, which can otherwise result in unnecessarily prolonged attitude maneuvers.  

Closely related, the authors in \cite{zongcheng2024improved} proposed an enhanced geometric controller for quadrotor position tracking incorporating a novel attitude extraction method that combines unit quaternion representation with rotation axis-angle vector conversion. This approach mitigates the classical geometric control's susceptibility to a 180-degree turnover singularity while also preventing the unwinding phenomenon. The controller follows a cascaded structure, where a position loop and an attitude loop geometrically derive the desired attitude from the required thrust direction. The proposed Improved Geometric Controller (IGC) is proven to be globally asymptotically stable, effectively avoiding singularities even when the quadrotor experiences large initial attitude and position errors.

\medskip

Within the framework that GA provides, several different approaches may be taken for modeling a particular geometry, including Vector Space GA (VSGA), Projective/Plane-based GA (PPGA), and Conformal GA (CGA)\cite{corrochano2001geometric, doran2003geometric,sommer2013geometric,bayro2018geometric, bayro2020geometric}. A PPGA approach to describing the kinematics of multibody systems was presented in \cite{Bau24a}, while a similar approach but for describing rigid bodies, flexible-joints, geometrically exact beams, and lower-pair joints was presented in \cite{Bau24b}. Our work, on the other hand focuses on the VSGA approach. 
The developed method stems from previous work~\cite{escamilla2023stabilization} on GA-based modeling and tracking stabilization of a subactuated UAV where a coordinate-free model and two PID controllers were developed -- one of them for stabilizing the attitude subsystem and the second one for stabilizing the translational subsystem. This result, however, lacked a stability proof addressing the closed-loop of the model with both controllers.

\subsection{Main Results}

This work introduces a comprehensive model for the translational and rotational dynamics of a body in $3$D space using GA, along with an extension for underactuated systems. 
%
The advantages of using this type of model for trajectory tracking are demonstrated through the development of a simple and efficient control strategy. This approach analytically derives rotational references from translational ones, thereby eliminating the need for online differentiation algorithms. The resulting control law has a low computational cost, relying only on basic operations, with no memory requirements or iterative search loops. It is important to emphasize that the primary goal of this work is to provide a proof of concept illustrating how GA tools can streamline trajectory tracking control design for underactuated systems, rather than to propose an advanced control methodology.


The main contributions of this work can be summarized as:
\begin{itemize}
    \item Development of a full model for the translational and rotational dynamics of a body in $3$D space using GA, including underactuated systems and perturbations, corresponding to translational forces.
    \item Elimination of differentiation algorithms by analytically obtaining rotational references from translational ones, a crucial feature when dealing with underactuated systems.
    \item Design of a low-computational-cost controller, requiring no memory. 
    \item Proof of closed-loop stability based on input-to-state stability-related tools. 
    \item Presentation of numerical simulations of a quad tilt-rotorcraft performing trajectory tracking in a windy environment validate the GA controller’s stability and performance.
\end{itemize}
The remainder of the manuscript is organized as follows. 
The basic notation and terminology necessary for GA is presented in Section~\ref{notation}. Section~\ref{model} introduces the GA-based model that describes the translational and rotational dynamics of an object in $3$D space. The derivation of the proposed controller for an underactuated system with its corresponding Lyapunov stability proof is presented in Section~\ref{control}. Simulation results of the implementation of the proposed controller 
are provided in Section~\ref{simulation}. A comparative analysis between the proposed controller and six representative, state-of-the-art control algorithms is outlined in Section~\ref{Discussion}. Finally, Section~\ref{conclusions} highlights our main findings and conclusions, and outlines future research directions.

\section{NOTATION AND TERMINOLOGY FOR GA}\label{notation}

The key innovation of GA is an associative vector product termed the \textit{Geometric Product} denoted by juxtaposition as
\begin{IEEEeqnarray}{c} 
		\mathbf{u}\mathbf{v} = \mathbf{u}\cdot\mathbf{v} + \mathbf{u}\wedge\mathbf{v}
\end{IEEEeqnarray}
where $\cdot$ is the usual vector dot product (more generally, the left contraction) and $\wedge$ is the wedge product---which is closely related (three dimensions) to the usual cross product\cite{macdonald2010linear}. 

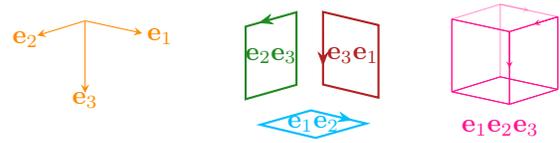
\begin{figure}
    \centering
    \hfill\begin{tikzpicture}[draw=white,tdplot_main_coords,font=\sffamily,scale=1.0]
    \node[inner sep=0pt,minimum size=0pt] (E2) at (1,0,0)  {};
    \node[inner sep=0pt,minimum size=0pt] (E1) at (0,1,0)  {};
    \node[inner sep=0pt,minimum size=0pt] (E3) at (0,0,1)  {};


    \draw[myDarkOrange,-stealth] (0,0,0)-- (E1) node[pos=1.25] {$~\mathbf{e}_1$};
    \draw[myDarkOrange,-stealth] (0,0,0) -- (E2) node[pos=1.25] {$\mathbf{e}_2$};
    \draw[myDarkOrange,-stealth] (0,0,0) -- (E3) node[pos=1.1] {$\mathbf{e}_3$};

\end{tikzpicture}\hfill\tdplotsetmaincoords{285}{132.5}
\begin{tikzpicture}[baseline={([yshift=-3.5ex]current bounding box.center)}, draw=white,tdplot_main_coords,font=\sffamily,scale=1.0]
    \node[inner sep=0pt,minimum size=0pt] (E2) at (1,0,0)  {};
    \node[inner sep=0pt,minimum size=0pt] (E1) at (0,1,0)  {};
    \node[inner sep=0pt,minimum size=0pt] (E3) at (0,0,1)  {};

        \begin{scope}[canvas is xy plane at z=1.45]
        \pgfmathsetmacro{\normValue}{1}
        \pgfmathsetmacro{\side}{sqrt(\normValue)}
        \node[inner sep=0pt,minimum size=0pt] (Origin) at (0.9,0.9) {};
        \draw[thick,myDeepSkyBlue] ($(Origin)+(-\side/2,-\side/2)$)  rectangle ($(Origin)+(\side/2,\side/2)$) node[pos=.5] {$\mathbf{e}_1\mathbf{e}_2$};
        \draw[{Stealth[length=2mm]}-, myDeepSkyBlue]($(Origin)+(-\side/2,\side/4)$) -- ($(Origin)+(-\side/2,-\side/4)$);
    \end{scope} 
 
     \begin{scope}[canvas is yz plane at x=0]
        \pgfmathsetmacro{\normValue}{1}
        \pgfmathsetmacro{\side}{sqrt(\normValue)}
        \node[inner sep=0pt,minimum size=0pt] (Origin) at (0.75,0.7) {};
        \draw[thick,myFireBrick] ($(Origin)+(-\side/2,-\side/2)$)  rectangle ($(Origin)+(\side/2,\side/2)$) node[pos=.5] {$\,\mathbf{e}_3\mathbf{e}_1$};
        \draw[{Stealth[length=2mm]}-, myFireBrick]($(Origin)+(-\side/2,\side/4)$) -- ($(Origin)+(-\side/2,-\side/4)$);
    \end{scope} 
 
     \begin{scope}[canvas is xz plane at y=0]
        \pgfmathsetmacro{\normValue}{1}
        \pgfmathsetmacro{\side}{sqrt(\normValue)}
        \node[inner sep=0pt,minimum size=0pt] (Origin) at (0.75,0.7) {};
        \draw[thick,myForestGreen] ($(Origin)+(-\side/2,-\side/2)$)  rectangle ($(Origin)+(\side/2,\side/2)$) node[pos=.5] {$\mathbf{e}_2\mathbf{e}_3$};
        \draw[{Stealth[length=2mm]}-, myForestGreen]($(Origin)+(\side/4,-\side/2)$) -- ($(Origin)+(-\side/4,-\side/2)$);
    \end{scope}
    
\end{tikzpicture}
\hfill~\tdplotsetmaincoords{285}{130}\begin{tikzpicture}[baseline={([yshift=-3.5ex]current bounding box.center)},scale=1.0,draw=white,tdplot_main_coords,font=\sffamily]
\begin{scope}[shift={(0,0,0)}]
    \pgfmathsetmacro{\normValue}{1}
    \pgfmathsetmacro{\sideLen}{(\normValue)^(1/3)}
    \pgfmathsetmacro{\sideLenHalf}{\sideLen/2}
    \pgfmathsetmacro{\labHeight}{\sideLen+0.7}

    \draw[myDeepPink,-stealth,ultra thin,opacity=0.25] (0,0,0) -- (0,\sideLenHalf,0)  {};
    \draw[myDeepPink,-stealth,ultra thin] (0,\sideLen,0) -- (\sideLenHalf,\sideLen,0)  {};
    \draw[myDeepPink,-stealth,ultra thin] (\sideLen,\sideLen,0) -- (\sideLen,\sideLen,\sideLenHalf) {};

    \draw[myDeepPink,opacity=0.25] (0,0,0)-- (0,\sideLen,0) {};
    \draw[myDeepPink,opacity=0.25] (0,0,0)-- (\sideLen,0,0) {};
    \draw[myDeepPink] (0,\sideLen,0)-- (\sideLen,\sideLen,0) {};
    \draw[myDeepPink] (\sideLen,0,0)-- (\sideLen,\sideLen,0) {};

    \draw[myDeepPink,opacity=0.25] (0,0,0)-- (0,0,\sideLen) node[pos=1.15] {};
    \draw[myDeepPink,opacity=0.25] (0,0,0)-- (\sideLen,0,0) {};
    \draw[myDeepPink] (0,0,\sideLen)-- (\sideLen,0,\sideLen) {};
    \draw[myDeepPink] (\sideLen,0,0)-- (\sideLen,0,\sideLen) {};

    \draw[myDeepPink,opacity=0.25] (0,0,0)-- (0,0,\sideLen) {};
    \draw[myDeepPink,opacity=0.25] (0,0,0)-- (0,\sideLen,0) {};
    \draw[myDeepPink] (0,0,\sideLen)-- (0,\sideLen,\sideLen) {};
    \draw[myDeepPink] (0,\sideLen,0)-- (0,\sideLen,\sideLen) {};

    \draw[myDeepPink] (0,0,\sideLen)-- (0,\sideLen,\sideLen) {};
    \draw[myDeepPink] (0,0,\sideLen)-- (\sideLen,0,\sideLen) {};
    \draw[myDeepPink] (0,\sideLen,\sideLen)-- (\sideLen,\sideLen,\sideLen) {};
    \draw[myDeepPink] (\sideLen,0,\sideLen)-- (\sideLen,\sideLen,\sideLen) {};
    
    \draw[myDeepPink] (0,\sideLen,0)-- (0,\sideLen,\sideLen) {};
    \draw[myDeepPink] (0,\sideLen,0)-- (\sideLen,\sideLen,0) {};
    \draw[myDeepPink] (0,\sideLen,\sideLen)-- (\sideLen,\sideLen,\sideLen) {};
    \draw[myDeepPink] (\sideLen,\sideLen,0)-- (\sideLen,\sideLen,\sideLen) {};

    \draw[myDeepPink] (\sideLen,0,0)-- (\sideLen,0,\sideLen) {};
    \draw[myDeepPink] (\sideLen,0,0)-- (\sideLen,\sideLen,0) {};
    \draw[myDeepPink] (\sideLen,0,\sideLen)-- (\sideLen,\sideLen,\sideLen) {};
    \draw[myDeepPink] (\sideLen,\sideLen,0)-- (\sideLen,\sideLen,\sideLen) {};

    \node[myDeepPink] at (0,0,\labHeight) {$\mathbf{e}_1\mathbf{e}_2\mathbf{e}_3$};
\end{scope}
\end{tikzpicture}\hfill~~~
    \caption{Basis elements for the vectors ($\mathbb{L}^3$), bivectors ($\mathbb{B}^3$), and trivectors ($\mathbb{T}^3$) in $3$D Euclidean space ($\mathbb{G}^3$).}
    \label{fig:basisElements}
\end{figure}

In GA, only real numbers are associated with ``scalar'' quantities which are denoted here using lowercase italic font i.e., $b,k,\ell, x, \alpha, \iota,\theta, \tau\in\mathbb{R}$. Traditional (polar) vectors are associated with directed lengths (here in Euclidean 3-space) and denoted using lowercase upright bold font i.e., $\bv, \e,\mathbf{u}, \mathbf{w}, \mathbf{x}, \mathbf{v}, \mathbf{f}\in\mathbb{L}^3$. Traditional axial vectors are closely related with directed areas (bivectors) in Euclidean 3-space and denoted using uppercase upright bold font i.e., $\mathbf{B},\mathbf{E},\Tau,\bm{\Omega}\in\mathbb{B}^3$. Trivectors in 3-space are pseudoscalars that can be represented as scaled versions of a unit directed volume denoted with $\mathbf{I}$, as $c\mathbf{I}\in\mathbb{T}^3$. One exception to the notational rules provided above is that unit bivectors $\mathbf{j}\in\{\mathbf{B}\in\mathbb{B}^3 : \norm{\mathbf{B}}=1\}$ are denoted with \textit{lowercase} upright bold font $\mathbf{j}$ because they have the property 
\begin{IEEEeqnarray}{c} 
		\mathbf{j}^2 = -1
\end{IEEEeqnarray}
and furthermore, they play a role similar to that of the traditional imaginary numbers (as a generator of rotations). Finally, a rotor is a unit norm multivector used to represent a rotation. In  Euclidean 3-space, a rotor $R$ is formed as the sum of a scalar and bivector $R\in \{M\in\mathbb{R}\oplus\mathbb{B}^3 : \norm{M}=1\}$ and is isomorphic to the traditional unit quaternion. Every multivector $M$ in  $\mathbb{G}^3$ is a unique sum of $k$-vector parts $\langle M \rangle_k$
\begin{IEEEeqnarray}{rCl}\IEEEyesnumber
    M = \langle M \rangle_0 + \langle M \rangle_1 +  \langle M \rangle_2 + \langle M \rangle_3
\end{IEEEeqnarray}
A general multivector in Euclidean 3-space $A,B,M,R\in\mathbb{G}^3$ is the direct sum of the scalars (signed line segments), vectors (directed line segments), bivectors (directed areas), and trivectors (directed volumes) in Euclidean 3-space (illustrated in Fig.~\ref{fig:basisElements})
\begin{IEEEeqnarray}{c} 
		\mathbb{G}^3 = \mathbb{R} \oplus \mathbb{L}^3 \oplus \mathbb{B}^3 \oplus \mathbb{T}^3
\end{IEEEeqnarray}
and can be represented in terms of an orthonormal basis
\begin{equation}
		\{1,~\e_1,~\e_2,~\e_3,~\e_1\e_2,~\e_2\e_3,~\e_3\e_1,~\e_1\e_2\e_3\}
	\end{equation}
as
\begin{IEEEeqnarray}{rClr} 
	M   &=& (m_0)(1) &(\text{scalar})\IEEEnonumber\\
            && +\> m_1\e_1 + m_2\e_2 + m_{3}\e_3 &(\text{vector}) \IEEEnonumber\\
            && +\>  m_4 \e_1\e_2 + m_5\e_2\e_3+m_6\e_3\e_1 ~~~~&(\text{bivector}) \IEEEnonumber\\
            && +\>  m_7 \e_1\e_2\e_3. &(\text{trivector})
\end{IEEEeqnarray}
A $k$-blade is any multivector that is expressible as the outer product of $k$ linearly independent $1$-vectors
\begin{IEEEeqnarray}{rCl}\IEEEyesnumber
    \mathbf{B} = \mathbf{u}_1\wedge\mathbf{u}_2 \wedge\cdots \wedge \mathbf{u}_k.
\end{IEEEeqnarray}
In $\mathbb{G}^3$, all non-zero $k$-vectors are $k$-blades. Let $\mathbf{B} = \mathbf{u}_1\mathbf{u}_2\cdots\mathbf{u}_k$ be a geometric product of orthogonal vectors. Then the reverse of $\mathbf{B}$ is 
\begin{IEEEeqnarray}{rCl}\IEEEyesnumber
    \mathbf{B}^\dagger = \mathbf{u}_k\mathbf{u}_{k-1}\cdots\mathbf{u}_1.
\end{IEEEeqnarray}
The \textit{norm} of a multivector $M = \langle M \rangle_0 + \langle M \rangle_1 + \langle M \rangle_2 + \langle M \rangle_3$ is given by 
\begin{IEEEeqnarray}{rCl}\IEEEyesnumber
		~~~~~~~~~~\norm{M}  &\triangleq& \sqrt{\langle M M^\dagger\rangle_0}
\end{IEEEeqnarray}
Suppose we have a vector $\mathbf{a}$ and a $k$-blade $\mathbf{B}$. Then the geometric product may be expressed as 
\begin{IEEEeqnarray}{c}
    {\mathbf{a}\mathbf{B}} = {\mathbf{a}\cdot\mathbf{B}} + {\mathbf{a}\wedge\mathbf{B}}
\end{IEEEeqnarray} 
where the inner (left contraction) and outer products are 
\begin{IEEEeqnarray}{c}
    {\mathbf{a}\cdot \mathbf{B}} = \langle {\mathbf{a}}{\mathbf{B}} \rangle_{{k}{-}{1}}
\end{IEEEeqnarray} 
and
\begin{IEEEeqnarray}{c}
    {\mathbf{a}\wedge \mathbf{B}} = \langle {\mathbf{a}}{\mathbf{B}} \rangle_{{1}{+}{k}}.
\end{IEEEeqnarray} 
This may also be written in terms of symmetrical parts as
\begin{equation}
    {\mathbf{u}\cdot \mathbf{B}} = \frac{1}{2}\left({\mathbf{u}}{\mathbf{B}} -{(-1)}^{{k}}{\mathbf{B}} {\mathbf{u}}\right)
\end{equation}
and 
\begin{equation}
    {\mathbf{u}\wedge \mathbf{B}} = \frac{1}{2}\left({\mathbf{u}}{\mathbf{B}} +{(-1)}^{{k}}{\mathbf{B}} {\mathbf{u}}\vphantom{0^0}\vphantom{0^0}\right).
\end{equation}
The rotation of the {multivector $M$} by bivector angle $\mathbf{j}\theta$ is 
    \begin{IEEEeqnarray}{c}\IEEEnonumber 
        {\mathsf{R}_{{\mathbf{j}\theta}}({M})} =  {\mathrm{e}^{-\mathbf{j}\theta/2}}{M} {\mathrm{e}^{\,\mathbf{j}\theta/2}}		
    \end{IEEEeqnarray}
The projection of the multivector $M$ onto the blade $\mathbf{B}$ is 
\begin{IEEEeqnarray}{c} \IEEEyesnumber
    {\mathsf{P}_{{\mathbf{B}}}({M})} = ({M}\cdot{\mathbf{B}}){\mathbf{B}^{-1}}
    \label{eq:projection}
\end{IEEEeqnarray}
The reader interested in GA is referred to Macdonald's works~\cite{macdonald2010linear, macdonald2012vector, macdonald2017survey} for a more in depth treatment.

\tdplotsetmaincoords{285}{130}
\section{DEVELOPMENT OF A GA-BASED MODEL}\label{model}
The following assumptions are made to facilitate the development of the GA-based model for control.

\begin{assumption}
    Masses and principal moments of inertia are known.
\end{assumption}

\begin{assumption}
    Linear positions, velocities, accelerations and jerk  with respect to (w.r.t.) the inertial frame are measured with sufficient precision. 
\end{assumption}

\begin{assumption}
    Angular positions and velocities w.r.t. the inertial frame are measured with sufficient precision.
\end{assumption}

\begin{assumption}
    Actuation for the linear and rotational movement is provided by:
    \begin{itemize}
        \item A main thrust whose magnitude can be controlled. The thrust direction can be time-varying and is known, also the first and second derivative are known by the controller at all times, but the controller has no influence over this thrust's direction.
        \item Additional thrusts to generate torques in the three degrees of rotation.
    \end{itemize}
\end{assumption}
\begin{assumption}\label{assump:Ref}
    A reference trajectory, along with its first four time derivatives are known, and its second derivative is bounded uniformly in time.
\end{assumption}

\begin{table}[!ht]
    \centering
    \caption{Definition of the symbols used through this work}
    \scalebox{0.9}{\begin{tabular}{cc}
    \toprule
    \textbf{Symbol} & \textbf{Description} \\
    \midrule
    $\{\e_1,\e_2,\e_3\}$        & inertial frame\\
    $\{\bv_1,\bv_2,\bv_3\}$     & body frame\\
    $m$                   & mass of body \\
    $g$                   & coefficient of acceleration due to gravity \\
            $\turnt $                   & direction of thrust relative to reference copy\\
        $\bft= R \turnt R^{\dagger}$                  & direction of thrust \\
    $\thrust$                   & coefficient of thrust \\
    $\wind$                   & additive perturbation \\
        $\Tau$      & total torque\\
    $\bfxi,\xiDot,\xiDDot$      & translational position, velocity, acceleration\\
    $\bfxi_\mathrm{d},\xiDot_\mathrm{d},\xiDDot_\mathrm{d}$      & desired transl. position, velocity, acceleration\\
    $\bm{\ju}\theta,\bfOmega,\bfOmegaDot$    & angular position, velocity, acceleration\\
    $\bm{\ju}_\mathrm{d}\theta_\mathrm{d},\bfOmega_\mathrm{d},\bfOmegaDot_\mathrm{d}$    & desired angular position, velocity, acceleration\\
    $\bm{\ju}_\mathrm{e}\theta_\mathrm{e}$,     & angular error\\
    $R =\exp({-\bm{\ju}\theta/2})$                & rotor representing orientation\\
    $R_\mathrm{d} =\exp({-\bm{\ju}_\mathrm{d}\theta_\mathrm{d}/2})$      &  rotor representing desired orientation\\
    $R_\mathrm{e} =\exp({-\bm{\ju}_\mathrm{e}\theta_\mathrm{e}/2})$ & rotor representing orientation error\\
    $R_\mathrm{p} =\exp({-\bm{\ju}_\mathrm{p}\theta_\mathrm{p}/2})$ & rotor representing preferred orientation\\
    \bottomrule
    \end{tabular}}
    \label{tab:my_label}
\end{table}

\subsection{GA Model}

\subsubsection{Conventions}

To construct a coordinate-free model using GA, we take the approach of choosing a reference state, then describe other states relative to this reference. In particular, we imagine a ``reference copy'' of the body at the points in space that align the inertial and body frames. Geometric modeling is then done with respect to this reference copy of the body. The reference copy also allows for a further level of abstraction---we can think of the GA objects (such as rotors, force vectors, torque bivectors, ect.) as acting on the reference copy of the body, rather than acting on the frame.

The inertial frame is represented using the set of orthonormal vectors, $\{\e_1,\e_2,\e_3\}$. The frame is oriented in a North-East-Down (NED) configuration, where $\e_1$ points North, $\e_2$ points East, and $\e_3$ points Down. The body frame is represented using another set of orthonormal vectors, $\{\bv_1,\bv_2,\bv_3\}$ where
\begin{IEEEeqnarray}{c} 
    \bv_k = R\e_k R^{\dagger},~~k= 1,2,3.
\end{IEEEeqnarray}
When the frames are perfectly aligned, the body frame is in a NED orientation. 
In the following, we will use symbols summarized in Table~\ref{tab:my_label}.


\subsubsection{Equation of Translational Motion in $3$D Space}
The equation of motion of an object of mass $m$ in a $3$D space subjected to a gravity acceleration $g$, which is propelled by thrust $\thrust\bft$ and perturbed by an additive perturbation $\wind$, is given by the differential equation
\begin{equation}
    m\xiDDot =  m g \e_3 - \thrust\bft {+\wind}.
    \label{eq:TranslationalDynamicsGeneral}
\end{equation}
where $\bft= R \turnt R^\dagger$ is a unit vector in the body frame with $\turnt$ a fixed unit vector in the inertial frame and $\thrust$ is the coefficient of thrust.

For control purposes, it is relevant to obtain the corresponding system propulsion from the desired acceleration as
\begin{equation}
    m\xiDDot_\mathrm{d} =  m g \e_3 - \thrust_\mathrm{d}\bft_\mathrm{d}
    \label{eq:TranslationalDynamicsine2}
\end{equation}
where subscript $\mathrm{d}$ is for the desired variables. The above relation can be rewritten w.r.t. the inertial frame as 
\begin{equation}
    m\ddot{\bfxi}_\mathrm{d} =  m g \e_3 - \thrust_\mathrm{d} R_\mathrm{d} \turnt R_\mathrm{d}^\dagger.
    \label{eq:TranslationalDynamicsine3}
\end{equation}
The coefficient of thrust $\thrust$ can be found by rearranging the terms in \eqref{eq:TranslationalDynamicsine3}, applying the norm to both sides, and recognizing the rotations do not affect the norm
\begin{IEEEeqnarray}{rCl}\IEEEyesnumber
    -m(\ddot{\bfxi}_\mathrm{d} - g \e_3)         &=&  \thrust_\mathrm{d} R_\mathrm{d} \turnt R_\mathrm{d}^\dagger\IEEEyessubnumber\label{eq: computerotref}\\
    \norm{-m(\ddot{\bfxi}_\mathrm{d} - g \e_3)}  &=&  \norm{\thrust_\mathrm{d} R_\mathrm{d} \turnt R_\mathrm{d}^\dagger}\IEEEyessubnumber\\
     m\norm{\ddot{\bfxi}_\mathrm{d} - g \e_3}    &=& \thrust_\mathrm{d} \IEEEyessubnumber.\label{eq:ThrustDesired} 
\end{IEEEeqnarray}  
Then, from \eqref{eq: computerotref} we obtain the orientation and direction of the desired propulsion in the inertial frame. If $\thrust_\mathrm{d}\neq 0$ then
\begin{equation}
    \frac{ -m}{\thrust_\mathrm{d}}(\xiDDot_\mathrm{d} - g\e_3 ) = R_\mathrm{d} \turnt R_\mathrm{d}^\dagger
\end{equation}
and when $\thrust_\mathrm{d}=0$ any $R_\mathrm{d}\in\mathbb{R}\oplus\mathbb{B}^3$ is valid. 

\begin{figure*}
    \centering
    \begin{subcaptionblock}{.4\textwidth}
        \centering
        {    \tdplotsetmaincoords{105}{35}
{\scalebox{0.7}{\begin{tikzpicture}[draw=white,tdplot_main_coords,font=\sffamily,scale=2.25]
   
    \node[inner sep=0pt,minimum size=0pt] (E1) at (1,0,0)  {};
    \node[inner sep=0pt,minimum size=0pt] (E2) at (0,1,0)  {};
    \node[inner sep=0pt,minimum size=0pt] (E3) at (0,0,1)  {};


    \node[inner sep=0pt,minimum size=0pt] (eFrame) at (-0.25,0,0.25)  {Reference State};
    \draw[opacity = 0.6,myDarkOrange,-stealth,line width=1.] (0,0,0)-- (1,0,0) node[pos=1.1] {$\textcolor{myDarkOrange}{\mathbf{e}_1}$};
    \draw[opacity = 0.6,myDarkOrange,-stealth,line width=1.] (0,0,0) -- (0,1,0) node[pos=1.2] {$\textcolor{myDarkOrange}{\mathbf{e}_2}$};
    \draw[opacity = 0.6,myDarkOrange,-stealth,line width=1.] (0,0,0) -- (0,0,-1) node[pos=1.1] {$\textcolor{myDarkOrange}{\mathbf{e}_3}$};
    \begin{scope}[rotate around x=-55]
        \draw[opacity = 0.6,myDarkOrange,dashed,-stealth] (0,0,0) -- (0,0,-1) node[pos=1.1] {$\textcolor{myDarkOrange}{\turnt}\textcolor{black}{=}\textcolor{black}{\alpha_1}\textcolor{myDarkOrange}{\mathbf{e}_1} \textcolor{black}{+}\textcolor{black}{\alpha_2}\textcolor{myDarkOrange}{\mathbf{e}_2}\textcolor{black}{+}\textcolor{black}{\alpha_3}\textcolor{myDarkOrange}{\mathbf{e}_3}$};
    \end{scope}


\begin{scope}[shift={(2,.5,0)}]
    \begin{scope}[rotate around x=10,rotate around y=-10]
        \begin{scope}[rotate around z=10]
            \node[inner sep=0pt,minimum size=0pt] (eFrame) at (0.25,0,0.3)  {Desired State};
            \draw[opacity = 0.6, myCrimson,-stealth,line width=1.] (0,0,0)-- (1,0,0) node[pos=1.2] {$\mathbf{b}_{1\mathrm{d}}$};
            \draw[opacity = 0.6,myCrimson,-stealth,line width=1.] (0,0,0) -- (0,1,0) node[pos=1.4] {$\mathbf{b}_{2\mathrm{d}}$};
            \draw[opacity = 0.6,myCrimson,-stealth,line width=1.] (0,0,0) -- (0,0,-1) node[pos=1.1] {$\mathbf{b}_{3\mathrm{d}}$};
            \begin{scope}[rotate around x=-55]
                \draw[opacity = 0.6,myCrimson,dashed,-stealth] (0,0,0) -- (0,0,-1) node[pos=1.1] {$\bft_{\mathrm{d}}
                \textcolor{black}{=}
                \textcolor{black}{\alpha_1}\mathbf{b}_{1\mathrm{d}} \textcolor{black}{+}\textcolor{black}{\alpha_2}\mathbf{b}_{2\mathrm{d}}\textcolor{black}{+}\textcolor{black}{\alpha_3}\mathbf{b}_{3\mathrm{d}}$};
            \end{scope}
        \end{scope} 
    \end{scope}      
\end{scope}


\begin{scope}[shift={(0,-.5,1.5)}]
 \begin{scope}[rotate around x=-15,rotate around y=0]
        \begin{scope}[rotate around z=10]
        \node[inner sep=0pt,minimum size=0pt] (eFrame) at (0,0,0.25)  {Actual State};
            \draw[opacity = 0.6,myDodgerBlue,-stealth,line width=1.] (0,0,0)-- (1,0,0) node[pos=1.1] {$\textcolor{myDodgerBlue}{\mathbf{b}_1}$};
            \draw[opacity = 0.6,myDodgerBlue,-stealth,line width=1.] (0,0,0) -- (0,1,0) node[pos=1.2] {$\textcolor{myDodgerBlue}{\mathbf{b}_2}$};
            \draw[opacity = 0.6,myDodgerBlue,-stealth,line width=1.] (0,0,0) -- (0,0,-1) node[pos=1.1] {$\textcolor{myDodgerBlue}{\mathbf{b}_3}$};
            \begin{scope}[rotate around x=-55]
                \draw[opacity = 0.6,myDodgerBlue,dashed,-stealth] (0,0,0) -- (0,0,-1) node[pos=1.9] {$\bft\textcolor{black}{=}\textcolor{black}{\alpha_1}\textcolor{myDodgerBlue}{\mathbf{b}_1} \textcolor{black}{+}\textcolor{black}{\alpha_2}\textcolor{myDodgerBlue}{\mathbf{b}_2}\textcolor{black}{+}\textcolor{black}{\alpha_3}\textcolor{myDodgerBlue}{\mathbf{b}_3}$};
            \end{scope}
        \end{scope} 
    \end{scope}       
\end{scope}



\draw[ black,-stealth,line width=1.5] ($(0,0,0)!.001!(2,.5,0)$)--($(0,0,0)!.999!(2,.5,0)$) ;
\draw[ myCrimson,-stealth,line width=1.1] ($(0,0,0)!.001!(2,.5,0)$)--($(0,0,0)!.9975!(2,.5,0)$) node[pos=0.5,below] {$\bfxi_\mathrm{d}$};

\draw[ black,-stealth,line width=1.5] ($(0,-.5,1.5)!.001!(2,.5,0)$)--($(0,-.5,1.5)!.999!(2,.5,0)$);
\draw[ myPurple,-stealth,line width=1.1] ($(0,-.5,1.5)!.001!(2,.5,0)$)--($(0,-.5,1.5)!.9975!(2,.5,0)$) node[pos=0.5,above] {$\bfxi_\mathrm{e}$};

\draw[ black,-stealth,line width=1.5] ($(0,0,0)!.001!(0,-.5,1.5)$)--($(0,0,0)!.999!(0,-.5,1.5)$) ;
\draw[ myDodgerBlue,-stealth,line width=1.1] ($(0,0,0)!.001!(0,-.5,1.5)$)--($(0,0,0)!.995!(0,-.5,1.5)$) node[pos=0.5,left] {$\bfxi$};

\end{tikzpicture}}}}
        \caption{}\label{fig:position}
    \end{subcaptionblock}%
    \begin{subcaptionblock}{.6\textwidth}
        \centering
        {    \tdplotsetmaincoords{105}{35}
{\scalebox{.7}{\begin{tikzpicture}[draw=white,tdplot_main_coords,font=\sffamily,scale=2.25]
   
    \node[inner sep=0pt,minimum size=0pt] (E1) at (1,0,0)  {};
    \node[inner sep=0pt,minimum size=0pt] (E2) at (0,1,0)  {};
    \node[inner sep=0pt,minimum size=0pt] (E3) at (0,0,1)  {};


    \node[inner sep=0pt,minimum size=0pt] (eFrame) at (0,0,0.25)  {Reference State};
    \draw[myDarkOrange,-stealth,line width=1.] (0,0,0)-- (1,0,0) node[pos=1.1] {$\textcolor{myDarkOrange}{\mathbf{e}_1}$};
    \draw[myDarkOrange,-stealth,line width=1.] (0,0,0) -- (0,1,0) node[pos=1.2] {$\textcolor{myDarkOrange}{\mathbf{e}_2}$};
    \draw[myDarkOrange,-stealth,line width=1.] (0,0,0) -- (0,0,-1) node[pos=1.1] {$\textcolor{myDarkOrange}{\mathbf{e}_3}$};
    \begin{scope}[rotate around x=-55]
        \draw[myDarkOrange,dashed,-stealth] (0,0,0) -- (0,0,-1) node[pos=1.1] {$\textcolor{myDarkOrange}{\turnt}$};
    \end{scope}
    

\begin{scope}[shift={(3,0.5,-0.3)}]%
    \begin{scope}[rotate around x=10,rotate around y=-10]
        \begin{scope}[rotate around z=10]
        \node[inner sep=0pt,minimum size=0pt] (eFrame) at (0,0,0.25)  {Desired State};
            \draw[myCrimson,-stealth,line width=1.] (0,0,0)-- (1,0,0) node[pos=1.5] {$\mathbf{b}_{1\mathrm{d}}\textcolor{black}{=} \textcolor{myGray}{R_\mathrm{d}}\textcolor{myDarkOrange}{\mathbf{e}_1}\textcolor{myGray}{R_\mathrm{d}^\dagger}\textcolor{black}{=} \textcolor{myGray}{\textcolor{myGray}{R_\mathrm{e}}}\textcolor{myDodgerBlue}{\mathbf{b}_1}\textcolor{myGray}{\textcolor{myGray}{R_\mathrm{e}^\dagger}}\textcolor{black}{=} \textcolor{myGray}{\textcolor{myGray}{R_\mathrm{a}}}\textcolor{myDeepPink}{\mathbf{b}_{1\mathrm{p}}}\textcolor{myGray}{\textcolor{myGray}{R_\mathrm{a}^\dagger}}$};
            \draw[myCrimson,-stealth,line width=1.] (0,0,0) -- (0,1,0) node[pos=2.5] {~~~~~~~~~~~~~~~~$\mathbf{b}_{2\mathrm{d}}\textcolor{black}{=} \textcolor{myGray}{R_\mathrm{d}}\textcolor{myDarkOrange}{\mathbf{e}_2}\textcolor{myGray}{R_\mathrm{d}^\dagger}\textcolor{black}{=} \textcolor{myGray}{\textcolor{myGray}{R_\mathrm{e}}}\textcolor{myDodgerBlue}{\mathbf{b}_2}\textcolor{myGray}{\textcolor{myGray}{R_\mathrm{e}^\dagger}}\textcolor{black}{=} \textcolor{myGray}{\textcolor{myGray}{R_\mathrm{a}}}\textcolor{myDeepPink}{\mathbf{b}_{2\mathrm{p}}}\textcolor{myGray}{\textcolor{myGray}{R_\mathrm{a}^\dagger}}$};
            \draw[myCrimson,-stealth,line width=1.] (0,0,0) -- (0,0,-1) node[pos=1.1] {$\mathbf{b}_{3\mathrm{d}}\textcolor{black}{=} \textcolor{myGray}{R_\mathrm{d}}\textcolor{myDarkOrange}{\mathbf{e}_3}\textcolor{myGray}{R_\mathrm{d}^\dagger}\textcolor{black}{=} \textcolor{myGray}{\textcolor{myGray}{R_\mathrm{e}}}\textcolor{myDodgerBlue}{\mathbf{b}_3}\textcolor{myGray}{\textcolor{myGray}{R_\mathrm{e}^\dagger}}\textcolor{black}{=} \textcolor{myGray}{\textcolor{myGray}{R_\mathrm{a}}}\textcolor{myDeepPink}{\mathbf{b}_{3\mathrm{p}}}\textcolor{myGray}{\textcolor{myGray}{R_\mathrm{a}^\dagger}}$};
            \begin{scope}[rotate around x=-55]
                \draw[myCrimson,dashed,-stealth] (0,0,0) -- (0,0,-1) node[pos=1.25] {$\bft_{\mathrm{d}}
                \textcolor{black}{=} \textcolor{myGray}{R_\mathrm{d}}\textcolor{myDarkOrange}{\turnt} \textcolor{myGray}{R_\mathrm{d}^\dagger}
                \textcolor{black}{=} \textcolor{myGray}{R_\mathrm{e}}\textcolor{myDodgerBlue}{\bft} \textcolor{myGray}{R_\mathrm{e}^\dagger}\textcolor{black}{=} \textcolor{myGray}{\textcolor{myGray}{R_\mathrm{a}}}\textcolor{myDeepPink}{\bft_{\mathrm{p}}}\textcolor{myGray}{\textcolor{myGray}{R_\mathrm{a}^\dagger}}$};
            \end{scope}
        \end{scope} 
    \end{scope}      
\end{scope}


\begin{scope}[shift={(0,0,2)}]
 \begin{scope}[rotate around x=-15,rotate around y=0]
        \begin{scope}[rotate around z=10]
        \node[inner sep=0pt,minimum size=0pt] (eFrame) at (0,0,0.25)  {Actual State};
            \draw[myDodgerBlue,-stealth,line width=1.] (0,0,0)-- (1,0,0) node[pos=1.55] {$\textcolor{myDodgerBlue}{\mathbf{b}_1} \textcolor{black}{=} \textcolor{myGray}{R}\textcolor{myDarkOrange}{\mathbf{e}_1}\textcolor{myGray}{R^\dagger}$};
            \draw[myDodgerBlue,-stealth,line width=1.] (0,0,0) -- (0,1,0) node[pos=1.25] {$\textcolor{myDodgerBlue}{\mathbf{b}_2} \textcolor{black}{=} \textcolor{myGray}{R}\textcolor{myDarkOrange}{\mathbf{e}_2}\textcolor{myGray}{R^\dagger}$};
            \draw[myDodgerBlue,-stealth,line width=1.] (0,0,0) -- (0,0,-1) node[pos=1.1] {$\textcolor{myDodgerBlue}{\mathbf{b}_3} \textcolor{black}{=} \textcolor{myGray}{R}\textcolor{myDarkOrange}{\mathbf{e}_3}\textcolor{myGray}{R^\dagger}$};
            \begin{scope}[rotate around x=-55]
                \draw[myDodgerBlue,dashed,-stealth] (0,0,0) -- (0,0,-1) node[pos=1.9] {~~~~~~~~$\bft\textcolor{black}{=} \textcolor{myGray}{R}\textcolor{myDarkOrange}{\turnt} \textcolor{myGray}{R^\dagger}$};
            \end{scope}
        \end{scope} 
    \end{scope}       
\end{scope}


\begin{scope}[shift={(3,0.25,1.75)}]
 \begin{scope}[rotate around x=0,rotate around y=0]
        \begin{scope}[rotate around z=0]
                \node[  inner sep=0pt,minimum size=0pt] (eFrame) at (0,0,0.25)  {Preferred Orientation};
            \draw[  myDeepPink,-stealth,line width=1.] (0,0,0)-- (1,0,0) node[pos=1.5] {$\mathbf{b}_{1\mathrm{p}}\textcolor{black}{=} \textcolor{myGray}{R_\mathrm{p}}\textcolor{myDarkOrange}{\mathbf{e}_1}\textcolor{myGray}{R_\mathrm{p}^\dagger}$};
            \draw[  myDeepPink,-stealth,line width=1.] (0,0,0) -- (0,1,0) node[pos=1.5] {$\mathbf{b}_{2\mathrm{p}}\textcolor{black}{=} \textcolor{myGray}{R_\mathrm{p}}\textcolor{myDarkOrange}{\mathbf{e}_1}\textcolor{myGray}{R_\mathrm{p}^\dagger}$};
            \draw[  myDeepPink,-stealth,line width=1.] (0,0,0) -- (0,0,-1) node[pos=1.1] {$\mathbf{b}_{3\mathrm{p}}\textcolor{black}{=} \textcolor{myGray}{R_\mathrm{p}}\textcolor{myDarkOrange}{\mathbf{e}_1}\textcolor{myGray}{R_\mathrm{p}^\dagger}$};
            \begin{scope}[rotate around x=-55]
                \draw[  myDeepPink,dashed,-stealth] (0,0,0) -- (0,0,-1) node[pos=1.1] {$\bft_{\mathrm{p}}$};
            \end{scope}
        \end{scope} 
    \end{scope}  
\end{scope}


\begin{scope}[shift={(0,-0.75,0.25)}]
    \draw [myDodgerBlue, -Straight Barb] (0,0,0) to [bend left=45] node[above, sloped] {$R$} (0,0,1);
    \draw [test={0.4pt}{myDarkOrange}{myDodgerBlue}] (0,0,0) to [bend left=45] node[above, sloped] {} (0,0,1);
\end{scope}

\begin{scope}[shift={(0,1.25,-0.25)}]
    \draw [myDeepPink, -Straight Barb] (3,0.25,1.75) to [bend left=45] node[above, sloped] {$R_\mathrm{a}$} (3,0.5,1);
    \draw [myDeepPink, -Straight Barb] (3,0.25,1.75) to [bend left=45] node[above, sloped] {$R_\mathrm{a}$} (3,0.5,1);
    \draw [myCrimson, -Straight Barb] (3,0.25,1.75) to [bend left=45] node[above, sloped] {} (3,0.5,1);
    \draw [test={0.4pt}{myDeepPink}{myCrimson}] (3,0.25,1.75) to [bend left=45] node[above, sloped] {} (3,0.5,1);
\end{scope}

\begin{scope}[shift={(0,0,0.4)}]
    \draw [myPurple] (0.75,0,1)to [bend left=33] node[above, sloped,very thin,pos=0.75] {$R_\mathrm{e}$} (2.25,0.5,0);
    \draw [myCrimson, -Straight Barb] (0.75,0,1)to [bend left=33] node[above, sloped] {} (2.25,0.5,0);
    \draw [test={0.4pt}{myDodgerBlue}{myCrimson}] (0.75,0,1)to [bend left=33] node[above, sloped] {} (2.25,0.5,0);
\end{scope}

\begin{scope}[shift={(-0.25,0,-0.25)}]
    \draw [myDeepPink, -Straight Barb] (1,0,0.5) to [bend right=33] node[above, sloped,pos=0.75] {$R_\mathrm{p}$} (2.5,0.25,2);
    \draw [test={0.4pt}{myDarkOrange}{myDeepPink}] (1,0,0.5)to [bend right=33] node[above, sloped] {} (2.5,0.25,2);
\end{scope}

\begin{scope}[shift={(0,0.3,0)}]
    \draw [myCrimson, -Straight Barb] (0.5,0,0) to [bend right=45] node[above, sloped] {$R_\mathrm{d}$} (2.25,0.5,0);
    \draw [test={0.4pt}{myDarkOrange}{myCrimson}] (0.5,0,0)to [bend right=45] node[above, sloped] {} (2.25,0.5,0);
\end{scope}

\end{tikzpicture}}}}
        \caption{}\label{fig:rotation}
    \end{subcaptionblock}%
\caption{Illustration of the relationship between system states and various elements for (a) translation and (b) rotation. All states are represented as rotated and translated copies of a reference state represented by the \textcolor{myDarkOrange}{$\{\e_k\}$} frame. In particular, the \textcolor{myDodgerBlue}{$\{\bv_k\}$} frame is a copy of the  \textcolor{myDarkOrange}{$\{\e_k\}$} frame rotated by \textcolor{myDodgerBlue}{$R$} and translated by \textcolor{myDodgerBlue}{$\bfxi$}. The \textcolor{myCrimson}{$\{\bv_{k\mathrm{d}}\}$} frame is a copy of the  \textcolor{myDarkOrange}{$\{\e_k\}$} frame rotated by \textcolor{myCrimson}{$R_\mathrm{d}$} and translated by \textcolor{myCrimson}{$\bfxi_\mathrm{d}$}, but is also represented as a copy of the \textcolor{myDodgerBlue}{$\{\bv_k\}$} frame rotated by \textcolor{myPurple}{$R_\mathrm{e}$} and translated by \textcolor{myPurple}{$\bfxi_\mathrm{e}$}. The \textcolor{myHotPink}{$\{\bv_{k\mathrm{p}}\}$} frame is a copy of the  \textcolor{myDarkOrange}{$\{\e_k\}$} frame rotated by \textcolor{myDeepPink}{$R_\mathrm{p}$} and translated by \textcolor{myCrimson}{$\bfxi_\mathrm{d}$}. The \textcolor{myDarkOrange}{$\{\e_k\}$} frame is chosen to align  \textcolor{myDarkOrange}{$\e_3$} with the direction of acceleration due to gravity (additionally, aligning \textcolor{myDarkOrange}{$\e_1$} with North forms a NED frame) and the \textcolor{myDodgerBlue}{$\{\bv_k\}$} frame is assumed to align with the the principal moments of inertia of the mass distribution of the body. 
}
\label{fig:GARotations}
\end{figure*}

\subsubsection{Equation of Rotational Motion in $3$D Space}
Defining the commutator product, denoted by $[\cdot,\cdot]$, as 
\begin{IEEEeqnarray}{c} 
    [A,B] \triangleq {\textstyle\frac{1}{2}}(AB-BA),\label{eq:commutator}
\end{IEEEeqnarray}
the rotational equation of motion can be expressed as  
\begin{IEEEeqnarray}{c} 
     R\Big(\mathcal{I}(\dot{\bm{\Omega}}_\mathrm{b}) - \big[\bm{\Omega}_\mathrm{b}, \mathcal{I}(\bm{\Omega}_\mathrm{b})\big]\Big)R^\dagger = \Tau 
\end{IEEEeqnarray}
where the bivector $\Tau$ is the total torque on the body, $\mathcal{I}(\mathbf{B})$ is the inertia tensor with respect to mass density $\rho$\cite{doran2003geometric}, with structure
\begin{IEEEeqnarray}{c} 
    \mathcal{I}(\mathbf{B}) = \int\dd^3x\rho\mathbf{x}\wedge(\mathbf{x}\cdot\mathbf{B}),
\end{IEEEeqnarray}
and $\bm{\Omega}_\mathrm{b}$ is the body angular velocity bivector given by
\begin{IEEEeqnarray}{c} 
    \bm{\Omega}_\mathrm{b} = R^\dagger\bm{\Omega} R.
\end{IEEEeqnarray}

The total external torque $\Tau$ may be expressed in terms of the inertial frame or projected onto the principal planes of the body as
\begin{IEEEeqnarray}{rCl} \IEEEyesnumber
    \Tau &=&  \smallT_{12}\e_1\e_2 + \smallT_{23} \e_2\e_3 + \smallT_{31} \e_3\e_1\IEEEyessubnumber\\
    &=& \tau_{12} \bv_1\bv_2 + \tau_{23} \bv_2\bv_3 + \tau_{31} \bv_3\bv_1.\IEEEyessubnumber
\end{IEEEeqnarray}

In terms of the body angular velocity $\bm{\Omega}_\mathrm{b}$, 
the time derivative of the rotor $R$ is
\begin{IEEEeqnarray}{rCl} 
    \dot{R} &=& -{\textstyle\frac{1}{2}}\bm{\Omega}R = -{\textstyle\frac{1}{2}}R\bm{\Omega}_\mathrm{b} 
\end{IEEEeqnarray}
and its reverse
\begin{IEEEeqnarray}{rCl} 
    \textstyle\dot{R}^\dagger &=&  -{\textstyle\frac{1}{2}} R^\dagger \bm{\Omega} = -{\textstyle\frac{1}{2}}\bm{\Omega}_\mathrm{b}R^\dagger. 
\end{IEEEeqnarray}
A rotor representing the orientation of the system may be expressed in terms of Euler angles $\roll$ (roll), $\pitch$ (pitch), and $\yaw$ (yaw) as \cite{doran2003geometric}  
\begin{IEEEeqnarray*}{c} 
    R = \exp(-\e_1\e_2\yaw/2)\exp(-\e_2\e_3\pitch/2)\exp(-\e_1\e_2\roll/2).\label{eq:RtoEuler}\IEEEyesnumber\IEEEeqnarraynumspace
\end{IEEEeqnarray*}

%
%

Solving in the equation of motion for rotational acceleration w.r.t the body angular acceleration $\dot{\bm{\Omega}}_\mathrm{b}$, yields 
\begin{IEEEeqnarray}{rCl} 
    \mathcal{I}(\dot{\bm{\Omega}}_\mathrm{b}) \!-\! \big[\bm{\Omega}_\mathrm{b}, \mathcal{I}(\bm{\Omega}_\mathrm{b})\big] &=& R^\dagger\Tau R  \IEEEyessubnumber\\
    \mathcal{I}(\dot{\bm{\Omega}}_\mathrm{b}) &=&   R^\dagger\Tau R + \big[\bm{\Omega}_\mathrm{b}, \mathcal{I}(\bm{\Omega}_\mathrm{b})\big]\IEEEyessubnumber\\
    \dot{\bm{\Omega}}_\mathrm{b} &=&  \mathcal{I}^{-1}\!\Big(\! R^\dagger\Tau R +\! \big[\bm{\Omega}_\mathrm{b}, \mathcal{I}(\bm{\Omega}_\mathrm{b})\big]\Big). \IEEEyessubnumber\label{eq:rotationaldynamics}~~~~~~
\end{IEEEeqnarray}

For symmetric systems with principle planes of rotation $\mathbf{b}_1\mathbf{b}_2$, $\mathbf{b}_2\mathbf{b}_3$, and $\mathbf{b}_3\mathbf{b}_1$, the effect of the inertial tensor $\mathcal{I}(\mathbf{B})$ on an arbitrary bivector $\mathbf{B}$ can be computed by decomposing $\mathbf{B}$ in terms of the principle planes of rotation 
\begin{IEEEeqnarray}{rCl} 
    \mathbf{B} &=&   b_{12}\mathbf{b}_1\mathbf{b}_2 + b_{23}\mathbf{b}_2\mathbf{b}_3 +  b_{31}\mathbf{b}_3\mathbf{b}_1
\end{IEEEeqnarray}
from where it follows that 
\begin{align}
    \mathcal{I}(\mathbf{B}) =&  \iota_{12} b_{12}\mathbf{b}_1\mathbf{b}_2 + \iota_{23}b_{23}\mathbf{b}_2\mathbf{b}_3 +  \iota_{31}b_{31}\mathbf{b}_3\mathbf{b}_1 \label{eq:GAinteralTensor}
\end{align}
and 
\begin{align}
    \mathcal{I}^{-1}(\mathbf{B}) =&   \frac{b_{12}}{\iota_{12}}\mathbf{b}_1\mathbf{b}_2 + \frac{b_{23}}{\iota_{23}}\mathbf{b}_2\mathbf{b}_3 +  \frac{b_{31}}{\iota_{31}}\mathbf{b}_3\mathbf{b}_1  \label{eq:momentsinertia}
\end{align}
where $\iota_{12}$, $\iota_{23}$, and $\iota_{31}$ are the principal moments of inertia.

\subsubsection{Determining the Error Rotor}

Because the system is underactuated, the desired thrust direction $\bft_\mathrm{d}$ from the translational tracking problem may conflict with the preferred orientation $R_\mathrm{p}$ provided as input. In this work, we prioritize the translational tracking problem, and, as a result, we define a desired rotation $R_\mathrm{d}$ which satisfies the translational tracking objective, but is also as close to the preferred orientation $R_\mathrm{p}$ as possible.

Therefore, given a thrust direction $\bft$, a desired thrust direction $\bft_\mathrm{d}$, and preferred orientation $R_\mathrm{p}$, we seek a desired rotation $R_\mathrm{d}$ and the corresponding error rotor $R_\mathrm{e}$. First, we find the direction the thrust \textit{would be} pointing in, if the body were described by the preferred orientation rotor $R_\mathrm{p}$ as 
\begin{IEEEeqnarray}{c}
   R_\mathrm{p} \turnt  R_\mathrm{p}^\dagger =  \bft_\mathrm{p} 
   \label{eqn:ThrustWouldBe}
\end{IEEEeqnarray}
and then compute the rotor which would align $\bft_\mathrm{p}$ to the desired thrust direction $\bft_\mathrm{d}$ as
\begin{IEEEeqnarray}{c}
    R_\mathrm{a} = (\bft_\mathrm{p} \bft_\mathrm{d})^{-1/2}
\end{IEEEeqnarray}
therefore
\begin{IEEEeqnarray}{c}
   R_\mathrm{a} \bft_\mathrm{p}  R_\mathrm{a}^\dagger =  \bft_\mathrm{d}.
\end{IEEEeqnarray}
Then, the desired body frame vectors can be expressed as 
\begin{IEEEeqnarray}{rCl}\IEEEyesnumber\IEEEyessubnumber
   R_\mathrm{a}R_\mathrm{p} \mathbf{e}_k  R_\mathrm{p}^\dagger R_\mathrm{a}^\dagger &=&  \bv_{k\mathrm{d}}\\ 
   &=&   R_\mathrm{d} \e_k  R_\mathrm{d}^\dagger \IEEEyessubnumber
\end{IEEEeqnarray}
and we can express the desired rotation $R_\mathrm{d}$ as a composition of rotations as
\begin{IEEEeqnarray}{c}
   R_\mathrm{a}R_\mathrm{p} =
   R_\mathrm{d}.
\end{IEEEeqnarray}
Finally, the error rotor is given by
\begin{IEEEeqnarray}{c}
   R_\mathrm{a}R_\mathrm{p}R^\dagger =
   R_\mathrm{e}
\end{IEEEeqnarray}
so that
\begin{IEEEeqnarray}{rCl}\IEEEyesnumber\label{eq:errorRot}\IEEEyessubnumber
   R_\mathrm{a}R_\mathrm{p}R^\dagger \bv_k R R_\mathrm{p}^\dagger R_\mathrm{a}^\dagger &=& \bv_{k\mathrm{d}}\\
   &=&   R_\mathrm{e} \bv_k  R_\mathrm{e}^\dagger \IEEEyessubnumber
\end{IEEEeqnarray}
Finally, the orientation error may be expressed as the exponential of angular error as
\begin{IEEEeqnarray}{rCl}\IEEEyesnumber
    R_\mathrm{e}                                        &=&\exp({-\bm{\ju}_\mathrm{e}\theta_\mathrm{e}/2})\label{eq:biVecAngle}.
\end{IEEEeqnarray}
Therefore, the desired rotation $R_\mathrm{d}$ is the rotation which aligns the thrust vector $\bft$ with the desired thrust vector $\bft_\mathrm{d}$, and which is as close to the preferred orientation $R_\mathrm{p}$ as possible. Moreover, it is achieved by performing the rotation described by $R_\mathrm{e}$ to the body. 

In the special case that a preferred orientation $R_\mathrm{p}$ is not given, the problem is under-determined, but one solution may be found by forming 
\begin{IEEEeqnarray}{rCl}
    R_\mathrm{d} &=& (\turnt \bft_\mathrm{d})^{-1/2}\\
    R_\mathrm{e} &=& (\bft\bft_\mathrm{d})^{-1/2}
\end{IEEEeqnarray}
so that
\begin{IEEEeqnarray}{rCl}
   R_\mathrm{d} \turnt  R_\mathrm{d}^\dagger &=&  \bft_\mathrm{d}\\
   R_\mathrm{e} \bft R_\mathrm{e}^\dagger &=&  \bft_\mathrm{d}
\end{IEEEeqnarray}
The aforementioned rotations are summarized in Fig.~\ref{fig:GARotations}.

\subsection{Derivatives of the Desired Rotation}
For controlling the rotational sub-system, we need to know the first and second order derivatives of the desired rotation in the inertial frame. Hereinafter, we show that both derivatives can be obtained analytically using GA properties.

\subsubsection{Desired Angular Velocities}
To determine the desired angular velocity $\bfOmega _\mathrm{d}$, take the analytical derivative of $R_\mathrm{d} =\exp({-\bm{\ju}_\mathrm{d}\theta_\mathrm{d}/2})$
\begin{IEEEeqnarray}{rCl}\IEEEyesnumber
                 \dot{R}_\mathrm{d}
                 &=&\frac{\dd}{\dd t}\left(\frac{-\bm{\ju}_\mathrm{d}\theta_\mathrm{d}}{2}\right)R_\mathrm{d}\label{eq:desiredRotor}
\end{IEEEeqnarray}
and define the desired angular velocity bivector as
\begin{IEEEeqnarray}{rCl}\IEEEyesnumber
    \bm{\Omega}_\mathrm{d}  &=&  \frac{\dd}{\dd t}\left(\bm{\ju}_\mathrm{d}\theta_\mathrm{d}\right).\label{eq:angularBivec}
\end{IEEEeqnarray}
Then, using \eqref{eq:angularBivec} in \eqref{eq:desiredRotor} yields
\begin{IEEEeqnarray}{rCl}\IEEEyesnumber
                 \dot{R}_\mathrm{d}
                 &=&-\frac{1}{2}\bm{\Omega}_\mathrm{d}R_\mathrm{d} \label{eq:derivativedesiredrotor}
\end{IEEEeqnarray}
and solving gives   
\begin{IEEEeqnarray}{rCl}\IEEEyesnumber
    \bm{\Omega}_\mathrm{d}                                      
                &=&-2\dot{R}_\mathrm{d}R_\mathrm{d}^\dagger \label{eq:Omegadesired}
\end{IEEEeqnarray}
where $R_\mathrm{d}$ is defined in \eqref{eq:TranslationalDynamicsine3} and $\dot{R}_\mathrm{d}$ is given by 
\begin{IEEEeqnarray}{rCl}
\dot{R}_\mathrm{d}&=&\dot{R}_\mathrm{a}R_\mathrm{p}+R_\mathrm{a}\dot{R}_\mathrm{p} \label{eq:dotRd}
    \end{IEEEeqnarray} 
with
    \begin{IEEEeqnarray}{rCl}\IEEEyesnumber
\dot{R}_\mathrm{a}&=&-\frac{1}{2}\frac{\dot{\bft}_\mathrm{p}\bft_\mathrm{d}+\bft_\mathrm{p}\dot{\bft}_\mathrm{d}}{(\bft_\mathrm{p}\bft_\mathrm{d})^{3/2}} \label{eq:dotRa}\\
\dot{\bft}_\mathrm{p}&=&\dot{R}_\mathrm{p}\turnt R_\mathrm{p}^\dagger+R_\mathrm{p}\turnt \dot{R}_\mathrm{p}^\dagger + R_\mathrm{p}\dot{\turnt} R_\mathrm{p}^\dagger\label{eq:dottp}
\end{IEEEeqnarray}
Finally, to compute $\dot{\bft}_\mathrm{d}$, take the derivative of \eqref{eq:TranslationalDynamicsine2}
\begin{IEEEeqnarray}{rCl}\IEEEyesnumber
    \frac{\dd (\thrust_\mathrm{d}\bft_\mathrm{d})}{\dd t}   &=&  -m\dddot{\bm{\xi}}_\mathrm{d}\IEEEyessubnumber \label{eq:3DerivDesiRef}\\
    \frac{\dd (\thrust_\mathrm{d}\bft_\mathrm{d})}{\dd t}   &=&  \dot{\thrust}_\mathrm{d}\bft_\mathrm{d}+\thrust_\mathrm{d}\dot{\bft}_\mathrm{d}\IEEEyessubnumber\\
     \dot{\bft}_\mathrm{d}                                   &=&\frac{\dd(\thrust_\mathrm{d}\bft_\mathrm{d})}{\dd t}\frac{1}{\thrust_\mathrm{d}}-\frac{\dot{\thrust}_\mathrm{d} \bft_\mathrm{d}}{\thrust_\mathrm{d}}\IEEEyessubnumber\label{eq:bVecDotDes}
\end{IEEEeqnarray} 
with $\thrust_\mathrm{d}$ from \eqref{eq:ThrustDesired} and $\dot{\thrust}_\mathrm{d}$ computed from the square root of the square of the norm as
   \begin{IEEEeqnarray}{rCl}\IEEEyesnumber 
    \dot{\thrust}_\mathrm{d}                                    &=&\frac{\dd}{\dd t} m\sqrt{\big(\norm{g\e_3-\ddot{\bm{\xi}}_\mathrm{d}}\big)^2}\IEEEyessubnumber\\
    &=&  m\frac{(\ddot{\bm{\xi}}_\mathrm{d}-g\e_3) \cdot \dddot{\bm{\xi}}_\mathrm{d}}{\norm{g\e_3-\ddot{\bm{\xi}}_\mathrm{d}}}.\IEEEyessubnumber \label{eq:bDotDes}
\end{IEEEeqnarray}

\subsubsection{Desired Angular Acceleration}
To determine the desired angular acceleration $\bfOmegaDot _\mathrm{d}$, take the derivative of \eqref{eq:Omegadesired} and note that differentiation and reversion commute.
    \begin{IEEEeqnarray}{rCl}\IEEEyesnumber
      \dot{\bm{\Omega}}_\mathrm{d}                                    &=&-2\ddot{R}_\mathrm{d}R_\mathrm{d}^\dagger-2\dot{R}_\mathrm{d}\dot{R}_\mathrm{d}^\dagger \label{eq:Omegaddotesired}
\end{IEEEeqnarray}
where $R_\mathrm{d}$ and $\dot{R}_\mathrm{d}$ are defined in \eqref{eq:TranslationalDynamicsine3} and \eqref{eq:dotRd}, and $\ddot{R}_\mathrm{d}$ is computed as
\begin{IEEEeqnarray}{rCl}
\ddot{R}_\mathrm{d}&=&\ddot{R}_\mathrm{a}R_\mathrm{p}+2\dot{R}_\mathrm{a}\dot{R}_\mathrm{p}+R_\mathrm{a}\ddot{R}_\mathrm{p} \label{eqddotRd}
    \end{IEEEeqnarray} 
and also
\begin{IEEEeqnarray}{rCl}\IEEEyesnumber
    \ddot{R}_\mathrm{a}                                             &=&-\frac{1}{2}\frac{\ddot{\bft}_\mathrm{p}\bft_\mathrm{d}+2\dot{\bft}_\mathrm{p}\dot{\bft}_\mathrm{d}+\bft_\mathrm{p}\ddot{\bft}_\mathrm{d}}{(\bft_\mathrm{p} \bft_\mathrm{d})^{3/2}} +\frac{3}{4}\frac{(\dot{\bft}_\mathrm{p}\bft_\mathrm{d}+\bft_\mathrm{p}\dot{\bft}_\mathrm{d})^2}{(\bft_\mathrm{p} \bft_\mathrm{d})^{5/2}}\label{eqRDDotdes}\\
    \ddot{\bft}_\mathrm{p}&=&\ddot{R}_\mathrm{p}\turnt R_\mathrm{p}^\dagger+R_\mathrm{p}\turnt \ddot{R}_\mathrm{p}^\dagger+R_\mathrm{p}\ddot{\turnt} R_\mathrm{p}^\dagger \nonumber\\
    &~& \quad + 2\dot{R}_\mathrm{p}\dot{\turnt} R_\mathrm{p}^\dagger+ 2R_\mathrm{p}\dot{\turnt} \dot{R}_\mathrm{p}^\dagger+ 2\dot{R}_\mathrm{p}\turnt \dot{R}_\mathrm{p}^\dagger\label{eqddottp}
\end{IEEEeqnarray}
To compute $\ddot{\thrust}_\mathrm{d}$ and $\ddot{\bft}_\mathrm{d}$, take the second derivative of \eqref{eq:TranslationalDynamicsine2}
\begin{IEEEeqnarray}{rCl}\IEEEyesnumber
    \frac{\dd^2 (\thrust_\mathrm{d}\bft_\mathrm{d})}{\dd t^2}   &=&-m\ddddot{\bm{\xi}}_\mathrm{d} \IEEEyessubnumber \label{eq:4DerivDesiRef}\\
    \frac{\dd^2 (\thrust_\mathrm{d}\bft_\mathrm{d})}{\dd t^2}   &=&\ddot{\thrust}_\mathrm{d}\bft_\mathrm{d}+2\dot{\thrust}_\mathrm{d}\dot{\bft}_\mathrm{d}+\thrust_\mathrm{d}\ddot{\bft}_\mathrm{d}.\IEEEyessubnumber
\end{IEEEeqnarray}
Then, solving gives
\begin{IEEEeqnarray}{rCl}\IEEEyesnumber
        \ddot{\bft}_\mathrm{d}                                      &=&\frac{\dd^2 (\thrust_\mathrm{d}\bft_\mathrm{d})}{\dd t^2}\frac{1}{\thrust_\mathrm{d}}-\frac{2\dot{\thrust}_\mathrm{d}\dot{\bft}_\mathrm{d}}{\thrust_\mathrm{d}}-\frac{\ddot{\thrust}_\mathrm{d}\bft_\mathrm{d}}{\thrust_\mathrm{d}}\label{eq:bVecDDotDes}
\end{IEEEeqnarray}
and  
\begin{IEEEeqnarray}{rCl}\IEEEyesnumber
    \frac{\ddot{\thrust}_\mathrm{d}}{m}  &=& \frac{\dddot{\bm{\xi}}_\mathrm{d}^2+(\ddot{\bm{\xi}}_\mathrm{d}-g\e_3)\cdot \ddddot{\bm{\xi}}_\mathrm{d}}{\norm{g\e_3 -\ddot{\bm{\xi}}_\mathrm{d}}} -\frac{\big((\ddot{\bm{\xi}}_\mathrm{d}-g\e_3)\cdot \dddot{\bm{\xi}}_\mathrm{d}\big)^2}{\norm{g\e_3-\ddot{\bm{\xi}}_\mathrm{d}}^3}\label{eq:bDDotDes}
\end{IEEEeqnarray}

\subsection{Rotational Error for Underactuated Systems}

When a system is not capable of generating the desired propulsion $\bft_\mathrm{d}$ in  \eqref{eq:TranslationalDynamicsine2} instantaneously, as in the case of underactuated systems, 
the desired rotation is governed by a dynamic process that must be followed to achieve the target orientation.
Thus, assuming an actual rotation different than the desired rotation 
\begin{IEEEeqnarray}{rCl}\IEEEyesnumber
    m\xiDDot_\mathrm{d} &=&  m g \e_3 - \thrust_\mathrm{d}\bft \IEEEyessubnumber\\
                        &=&m g \e_3 - \thrust_\mathrm{d}\bft +\thrust_\mathrm{d}\bft_\mathrm{d}-\thrust_\mathrm{d}\bft_\mathrm{d} \IEEEyessubnumber\\
                        &=&  m g \e_3 - \thrust_\mathrm{d}\bft_\mathrm{d}+\thrust_\mathrm{d}(\bft_\mathrm{d}-\bft) \IEEEyessubnumber \label{eq:rotationerrorinb}
\end{IEEEeqnarray}
where the bivector angle error 
\begin{IEEEeqnarray}{rCl}\IEEEyesnumber
    \bft_\mathrm{d}-\bft                       &=&R_\mathrm{d}\turnt R_\mathrm{d}^\dagger-R\turnt R^{\dagger}\IEEEyessubnumber\\
    R^\dagger(\bft_\mathrm{d}-\bft)R            &=&R^\dagger R_\mathrm{d}\turnt R_\mathrm{d}^\dagger R-\turnt \IEEEyessubnumber \label{eq:RotationError}\\
    \norm{R^\dagger(\bft_\mathrm{d}-\bft)R}     &=&\norm{R^\dagger R_\mathrm{d}\turnt R_\mathrm{d}^\dagger R-\turnt}\IEEEyessubnumber\\
    \norm{\bft_\mathrm{d}-\bft}                 &=&\norm{R_\mathrm{e}\turnt R_\mathrm{e}^\dagger-\turnt }\IEEEyessubnumber \label{eq:turnberror} \\
    R_\mathrm{e}                                        &=&R^\dagger R_\mathrm{d}\IEEEyessubnumber \\
    R_\mathrm{e}                                        &=&\exp({-\bm{\ju}_\mathrm{e}\theta_\mathrm{e}/2})  \IEEEyessubnumber \label{eq:similar}
\end{IEEEeqnarray}
In \eqref{eq:similar}, we obtain an expression for the rotational error similar to the one in~\cite{fresk2013full}, however, in this work we go further and we obtain the bivector $\bm{\ju}_\mathrm{e}\theta_\mathrm{e}$ representing  the plane and extent of error as
\begin{equation}
\bm{\ju}_\mathrm{e}\theta_\mathrm{e}=-2\,\mathrm{acos}\big(\grade{R_\mathrm{e}}_0 \big) \grade{R_\mathrm{e}}_2 \norm{\grade{R_\mathrm{e}}_2}^{-1}.
\label{eq:JthetaDesired}
\end{equation}

\subsubsection{Rotational Error Bounds}
We are interested in the rotational error bounds. Using \eqref{eq:turnberror} we can decomposed the vector $\turnt$ in the projections of the inertial frame $\turnt_i=\alpha_i\e_i$ for $i=1,2,3$ as
\begin{IEEEeqnarray}{rCl}\IEEEyesnumber
    \norm{\bft_\mathrm{d}-\bft}   &=&\norm{R_\mathrm{e}\turnt_1 R_\mathrm{e}^\dagger+ R_\mathrm{e}\turnt_2 R_\mathrm{e}^\dagger+R_\mathrm{e}\turnt_3 R_\mathrm{e}^\dagger \IEEEyessubnumber \nonumber\\
                                &&\quad-\turnt_1-\turnt_2-\turnt_3}\IEEEyessubnumber
\end{IEEEeqnarray}
Using the triangle inequality
\begin{IEEEeqnarray}{rCl}\IEEEyesnumber
    \norm{\bft_\mathrm{d}-\bft}   &\leq& \sum_{i} \norm{R_\mathrm{e}\turnt_i R_\mathrm{e}^\dagger-\turnt_i}\quad i=1,2,3\\
                                &\leq&3\norm{R_\mathrm{e}\turnt_\mathrm{m} R_\mathrm{e}^\dagger-\turnt_\mathrm{m}}=3\norm{\bar{a}_m}\norm{R_\mathrm{e}\e_\mathrm{m} R_\mathrm{e}^\dagger-\e_\mathrm{m}}  \IEEEnonumber
                                \label{berrorfunctm}
\end{IEEEeqnarray}
with $\norm{R_\mathrm{e}\turnt_\mathrm{m} R_\mathrm{e}^\dagger-\turnt_\mathrm{m}}=\max\left( \norm{R_\mathrm{e}\turnt_i R_\mathrm{e}^\dagger-\turnt_i}\right)$, ${\turnt_i=\bar{a}_i\e_i}$ for $i=1,2,3$ with scalars $\bar{a}_i$ and $\bar{a}_m=\max(\bar{a}_i)$. In Euclidean spaces, all unit vectors square to unity, thus we compute the rotation error as
\begin{IEEEeqnarray}{rCl}\IEEEyesnumber    
\norm{R_\mathrm{e}\e_i R_\mathrm{e}^\dagger-\e_i}^2&=&\norm{\eu^{-\bm{\ju}_\mathrm{e}\theta_\mathrm{e}/2}\e_i \eu^{\,\bm{\ju}_\mathrm{e}\theta_\mathrm{e}/2}-\e_i}^2\IEEEyessubnumber\\
                                                &=&2-2(\eu^{-\bm{\ju}_\mathrm{e}\theta_\mathrm{e}/2}\e_i \eu^{\,\bm{\ju}_\mathrm{e}\theta_\mathrm{e}/2}) \cdot \e_i \IEEEyessubnumber\\ 
                                                &=&2-2\mathsf{P}_{\e_i}\big(\eu^{-\bm{\ju}_\mathrm{e}\theta_\mathrm{e}/2}\e_i \eu^{\,\bm{\ju}_\mathrm{e}\theta_\mathrm{e}/2}\big)\e_i. \IEEEyessubnumber\label{eq:errorBound}
\end{IEEEeqnarray}
By decomposing the bivector $\bm{\ju}_\mathrm{e}\theta_\mathrm{e}/2$ as
\begin{IEEEeqnarray}{rCl}\IEEEyesnumber    
    \bm{\ju}_\mathrm{e}\theta_\mathrm{e}/2 &=& a_\mathrm{e}\e_1\e_2 + b_\mathrm{e}\e_2\e_3 + c_\mathrm{e}\e_3\e_1 
    \label{eq:errorBivectorProjections}
\end{IEEEeqnarray}
and the rotor $\eu^{-\bm{\ju}_\mathrm{e}\theta_\mathrm{e}/2}$ as
\begin{IEEEeqnarray}{rCl}\IEEEyesnumber       
        \exp({-\bm{\ju}_\mathrm{e}\theta_\mathrm{e}/2})&=&\varpi_0+\varpi_{12}\e_1\e_2+\varpi_{23}\e_2\e_3+\varpi_{31}\e_3\e_1 \IEEEeqnarraynumspace
\end{IEEEeqnarray}
Noting that for exponential of a $3$D bivector \cite{dargys2021exponentials} 
\begin{IEEEeqnarray}{rCl}\IEEEyesnumber\label{eq:errorRotorProjections}    
    \varpi_0    &=&\cos\!\big(\norm{\bm{\ju}_\mathrm{e}\theta_\mathrm{e}/2}\big)\IEEEyessubnumber\\
    \varpi_{12}    &=&-\frac{a_\mathrm{e}}{2}\frac{\sin\!\big(\norm{\bm{\ju}_\mathrm{e}\theta_\mathrm{e}/2}\big)}{\norm{\bm{\ju}_\mathrm{e}\theta_\mathrm{e}/2}}\IEEEyessubnumber\\
    \varpi_{23}    &=&-\frac{b_\mathrm{e}}{2}\frac{\sin\!\big(\norm{\bm{\ju}_\mathrm{e}\theta_\mathrm{e}/2}\big)}{\norm{\bm{\ju}_\mathrm{e}\theta_\mathrm{e}/2}}\IEEEyessubnumber\\
    \varpi_{31}    &=&-\frac{c_\mathrm{e}}{2}\frac{\sin\!\big(\norm{\bm{\ju}_\mathrm{e}\theta_\mathrm{e}/2}\big)}{\norm{\bm{\ju}_\mathrm{e}\theta_\mathrm{e}/2}}.\IEEEyessubnumber
\end{IEEEeqnarray}
For $\e_i=\e_3$, \eqref{eq:errorBound} can be bounded as 
\begin{IEEEeqnarray}{rCl}\IEEEyesnumber
    \norm{R_\mathrm{e}\e_3 R_\mathrm{e}^\dagger-\e_3} ^2 &=&2\big[1-(\varpi_0^2+\varpi_{12}^2-\varpi_{23}^2-\varpi_{31}^2)\big]\label{eq:errorBoundA}\IEEEeqnarraynumspace\IEEEyessubnumber\\ 
%
%
    %
    %
    &=&\!\left(\frac{2\sin^2\!\big(\norm{\bm{\ju}_\mathrm{e}\theta_\mathrm{e}/2}\big)}{\norm{\bm{\ju}_\mathrm{e}\theta_\mathrm{e}}^2}\right)\!\big(\norm{\bm{\ju}_\mathrm{e}\theta_\mathrm{e}}^2
    \!\nonumber\\
    &~&\quad-(a_\mathrm{e}^2-b_\mathrm{e}^2-c_\mathrm{e}^2)\big)\IEEEyessubnumber\\
    &=&\frac{1}{2}\left(\frac{\sin\!\big(\norm{\bm{\ju}_\mathrm{e}\theta_\mathrm{e}/2}\big)}{\norm{\bm{\ju}_\mathrm{e}\theta_\mathrm{e}/2}}\right)^{\!\!2} \!2(b_\mathrm{e}^2+c_\mathrm{e}^2).\IEEEyessubnumber
\end{IEEEeqnarray}
Finally, taking the square root
\begin{IEEEeqnarray}{rCl}\IEEEyesnumber
    \norm{R_\mathrm{e}\e_3 R_\mathrm{e}^\dagger-\e_3} &=&\left(\frac{\sin\!\big(\norm{\bm{\ju}_\mathrm{e}\theta_\mathrm{e}/2}\big)}{\norm{\bm{\ju}_\mathrm{e}\theta_\mathrm{e}/2}}\right)\sqrt{(b_\mathrm{e}^2+c_\mathrm{e}^2)}
\end{IEEEeqnarray}
knowing
\begin{align}
    \left(\frac{\sin(\norm{\bm{\ju}_\mathrm{e}\theta_\mathrm{e}/2})}{\norm{\bm{\ju}_\mathrm{e}\theta_\mathrm{e}/2}}\right)\leq 1
\end{align}
we can bound the rotation error as
\begin{IEEEeqnarray}{rCl}\IEEEyesnumber
\norm{R_\mathrm{e}\e_3 R_\mathrm{e}^\dagger-\e_3} &\leq&\norm{\bm{\ju}_\mathrm{e}\theta_\mathrm{e}}
\end{IEEEeqnarray}
Using the same procedure, for $\e_i=\e_1$ it can be shown that
\begin{IEEEeqnarray}{rCl}\IEEEyesnumber
    \norm{R_\mathrm{e}\e_1 R_\mathrm{e}^\dagger-\e_1}^2 &=& 2\big[1-(\varpi_0^2-\varpi_{12}^2+\varpi_{23}^2-\varpi_{31}^2)\big]\IEEEeqnarraynumspace\IEEEyessubnumber\\
    \norm{R_\mathrm{e}\e_1 R_\mathrm{e}^\dagger-\e_1} &=&\left(\frac{\sin\!\big(\norm{\bm{\ju}_\mathrm{e}\theta_\mathrm{e}/2}\big)}{\norm{\bm{\ju}_\mathrm{e}\theta_\mathrm{e}/2}}\right)\sqrt{(a_\mathrm{e}^2+c_\mathrm{e}^2)}\IEEEyessubnumber\\
    \norm{R_\mathrm{e}\e_1 R_\mathrm{e}^\dagger-\e_1} &\leq&\norm{\bm{\ju}_\mathrm{e}\theta_\mathrm{e}}\IEEEyessubnumber
\end{IEEEeqnarray}
and for $\e_i=\e_2$
\begin{IEEEeqnarray}{rCl}\IEEEyesnumber
    \norm{R_\mathrm{e}\e_2 R_\mathrm{e}^\dagger-\e_2}^2 &=& 2\big[1-(\varpi_0^2-\varpi_{12}^2-\varpi_{23}^2+\varpi_{31}^2)\big]\IEEEeqnarraynumspace\IEEEyessubnumber\\
    \norm{R_\mathrm{e}\e_2 R_\mathrm{e}^\dagger-\e_2} &=&\left(\frac{\sin\!\big(\norm{\bm{\ju}_\mathrm{e}\theta_\mathrm{e}/2}\big)}{\norm{\bm{\ju}_\mathrm{e}\theta_\mathrm{e}/2}}\right)\sqrt{(a_\mathrm{e}^2+b_\mathrm{e}^2)}\IEEEyessubnumber\\
    \norm{R_\mathrm{e}\e_2 R_\mathrm{e}^\dagger-\e_2} &\leq&\norm{\bm{\ju}_\mathrm{e}\theta_\mathrm{e}}\IEEEyessubnumber
\end{IEEEeqnarray}
concluding that \eqref{berrorfunctm} can be bounded as
\begin{IEEEeqnarray}{rCl}\IEEEyesnumber
\norm{\bft_\mathrm{d}-\bft}&\leq&3\norm{\bar{a}_m}\norm{\bm{\ju}_\mathrm{e}\theta_\mathrm{e}}.
\end{IEEEeqnarray}

\section{CASCADED CONTROL STRATEGY BASED ON GA} \label{control}

\begin{figure*}
\centering
    {\scalebox{0.675}{\begin{tikzpicture}
        
        \useasboundingbox (-3.3,-6) rectangle (23.5,5.8);
        
        \node[myDodgerBlue,, dspnodeopen,dsp/label=left]  at (0,5)    (xi) {Position $\bfxi$};
        \node[myCrimson, dspnodeopen,dsp/label=left]  at (0,4)    (xiDes) {Desired Position $\bfxi_\mathrm{d}$};
        \node[dspadder]  at (1,4.5)   (xiError) {};
        \draw [myptr]  (xi) -| (xiError);
        \draw [myptr]  (xiDes) -| (xiError) node[pos=0.3,below] {$-1\phantom{-}$}node[pos=0.3,gain] {};
    
        \node[myDodgerBlue,,dspnodeopen,dsp/label=left]  at (0,3)    (xiDot) {Velocity $\dot{\bfxi}$};
        \node[myCrimson,dspnodeopen,dsp/label=left]  at (0,2)    (xiDotDes) {Desired Velocity $\dot{\bfxi}_\mathrm{d}$};
        \node[dspadder]                    at (1,2.5)   (xiDotError) {};
        \draw [myptr]  (xiDot) -| (xiDotError);
        \draw [myptr]  (xiDotDes) -| (xiDotError) node[pos=0.3,below] {$-1\phantom{-}$}node[pos=0.3,gain] {};

        \node[dspfilter,drop shadow,fill=myPeachPuff,minimum size=0.8cm] at (2,3.5) (K1){$K_1$};
        \draw [myptr]  (xiError) -| node[pos=0.5,above,align=center,text width=2.6cm] {Position Error\\ $\bfxi_\mathrm{e}$}(K1);
        \draw [myptr]  (xiDotError) -| node[pos=0.5,below,align=center,text width=2.75cm] {$\dot{\bfxi}_\mathrm{e}$\\ Velocity Error}(K1);

        \node[myCrimson, dspnodeopen,dsp/label=left,]  at (0.625,0)    (xiDDotDes) {Desired Acceleration $\ddot{\bfxi}_\mathrm{d}$};
        \node[dspadder]  at (3.5,3.5)   (xiDDotError) {};
        \draw [myptr]  (K1) -- node[pos=0.5,below] {$-1\phantom{-}$} node[pos=0.5,gain] {} (xiDDotError);
        \draw [myptr]  (xiDDotDes) -| (xiDDotError);

        \node[dspnodeopen,dsp/label=above,]  at (4.5,4.75)    (gravity) {~};
        \node[align=center,text width=1.25cm]  at (4.5,5.25)    (gravityLab) {\textcolor{myGray}{Gravity}\\ $g\mathbf{e}_3$};
        
        \node[dspadder]  at (5,3.5)   (dxiv) {};
        \draw [myptr]  (gravity) -| node[pos=0.75,left] {$-1$}node[pos=0.75,gain,rotate=-90] {} (dxiv);
        \draw [myptr]  (xiDDotError) --  node[pos=1.0,below,align=left,text width=1.75cm] {$\dd  \dot{\bfxi}_{\mathrm{v}}$\\ Virtual\\ Desired\\ Acceleration}(dxiv);

        \node[dspfilter,drop shadow,fill=white,minimum size=0.8cm] at (7.25,3.5) (norm){$~\norm{\cdot}~$};
        \node[myForestGreen,dspnodeopen,dsp/label=right] at (9.5,3.5)   (thrust){Desired Thrust Magnitude $f_\mathrm{d}$};
        \draw [myptr]  (norm) -- (thrust);


        \node[dspnodefull,dsp/label=right] at (8,3.5)   (bd) {};
        \node[gain] at (5.75,3.5)   (cas) {};
        \draw  (dxiv) -- (cas);
        \draw [myptr]  (cas) -- node[pos=-0.15,above=2pt,align=center,text width=0.75cm] {\textcolor{myDodgerBlue}{Mass $m$}}(norm);
        
        \node[dspnodefull,dsp/label=right] at (8,2.25)   (bd2) {};

        \node[dspfilter,drop shadow,fill=white,minimum size=0.8cm] at (10.5,2.25) (timesA){$\times$};
        \node[dspfilter,drop shadow,fill=white,minimum size=0.8cm] at (9,2.25) (inv){~$(\cdot)^{-1}$~};
        \draw[myptr]  (bd2) -- (inv);
        \draw[myptr]  (inv) -- (timesA);

        \node[dspnodefull,dsp/label=right] at (6.5,3.5)   (tap) {};
        \node[gain] at (9.5,1.25)   (neg) {};
        \draw  (tap) |- (neg);
        \draw[myptr]  (neg) -| node[pos=-0,above] {$-1\phantom{-}$}(timesA);
        

        \node[myDodgerBlue, dspnodeopen,dsp/label=left,]  at (0.625,1)    (xiDDot) {Acceleration $\ddot{\bfxi}$};

        \node[myDodgerBlue, dspnodeopen,dsp/label=left]  at (0.625,-2)    (xiDDDot) { Jerk $\dddot{\bfxi}$};

        \node[myCrimson, dspnodeopen,dsp/label=left]  at (0.625,-3.5)    (xiDDDotDes) {Desired Jerk $\dddot{\bfxi}_\mathrm{d}$};
        \node[myCrimson, dspnodeopen,dsp/label=left]  at (0.625,-5)    (xiDDDDotDes) {Desired Snap $\ddddot{\bfxi}_\mathrm{d}$};

        \node[dspfilter,drop shadow,fill=white] at (8,-3.5) (eq36to39){~Eqs.\,(\ref{eq:Omegadesired}-\ref{eq:dottp},\ref{eq:bVecDotDes},\ref{eq:bDotDes})~};
        \node[dspfilter,drop shadow,fill=white] at (5.75,-5) (eq40to44){~Eqs.\,(\ref{eq:Omegaddotesired}-\ref{eqddottp},\ref{eq:bVecDDotDes},\ref{eq:bDDotDes})~};
        \draw [myptr]  (xiDDDotDes) -- (eq36to39);
        \draw [myptr]   (bd) -- (eq36to39);

        \node[myDodgerBlue,, dspnodeopen,dsp/label=left]  at (11.5,-1)    (Rdagger) {Orientation $R$};
        \node[myDodgerBlue,, dspnodeopen,dsp/label=left]  at (12.4,-2)    (omega) {Angular Velocity $\bm{\Omega}$};

        
        \draw [myptr]   (xiDDDDotDes) -- (eq40to44.180);

        \node[dspadder]  at (14,-3.5)   (OmegaError) {};
        \draw [myptr]  (omega) -| (OmegaError);
        \draw [myptr]  (eq36to39) -- node[pos=.5,above,align=center,text width=4cm]{Desired Angular Velocity\\ $\bm{\Omega}_\mathrm{d}$} node[pos=0.85,below] {$-1\phantom{-}$}node[pos=0.85,gain] {} (OmegaError);

        \node[dspfilter,drop shadow,fill=myPeachPuff,minimum size=0.8cm] at (16,-3.5) (K2){$K_2$};
        \node[dspfilter,drop shadow,fill=white,minimum size=0.8cm] at (13,-1) (timesRdagger){~Eq.\,\eqref{eq:biVecAngle}~};
        \node[dspfilter,drop shadow,fill=white,minimum size=0.8cm] at (16,-1) (eq47){~Eq.\,\eqref{eq:JthetaDesired}~};
        \node[dspadder]  at (16,-5)   (OmegaDotError) {};
        \draw [myptr]  (OmegaError) -- (K2);
        \draw [myptr]  (K2) -- node[pos=0.5,left] {$-1$} node[pos=0.5,gain,rotate=-90] {}(OmegaDotError);
        \draw [myptr]  (eq40to44) -- node[pos=0.5,above]{Desired Angular Acceleration $\dot{\bm{\Omega}}_\mathrm{d}$} (OmegaDotError);
        \draw [myptr]  (timesRdagger) --node[pos=0.5,above,align=center,text width=2cm] {Orientation\\ Error\\ $R_\mathrm{e}$} (eq47);
        \draw [myptr]  (eq47) --node[pos=0.5,right,align=center,text width=1cm] {Angular Error\\ $\mathbf{j}_\mathrm{e}\theta_\mathrm{e}$} (K2);
        \draw [myptr]  (Rdagger) -- (timesRdagger);

        \node[dspfilter,drop shadow,fill=white] at (20,-5) (eq75){~Eq.\,\eqref{eq:torque}~};
        \node[myForestGreen,dspnodeopen,dsp/label=right,align=center,text width=1.0cm] at (22,-5)   (torque){Desired\\ Torque\\ $\Tau$};
        \draw [myptr]  (OmegaDotError) -- node[pos=0.5,above,align=center,text width=2.25cm] {Virtual Desired\\ Angular\\ Acceleration\\ $\dot{\bm{\Omega}}_{\mathrm{v}}$}(eq75);
        \draw [myptr]  (eq75) -- (torque);

        \node[myDodgerBlue,align=center,text width=1.5cm]  at (20,-2.75)    (moments) {Principal\\ Moments\\ of Inertia\\ $\{\iota_1,\iota_2,\iota_3\}$};
        \node[myDodgerBlue, dspnodeopen,dsp/label=above,align=center,text width=1.5cm]  at (20,-3.75)    (moments) {~};
        \draw [myptr]  (moments) -- (eq75);

        \node[dspfilter,drop shadow,fill=white,minimum size=0.8cm] at (13,1) (eq23to29){~Eqs.\,\eqref{eqn:ThrustWouldBe}-\eqref{eq:errorRot}~};
        \draw [myptr]  (eq23to29) -- node[pos=0.5,right,align=center,text width=2.5cm] {}(timesRdagger);
        \draw[myptr]  (timesA) -| node[pos=0.3,above,align=center,text width=2.5cm] {Desired Thrust\\ Direction $\bft_\mathrm{d}$}(eq23to29.120);

    \node[myDeepPink, dspnodeopen,dsp/label=right,align=center,text width=2cm]  at (14,2)    (Rh) {Preferred Orientation\\ $R_\mathrm{p}$};
    \draw[myptr]  (Rh) -| (eq23to29.60);
    
    \node[dspnodefull,dsp/label=right] at (13,-0.25)   (tap2) {};
    \draw[myptr]  (tap2) -| node[pos=0.225,above] {Desired Orientation
        $R_\mathrm{d}$} (eq36to39.25);

    \node[dspnodefull,dsp/label=right] at (11.5,2.25)   (tap3) {};
    \draw[myptr]  (tap3) -- ++(0,-1.75) -- ++(-3,0) -- (eq36to39.43);

    \node[dspnodefull,dsp/label=right] at (8,-2.75)   (tap4) {};
    \draw[myptr]  (tap4) -| (eq40to44.60);

    \node[dspnodefull,dsp/label=right] at (8.45,-2.5)   (tap5) {};
    \draw[myptr]  (tap5) -| (eq40to44.90);
    
    \node[dspnodefull,dsp/label=right] at (8.88,-2.25)   (tap6) {};
    \draw[myptr]  (tap6) -| (eq40to44.120);

    \draw [myptr]  (xiDDot) -| (eq36to39.130);
    \draw [myptr]  (xiDDDot) -| (eq40to44.140);

   \begin{scope}[on background layer]
        \node at (9.5,0.5) (pt){};
        \draw[myChocolate, thick, dashed] ($(pt)+(-8.5,-6)$) to [bend left=8] ($(pt)+(9.5,4.5)$);
        \node[myChocolate] at (16.5,4.5) {Translation};
        \node[myChocolate] at (18,4) {Rotation~~~};        
    \end{scope}

    \end{tikzpicture}}}
\caption{Block diagram of the control architecture highlighting desired/preferred variables~(\textcolor{myCrimson}{\solidLrule[3.5mm]}/\textcolor{myDeepPink}{\solidLrule[3.5mm]}), actual system descriptors~(\textcolor{myDodgerBlue}{\solidLrule[3.5mm]}),  and controller outputs~(\textcolor{myForestGreen}{\solidLrule[3.5mm]}). From the upper left portion of the diagram, using the translational positions and velocities, together with corresponding errors, and the third order derivative of the translation reference, it is possible to compute the desired thrust magnitude and direction. Then, using the preferred orientation, third and fourth  derivatives of the translational reference, together with the rotational errors, the desired torque is computed. 
This diagram highlights the cascaded structure of the control strategy, where the translational controller needs a desired rotation while the rotational controller obtain the references from the translational error.}
\label{fig:ControlSchema}
\end{figure*}
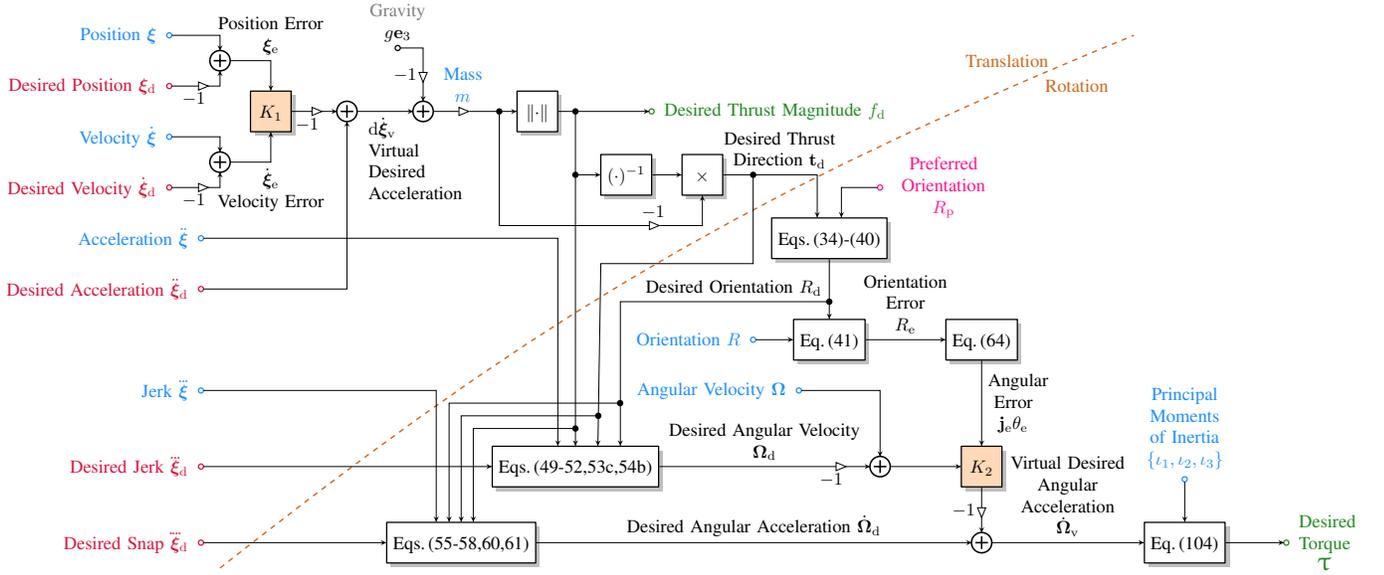 


\subsection{State Space representation}
Using the standard formalism of control theory, we represent vectors in $\mathbb{L}^3$ and bivectors in $\mathbb{B}^3$  using column matrices  with three real entries in $\mathcal{M}_{3,1}(\mathbb{R})$. We begin by expanding the vectors $\bfxi$, $\turnt$ and $\wind$ with respect to the inertial frame $\{\mathbf{e}_1, \mathbf{e}_2, \mathbf{e}_3\}$ 
\begin{IEEEeqnarray}{rCl}
    \bfxi &=& x\mathbf{e}_1 + y\mathbf{e}_2 + z\mathbf{e}_3\\
    \turnt &=& \nturnt_1\mathbf{e}_1 + \nturnt_2\mathbf{e}_2 + \nturnt_3\mathbf{e}_3\\
    {\wind} &=&  {\wind_1\mathbf{e}_1 + \wind_2\mathbf{e}_2 + \wind_3\mathbf{e}_3}
\end{IEEEeqnarray}
and represent the corresponding vectors with column matrices 
\begin{IEEEeqnarray}{rCl}
        \xiCol&=&[x,y,z]^\transpose\\
        \underline{\turnt}&=&[\nturnt_1,\nturnt_2,\nturnt_3]^\transpose\\
        {\underline{\wind}}&=&{[\wind_1,\wind_2,\wind_3]^\transpose}
\end{IEEEeqnarray}
The vectors $\mathbf{t}$, $\mathbf{t}_\mathrm{d}$, and the left hand side of \eqref{eq:RotationError}, i.e.,
\begin{IEEEeqnarray}{rCl}
    \mathbf{t}  = R\turnt R^\dagger  &=& t_1\mathbf{e}_1 + t_2\mathbf{e}_2 + t_3\mathbf{e}_3\\
    \mathbf{t}_\mathrm{d}  = R_\mathrm{d}\turnt R_\mathrm{d}^\dagger  &=& t_{\mathrm{d}1}\mathbf{e}_1 + t_{\mathrm{d}2}\mathbf{e}_2 + t_{\mathrm{d}3}\mathbf{e}_3\\
    \mathbf{t}_\mathrm{e}= R^\dagger(\bft_\mathrm{d}-\bft)R&=& t_{\mathrm{e}1}\mathbf{e}_1 + t_{\mathrm{e}2}\mathbf{e}_2 + t_{\mathrm{e}3}\mathbf{e}_3
\end{IEEEeqnarray}
respectively, are represented as 
\begin{IEEEeqnarray}{rCl}
        \underline{\mathbf{t}}&=&[t_1,t_2,t_3]^\transpose\\
        \underline{\mathbf{t}_\mathrm{d}}&=&[t_{\mathrm{d}1},t_{\mathrm{d}2},t_{\mathrm{d}3}]^\transpose\\
        \underline{\mathbf{t}_\mathrm{e}}&=&[t_{\mathrm{e}1},t_{\mathrm{e}2},t_{\mathrm{e}3}]^\transpose.
\end{IEEEeqnarray}
Expanding the bivectors $\mathbf{j}\theta$, $\bm\Omega_\mathrm{b}$, and $\Tau$ w.r.t. the canonical bivectors of the body frame $\{\mathbf{b}_1\mathbf{b}_2, \mathbf{b}_2\mathbf{b}_3, \mathbf{b}_3\mathbf{b}_1\}$ yields
\begin{IEEEeqnarray}{rCl}
    {\mathbf{j}\theta} &=& \vartheta_{12} \mathbf{b}_1\mathbf{b}_2 + \vartheta_{23}\mathbf{b}_2\mathbf{b}_3 + \vartheta_{31}\mathbf{b}_3\mathbf{b}_1\\
     {\bm\Omega}_\mathrm{b} &=& \Omega_{12} \mathbf{b}_1\mathbf{b}_2 + \Omega_{23}\mathbf{b}_2\mathbf{b}_3 + \Omega_{31}\mathbf{b}_3\mathbf{b}_1\\
     {\Tau} &=& \tau_{12} \mathbf{b}_1\mathbf{b}_2 + \tau_{23}\mathbf{b}_2\mathbf{b}_3 + \tau_{31}\mathbf{b}_3\mathbf{b}_1     
\end{IEEEeqnarray}
and bivectors are represented with the column matrices\footnote{We drop the subscript $\mathrm{b}$ when representing body angular velocity $\underline{\bm\Omega}_\mathrm{b}$ with the matrix $\underline{\bm{\Omega}}$.}
\begin{IEEEeqnarray}{rCl}\IEEEyesnumber
    \underline{\bm{\ju}\theta}&=&[\vartheta_{12},\vartheta_{23},\vartheta_{31}]^\transpose \\
    \underline{\bm{\Omega}}&=&[{\Omega}_{12},{\Omega}_{23},{\Omega}_{31}]^\transpose \\
    \underline{\Tau}&=&[\tau_{12},\tau_{23},\tau_{31}]^\transpose
\end{IEEEeqnarray}
In matrix form, the translation equation in \eqref{eq:TranslationalDynamicsGeneral} can be written in a set of first order differential equations as a function of the state space variables as
\begin{IEEEeqnarray}{rCl}\IEEEyesnumber \label{eq:FullSystemA}
    \xiDotCol&=&\underline{\dd\!\!\:\bfxi} \IEEEyesnumber\\
    \underline{\dot{\dd\!\!\:\bfxi}}&=&g\underline{\e_3}-\frac{\thrust}{m}\underline{\mathbf{t}}+{\underline{\wind}}  \IEEEyesnumber \label{eq:SystA}
\end{IEEEeqnarray}
Finally, the rotational dynamics in \eqref{eq:rotationaldynamics} can be expressed as
\begin{IEEEeqnarray}{rCl}\IEEEyesnumber \label{eq:FullSystemB}
    \underline{\dot{\bm{\ju}\theta}}&=&\underline{\bm{\Omega}} \IEEEyesnumber\\
    \underline{\dot{\bm{\Omega}}}&=&\underline{\mathcal{I}}^{-1}\underline{M}+\underline{\mathcal{I}}^{-1}\underline{\Tau} \IEEEyesnumber \label{eq:SystB} 
\end{IEEEeqnarray}
where the expressions in \eqref{eq:commutator} and \eqref{eq:GAinteralTensor} are represented in matrix form as
\begin{IEEEeqnarray}{rCl}
\underline{M}(\underline{\bm{\Omega}};\underline{\mathcal{I}}) =&\begin{bmatrix}
    {\Omega}_{23} {\Omega}_{31}(\iota_{12}-\iota_{31}) \\
    {\Omega}_{31} {\Omega}_{12}(\iota_{31}-\iota_{12})\\
    {\Omega}_{12} {\Omega}_{23}(\iota_{12}-\iota_{23})
\end{bmatrix}\!,~\underline{\mathcal{I}}=\begin{bmatrix}
    \iota_{12}&0&0\\
    0&\iota_{23}&0\\
    0&0&\iota_{31}\\
\end{bmatrix}\IEEEeqnarraynumspace
\end{IEEEeqnarray}
with $\iota_{ij}$ is the moment of inertia in the plane defined by $\mathbf{e}_i\mathbf{e}_j$ for $i,j=1,2,3$ and $i\neq j$.

\subsubsection{Virtual Inputs} 
As a first step we define two virtual inputs, the first one $\underline{\dd\xiDot}_{\mathrm{v}}\in \mathcal{M}_{3,1}(\mathbb{R})$ with coordinates taken from the inertial frame, and the second one $\underline{\dot{\bm{\Omega}}_{\mathrm{v}}} \in \mathcal{M}_{3,1}(\mathbb{R})$ represented with coordinates from the body frame. This yields
\begin{IEEEeqnarray}{rCl}\IEEEyesnumber             \label{eq:VirtualInputSystem}
    \underline{\dot{\dd\!\!\:\bfxi}_{\!\!\:\mathrm{v}}} &=&   g \underline{\e_3} - \frac{\thrust}{m}\underline{\mathbf{t}_\mathrm{d}} \IEEEyesnumber \label{eq:virtualinput1}\\
\underline{\dot{\bm{\Omega}}_{\mathrm{v}}}&=&\underline{\mathcal{I}}^{-1}\underline{M} +\underline{\mathcal{I}}^{-1}\underline{\Tau}   \IEEEyesnumber \label{eq:virtualinput2}
\end{IEEEeqnarray}
The systems inputs $\thrust$ and $\underline{\Tau}$ can be recovered from $\underline{\dot{\dd\!\!\:\bfxi}_{\!\!\:\mathrm{v}}}$ and $\underline{\dot{\bm{\Omega}}_{\mathrm{v}}}$, respectively, as 
\begin{IEEEeqnarray}{rCl}\IEEEyesnumber \label{eq:InputSystem}
    \thrust&=& m\norm{\underline{\dot{\dd\!\!\:\bfxi}_{\!\!\:\mathrm{v}}} - g \underline{\e_3}}    \IEEEyesnumber \\
    \underline{\Tau}&=&\underline{\mathcal{I}}\underline{\dot{\bm{\Omega}}_{\mathrm{v}}}-\underline{M}    \IEEEyesnumber\label{eq:torque}
\end{IEEEeqnarray}
As previously noted, due to the dynamics of the rotational subsystem, the desired angles ${\bm{\ju}_\mathrm{d}\theta_\mathrm{d}}$ cannot be supplied instantaneously.


\subsubsection{Tracking errors} 
Define the following tracking errors
\begin{IEEEeqnarray}{rCl}\IEEEyesnumber
    \underline{\bfxi_{\!\!\:\mathrm{e}}}          &=& \xiCol-\underline{\bfxi_{\!\!\:\mathrm{d}}} \IEEEyesnumber\\   
    \underline{\dd\!\!\:\bfxi_{\!\!\:\mathrm{e}}} &=& \underline{\dd\!\!\:\bfxi}-\underline{\dd\!\!\:\bfxi_{\!\!\:\mathrm{d}}} \IEEEyesnumber\\
    \underline{\dot{\dd\!\!\:\bfxi}_{\!\!\:\mathrm{e}}} &=& \underline{\dot{\dd\!\!\:\bfxi}}-\underline{\dot{\dd\!\!\:\bfxi}_{\!\!\:\mathrm{d}}} \IEEEyesnumber\\
    \underline{\ddot{\dd\!\!\:\bfxi}_{\!\!\:\mathrm{e}}} &=& \underline{\ddot{\dd\!\!\:\bfxi}}-\underline{\ddot{\dd\!\!\:\bfxi}_{\!\!\:\mathrm{d}}} \IEEEyesnumber\\ 
    \underline{\bm{\Omega}_\mathrm{e}}            &=& \underline{\bm{\Omega}}-\underline{\bm{\Omega}_\mathrm{d} }\IEEEyesnumber\\
    \underline{\dot{\bm{\Omega}}_\mathrm{e}}      &=& \underline{\dot{\bm{\Omega}}}-\underline{\dot{\bm{\Omega}}_\mathrm{d} }\IEEEyesnumber
\end{IEEEeqnarray}
with $\underline{\bm{\Omega}}_\mathrm{d}$ and $\underline{\dot{\bm{\Omega}}}_\mathrm{d}$ are computed in \eqref{eq:Omegadesired} and \eqref{eq:Omegaddotesired}, respectively. The tracking errors dynamics can be described as 
\begin{IEEEeqnarray}{rCl}\IEEEyesnumber 
    \underline{\dot{\bfxi}_{\!\!\:\mathrm{e}}} &=& \underline{\dd\!\!\:\bfxi_{\!\!\:\mathrm{e}}} \IEEEyesnumber\\
    \underline{\dot{\dd\!\!\:\bfxi}_{\!\!\:\mathrm{e}}} &=&g\underline{\e_3}-\frac{\thrust}{m}\underline{\mathbf{t}} { + \underline{\wind}} -\underline{\dot{\dd\!\!\:\bfxi}_{\!\!\:\mathrm{d}}} \nonumber\\
    &=& {\underline{\wind}}+ \underline{\dot{\dd\!\!\:\bfxi}_{\!\!\:\mathrm{v}}} -\underline{\dot{\dd\!\!\:\bfxi}_{\!\!\:\mathrm{d}}}+\frac{\thrust}{m}(\underline{\mathbf{t}_\mathrm{d}}-\underline{\mathbf{t}})~~~\\
    \underline{\dot{\bm{\ju}_{\mathrm{e}}\theta}_\mathrm{e}}&=&\underline{\bm{\Omega}_\mathrm{e}} \IEEEyesnumber\\
    \underline{\dot{\bm{\Omega}}_\mathrm{e}}&=&\underline{\mathcal{I}}^{-1}\underline{M}+\underline{\mathcal{I}}^{-1}\underline{\Tau} -\underline{\dot{\bm{\Omega}}_\mathrm{d}}=\underline{\dot{\bm{\Omega}}_{\mathrm{v}}}-\underline{\dot{\bm{\Omega}}_\mathrm{d}}\IEEEyesnumber 
\end{IEEEeqnarray}
with $\underline{\bm{\ju}_\mathrm{e}\theta_\mathrm{e}}$ already defined in \eqref{eq:JthetaDesired}.
\subsubsection{GA Model in state space representation}
Regarding the variables defined in \eqref{eq:VirtualInputSystem} as inputs, the state-space representation of the translational and rotational dynamics becomes a cascade of two linear time-invariant (LTI) systems with a nonlinear time-varying interconnection. Define the state vectors $\underline{x_n}\in\mathcal{M}_{6,1}(\mathbb{R})$ with $n=\{1,2\}$ 
for translational and rotational dynamics, respectively, as
\begin{align}
\underline{x_1}=&\left[\underline{\bfxi_\mathrm{e}}^\transpose,\underline{\dd\!\!\:\bfxi_{\!\!\:\mathrm{e}}}^\transpose\right]^\transpose,~~\underline{x_2}=\left[\underline{\bm{\ju}_\mathrm{e}\theta_\mathrm{e}}^\transpose,\underline{\bm{\Omega}_\mathrm{e}}^\transpose\right]^\transpose, \nonumber
\end{align}
define the input vectors $\underline{u_n}\in\mathcal{M}_{3,1}(\mathbb{R})$ from the virtual inputs \eqref{eq:VirtualInputSystem} as
\begin{align}    
    \underline{u_1}= \underline{\dot{\dd\!\!\:\bfxi}_{\!\!\:\mathrm{v}}} ,~~\underline{u_2}=\underline{\dot{\bm{\Omega}}_{\mathrm{v}}}.\nonumber
\end{align}
and the perturbation vector $\underline{\wind}(t)\in\mathcal{M}_{3,1}(\mathbb{R})$.
The tracking error dynamics can then be written as
\begin{subequations}
    \label{eq:StateSpaceRepresentation}
    \begin{align}
        \underline{\dot{x}_1}=&\underline{A_1}\underline{x_1}+\underline{B_1}\big(\underline{u_1}- \underline{\dot{\dd\!\!\:\bfxi}_{\!\!\:\mathrm{d}}} \big)+\underline{g}\big(t,\underline{x_1},\underline{x_2}\big){+\begin{bmatrix}0\\ \underline{\wind}(t)\end{bmatrix}}\\
        \underline{\dot{x}_2}=&\underline{A_2}\underline{x_2}+\underline{B_2}\big(\underline{u_2}-\underline{\dot{\bm{\Omega}}_\mathrm{d}}\big)
    \end{align}    
\end{subequations}
with matrices $\underline{A_n}\in\mathcal{M}_{6,6}(\mathbb{R})$, $\underline{B_n}\in\mathcal{M}_{6,3}(\mathbb{R})$
\begin{align}
    \underline{A_1}=\underline{A_2}=&\begin{bmatrix}
    \underline{0}&\underline{I}\\
    \underline{0}&\underline{0}
    \end{bmatrix}, \quad \underline{B_1}=\underline{B_2}=\begin{bmatrix}
    \underline{0}\\
    \underline{I}
    \end{bmatrix},
\end{align}
     and $\underline{g}\in\mathcal{M}_{6,1}(\mathbb{R})$ obtained from the rotation error~\eqref{eq:RotationError} after combining \eqref{eq:rotationerrorinb} with \eqref{eq:virtualinput1} using the tracking error variables
\begin{align}
    \label{eq:gtx-def}
    \underline{g}\big(t,\underline{x_1},\underline{x_2}\big)=&\begin{bmatrix}
    \underline{0}\\
    \frac{\thrust}{m}\underline{\mathbf{t}_\mathrm{e}}
    \end{bmatrix}
\end{align}

\subsection{Closed-Loop Control Design and Stability Analysis}
Given a reference trajectory $\underline{\bm{\xi}_\mathrm{d}}$, satisfying  \textbf{Assumption}~\ref{assump:Ref}, consider the following state-feedback controller
\begin{IEEEeqnarray}{rCl}\IEEEyesnumber\label{eq:ClosedLoopGains}
    \underline{u_1}=&-\underline{K_1}\underline{x_1}+\underline{\dot{\dd\!\!\:\bfxi}_{\!\!\:\mathrm{d}}}  \label{eq:u1input}\IEEEyessubnumber\\
    \underline{u_2}=&-\underline{K_2}\underline{x_2}+\underline{\dot{\bm{\Omega}}_\mathrm{d}} \IEEEyessubnumber
\end{IEEEeqnarray}

In Fig.~\ref{fig:ControlSchema} a block diagram of the control architecture is presented. In the closed loop, the virtual input  $\ddot{\bm{\xi}}_\mathrm{d}$ is computed using the translational reference and the translational controller compensation. The derivative of the virtual input 
\begin{IEEEeqnarray}{rCl}\IEEEyesnumber
    \underline{\dot{u}_1} =& \underline{\ddot{\dd\!\!\:\bfxi}_{\!\!\:\mathrm{v}}} = -\underline{K_1}\underline{\dot{x}_1}+\underline{\ddot{\dd\!\!\:\bfxi}_{\!\!\:\mathrm{d}}}, \: \underline{\dot{x}_1}=\left[\underline{\dd\!\!\:\bfxi_{\!\!\:\mathrm{e}}}^\transpose, \underline{\dot{\dd\!\!\:\bfxi}_{\!\!\:\mathrm{e}}}^\transpose \right]^\transpose  
     \\
    \underline{\ddot{u}_1} =& \underline{\dddot{\dd\!\!\:\bfxi}_{\!\!\:\mathrm{v}}} = -\underline{K_1}\underline{\ddot{x}_1}+\underline{\dddot{\dd\!\!\:\bfxi}_{\!\!\:\mathrm{d}}}, \: \underline{\ddot{x}_1}=\left[\underline{\dot{\dd\!\!\:\bfxi}_{\!\!\:\mathrm{e}}}^\transpose, \underline{\ddot{\dd\!\!\:\bfxi}_{\!\!\:\mathrm{e}}}^\transpose\right]^\transpose  \label{eq:u1input}
\end{IEEEeqnarray}
is required by \eqref{eq:3DerivDesiRef} and \eqref{eq:4DerivDesiRef}. We can see that computation of the third and forth derivatives of $\bm{\xi}_\mathrm{d}$ using \eqref{eq:u1input} implies the 
availability of the translational accelerations and jerk signals. Next, we will show that closed-loop stability implying appropriate reference tracking is guaranteed. A couple of definitions follow in order to define the precise stability notion pursued.
\color{black}
\begin{definition}
A continuous function $\alpha : \mathbb{R}_{\ge 0} \to \mathbb{R}_{\ge 0}$ is said to be of class $\mathcal{K}$ if it is strictly increasing and $\alpha(0)=0$, and of class $\mathcal{K}_\infty$ if it is of class $\mathcal{K}$ and unbounded. A function $\beta : \mathbb{R}_{\ge 0} \times \mathbb{R}_{\ge 0} \to \mathbb{R}_{\ge 0}$ is said to be of class $\mathcal{KL}$ if $\beta(\cdot,s)$ is of class $\mathcal{K}$ for every fixed $s$ and $\beta(r,\cdot)$ is decreasing and satisfies $\lim_{s\to\infty}\beta(r,s) = 0$ for every fixed $r$. 
\end{definition}
{
\begin{definition}[see~\cite{chaang_tac14}]\label{def:stability}
    A system of the form $\underline{\dot{x}} = f(t,\underline{x},\underline{\wind})$, with $f:\mathbb{R}_{\ge 0} \times \mathbb{R}^n \times \mathbb{R}^m \to \mathbb{R}^n$ is said to be: 
    \begin{itemize}
        \item Uniformly Globally Asymptotically Stable under zero input (0-UGAS) if there exists $\beta\in\mathcal{KL}$ such that every solution with $\underline{\wind}(t) \equiv 0$ satisfies
    \begin{IEEEeqnarray}{c}\IEEEyesnumber
        \label{eq:UGAS-def}
       \norm{\underline{x}(t)} \le \beta(\norm{\underline{x}(t_0)}, t-t_0 ) \quad \forall t\ge t_0\ge 0.\IEEEeqnarraynumspace
    \end{IEEEeqnarray}
        \item Input-to-State Stable (ISS) with respect to (wrt) small inputs if there exist $R>0$, $\beta\in\mathcal{KL}$, $\mu\in\mathcal{K}_\infty$ such that for all $t\ge t_0 \ge 0$
    \begin{multline*}
        \label{eq:ISS-def}
       \norm{\underline{\wind}(t)} < R \quad \Rightarrow \\
       \norm{\underline{x}(t)} \le \beta(\norm{\underline{x}(t_0)}, t-t_0 ) + \mu\left(\sup_{t_0 \le s \le t} \norm{\underline{\wind}(s)}\right)        
    \end{multline*}
        \item integral ISS (iISS) if there exist $\beta\in\mathcal{KL}$, $\mu_1,\mu_2 \in \mathcal{K}_\infty$, such that for all $t\ge t_0 \ge 0$
        \begin{align*}
            \norm{\underline{x}(t)} \le \beta(\norm{\underline{x}(t_0)}, t-t_0 ) + \mu_1 \int_{t_0}^t \mu_2(\norm{\underline{\wind}(s)}) ds 
        \end{align*}
        \item Strongly iISS, if it is ISS wrt small inputs and iISS.
    \end{itemize}
\end{definition}
All the properties of Definition~\ref{def:stability} are uniform with respect to initial time. The following theorem establishes stability and robustness of the proposed closed-loop system.
}

{
\begin{theorem}[Closed-loop Stability and Robustness]\label{thm:UGAS}
Let the double derivative of the desired position, $\underline{\dot{\dd\!\!\:\bfxi}_{\!\!\:\mathrm{d}}}$, be bounded uniformly in time, i.e. $\sup_{t\ge 0} \norm{\underline{\dot{\dd\!\!\:\bfxi}_{\!\!\:\mathrm{d}}}(t)} < \infty$. Then, the tracking error dynamics described by \eqref{eq:StateSpaceRepresentation}, with the (virtual) inputs selected according to \eqref{eq:ClosedLoopGains}, with matrices $\underline{K_1}$ and $\underline{K_2}$ such that $(\underline{A_1}-\underline{B_1}\underline{K_1})$ and $(\underline{A_2}-\underline{B_2}\underline{K_2})$ are Hurwitz, is Strongly iISS.
\end{theorem}
}
{
\begin{IEEEproof} From \eqref{eq:StateSpaceRepresentation} and \eqref{eq:ClosedLoopGains}, it follows that
    \begin{IEEEeqnarray}{rCl}\IEEEyesnumber \label{eq:clsysfull} 
        \underline{\dot{x}_1}&=&\underline{A_1^{C}}\underline{x_1}+\underline{g}(t,\underline{x_1},\underline{x_2}){+\begin{bmatrix}
            0\\ \underline{\wind}(t)
        \end{bmatrix}} \label{eq:clsys1}\\
        \underline{\dot{x}_2}&=&\underline{A_2^{C}}\underline{x_2} \label{eq:clsys2}\\
        \underline{A_i^{C}} &:=&  \underline{A_i}-\underline{B_i}\underline{K_i},\quad i=1,2. \notag
    \end{IEEEeqnarray}
    Since $\underline{A_i^C}$ is Hurwitz, there exists $\underline{P}_i = \underline{P}_i^\transpose \in \mathcal{M}_{6,6}(\mathbb{R})$ positive definite and such that 
    \begin{IEEEeqnarray}{rCl}\IEEEyesnumber
        \label{eq:Lyap-x1}
        \underline{A_i^{C}}^\transpose \underline{P}_i + \underline{P}_i \underline{A_i^{C}} = -\underline{I},
    \end{IEEEeqnarray}
    for $i=1,2$ and with  the identity matrix $\underline{I} \in \mathcal{M}_{6,6}(\mathbb{R})$.
    From~\eqref{eq:gtx-def}, it follows that 
    \begin{IEEEeqnarray}{rCl}\IEEEyesnumber
        &&\norm{g(t,\underline{x_1},\underline{x_2})}=\frac{\thrust}{m}\norm{\eu^{-\bm{\ju}_\mathrm{e}\theta_\mathrm{e}/2}\turnt\eu^{\,\bm{\ju}_\mathrm{e}\theta_\mathrm{e}/2}-\turnt}\IEEEnonumber\\
        &=&\norm{g\underline{e_3}-\underline{u}_1}\norm{\eu^{-\bm{\ju}_\mathrm{e}\theta_\mathrm{e}/2}\turnt\eu^{\,\bm{\ju}_\mathrm{e}\theta_\mathrm{e}/2}-\turnt}\IEEEnonumber\\
        &=&\norm{g\underline{e_3}+\underline{K_1}\underline{x_1}-\underline{\dot{\dd\!\!\:\bfxi}_{\!\!\:\mathrm{d}}}(t)}\norm{\eu^{-\bm{\ju}_\mathrm{e}\theta_\mathrm{e}/2}\turnt\eu^{\,\bm{\ju}_\mathrm{e}\theta_\mathrm{e}/2}-\turnt\IEEEnonumber}\\
        &\leq& (\mathsf{K}_1 \norm{\underline{x_1}}+\mathsf{K}_2)
        \mathsf{K}_3\norm{\underline{x_2}}\label{eq:gtx-bound}
    \end{IEEEeqnarray}
    with $\mathsf{K}_1=\norm{\underline{K_1}}$, $\mathsf{K}_2 = \sup_{t\ge 0} \norm{g\underline{e_3}-\underline{\dot{\dd\!\!\:\bfxi}_{\!\!\:\mathrm{d}}}(t)}$, and $\mathsf{K}_3=3\norm{\bar{a}_m}$. From the assumption on the boundedness of $\underline{\dot{\dd\!\!\:\bfxi}_{\!\!\:\mathrm{d}}}(t)$, we have that $\mathsf{K}_2 < \infty$. 
    Distinguish the cases (a) $\underline{x}_2(t_0) = 0$ and (b) $\underline{x}_2(t_0) \neq 0$. In case (a), from \eqref{eq:clsys2} it follows that $\underline{x}_2(t) = 0$ for all $t\ge t_0$. Then, \eqref{eq:clsys1} becomes linear time-invariant with $\underline{A}_1^C$ Hurwitz, and hence \eqref{eq:clsys1} is Strongly iISS while the state of \eqref{eq:clsys2} is constantly zero. 
    In case (b), from \eqref{eq:clsys2} then $\underline{x}_2(t) \neq 0$ for all $t\ge t_0$. 
    For $i=1,2$, define the quadratic functions 
    \begin{IEEEeqnarray}{rCl}\IEEEyesnumber
        V_i(\underline{x_i})&=&\underline{x_i}^T\underline{P}_i\underline{x_i},
    \end{IEEEeqnarray}
    which satisfy
   \begin{IEEEeqnarray}{rCl}\IEEEyesnumber
        \label{eq:sandwich}
        \lambda_{\min}(\underline{P}_i)\norm{\underline{x_i}}^2 &\leq V_i(\underline{x_i}) \leq \lambda_{\max}(\underline{P}_i)\norm{\underline{x_i}}^2.    
    \end{IEEEeqnarray}
    From \eqref{eq:clsys1}, \eqref{eq:Lyap-x1} and \eqref{eq:gtx-bound}, $V_1$ satisfies
    \begin{multline}
        \dot{V}_{1}(t,\underline{x_1},\underline{x_2},\underline{\wind}) \le
        -\norm{\underline{x_1}}^2 + \\
        2\norm{\underline{P}_1}\rho\big(\norm{\underline{x_1}}\big)\big[\mathsf{K}_3\norm{\underline{x_2}}+\norm{\underline{\wind}}\big]
    \end{multline}
    with the function $\rho$ defined as $\rho(s) :=\mathsf{K}_1 s^2+ \bar{\mathsf{K}}_2 s$ and $\bar{\mathsf{K}}_2 := \max\{\mathsf{K}_2,1\}$. Using the bounds \eqref{eq:sandwich}, it follows that
    \begin{IEEEeqnarray}{c}\IEEEyesnumber
        \rho\big(\norm{\underline{x_1}}\big)\le\mathsf{K}_1\frac{V_1}{\scriptstyle\lambda_{\min}(\underline{P}_1)}+\bar{\mathsf{K}}_2\sqrt{\frac{V_1}{\scriptstyle\lambda_{\min}(\underline{P}_1)}}=:\tilde{\rho}(V_1)\IEEEeqnarraynumspace
    \end{IEEEeqnarray}
    and, defining the positive constants $a_1 := 1/\lambda_{\max}(\underline{P}_1)$, $b_1 := 2 \norm{\underline{P}_1}$ and $c_1 := 2 \norm{\underline{P}_1} \mathsf{K}_3 / \lambda_{\min}(\underline{P}_2)$ yields
    \begin{IEEEeqnarray}{rCl}
        \dot{V}_{1} 
        &\le& - a_1 V_1 + \tilde{\rho}(V_1) \left[c_1 \sqrt{V_2} + b_1 \norm{\underline{\wind}} \right]\IEEEeqnarraynumspace
    \end{IEEEeqnarray}
    where the arguments of $V_1$ and $V_2$ have been omitted.
    Next, consider the class $\mathcal{K}_\infty$ function 
    \begin{IEEEeqnarray}{c}\IEEEyesnumber
        \chi(s)=\int_0^s \frac{\dd r}{1+\tilde{\rho}(r)}
    \end{IEEEeqnarray}
    and define the Lyapunov function $W$ as
    \begin{IEEEeqnarray}{c}\IEEEyesnumber
        W(\underline{x}_1,\underline{x}_2) := \chi(V_1) + 2k\sqrt{V_2}
    \end{IEEEeqnarray}
    with $k > c_1 \lambda_{\max}(\underline{P}_2)$.
    The function $W$ is continuous everywhere and differentiable everywhere except when $\underline{x}_2 = 0$. 
    The derivative of $W$ for $\underline{x}_2 \neq 0$ satisfies
    \begin{IEEEeqnarray}{rl}\IEEEnonumber
        &\dot W(t,\underline{x_1}, \underline{x_2}, \underline{\wind}) = \chi^\prime(V_1) \dot V_1 + \frac{k\dot V_2}{\sqrt{V_2}} \\
        &\le - \frac{a_1 V_1}{1 + \tilde\rho(V_1)} + \IEEEnonumber\\
        &\qquad \frac{\tilde\rho(V_1)}{1+\tilde\rho(V_1)} \left[ c_1 \sqrt{V_2} + b_1 \norm{\underline{\wind}} \right] - k\frac{\norm{\underline{x}_2}^2}{\sqrt{V_2}} \IEEEnonumber \\
        &\le - \frac{a_1 V_1}{1 + \tilde\rho(V_1)} - (k_2-c_1) \sqrt{V_2} + b_1 \norm{\underline{\wind}}
    \end{IEEEeqnarray}
    where \eqref{eq:Lyap-x1} and \eqref{eq:sandwich} were employed for $i=2$ and with $k_2 = k/\lambda_{\max}(\underline{P}_2)$. Since $k_2-c_1 > 0$, since the function $s \mapsto h(s) := \frac{a_1 s}{1+\tilde\rho(s)}$ is positive definite and satisfies $\lim_{s\to\infty} h(s) = a_1\lambda_{\min}(\underline{P}_1)/\mathsf{K}_1 > 0$, and taking into account \eqref{eq:sandwich}, there exists a continuous positive definite function $\eta$ such that
    \begin{IEEEeqnarray}{rCl}
        \dot W 
        &\le& -\eta(\underline{x}) + b_1 \norm{\underline{\wind}} \IEEEnonumber
    \end{IEEEeqnarray}
    and satisfying $\liminf_{\norm{\underline{x}}\to\infty} \eta(\underline{x}) > 0$.
    Recalling that in case (b) $\underline{x}_2(t) \neq 0$ for all $t\ge t_0$, according to \cite[Theorem~1]{chaang_tac14} the closed-loop system \eqref{eq:clsys1}--\eqref{eq:clsys2} is Strongly iISS. Combining the analysis of cases (a) and (b), the result is established.
\end{IEEEproof}
}
{
Theorem~\ref{thm:UGAS} establishes that the closed-loop system has guaranteed stability and robustness, as provided by the Strong iISS property. This entails the following: (i) in absence of external perturbations, i.e. when $\underline{\wind}(t) \equiv 0$, the closed-loop system is uniformly globally asymptotically stable, hence perfect trajectory tracking is asymptotically achieved; (ii) the smaller the amplitude of the perturbations, the closer the performance will be to the zero perturbation case; (iii) if the perturbation amplitude were unbounded but its energy remained bounded, then perfect trajectory tracking would also be asymptotically achieved. Feature (i) follows from the fact that Strong iISS implies 0-UGAS, (ii) from ISS wrt small inputs, and (iii) from iISS. From ISS wrt small inputs, it follows that stability is not guaranteed if the perturbations have large amplitudes and are sustained for sufficiently long times. 
The practical robustness and performance of this control strategy is tested in Section~\ref{simulation}. 
}

\section{CASE OF STUDY: SIMULATION OF A QUAD TILT-ROTOR AIRCRAFT}\label{simulation}
\begin{figure}[htb]
    \centering
    \includegraphics[width=\linewidth]{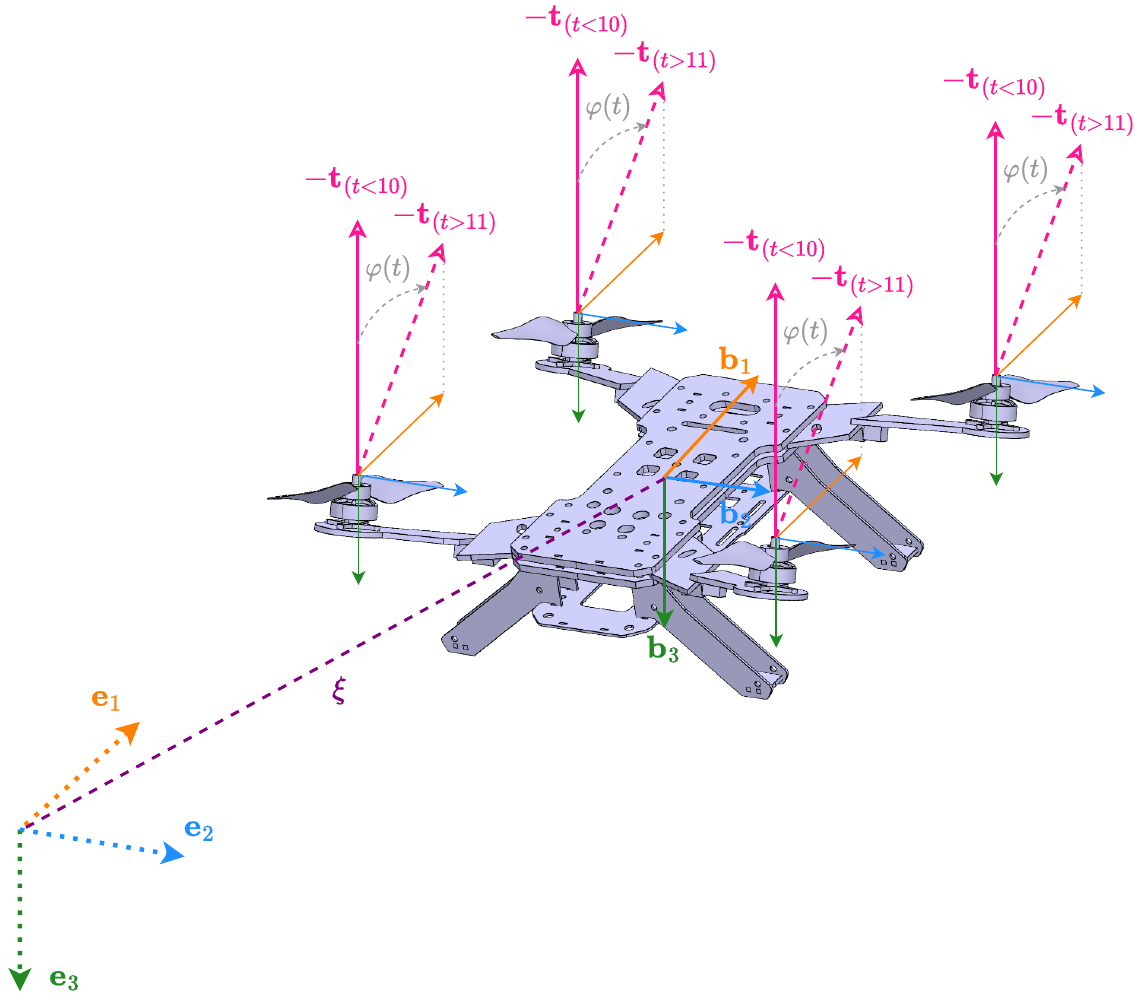}
    \caption{Quad tilt-rotor aircraft. The body frame is a rotated and translated copy of the inertial frame.  The trust vector $\bft$ produced by the four motors is expressed with a negative sign indicating its direction relative to the NED coordinate frame. The vector $-\mathbf{t}_{ (t<10)}$ denotes the trust direction for $t < 10$ sec, whereas $-\mathbf{t}_{ (t>11)}$ represents the trust direction resulting from the rotation of the four motors by the angle $\varphi=\pi/4$ for $t > 11$ sec, given in \eqref{eq:thrustvariantb}.}
    \label{fig:GA_dynamic_model}
\end{figure}

To assess the performance of the designed controller, simulations were performed using a variant of a quad tilt-rotor aircraft, as illustrated in Fig.~\ref{fig:GA_dynamic_model}. This type of quad rotorcraft can tilt its four motors by an angle $\varphi$, providing a main thrust whose magnitude can be controlled, but the controller has no influence over the thrust's direction, i.e. the thrust direction can be time-varying and is known by the controller at all times. 

For these tests, the aircraft mass is assumed to be $m=0.025$ kg and the principal moments of inertia are given by $\iota_{12}=\iota_{23}=4.856\times 10^{-3}, \iota_{31}=8.801\times 10^{-3}$ kg m$^2$. 


For simulation purposes, we consider the aircraft to be influenced by atmospheric disturbances, such as wind. The overall wind vector, relative to the inertial frame, consists of two components: a steady ambient component defined in the inertial frame and a stochastic component defined in the body frame, representing wind gusts and other turbulent effects~\cite{BMT12}. The wind gusts are modeled using the Dryden turbulence model given by~\cite{Yeager98}, considering parameters for low altitude and light turbulence given as $L_u=L_v=200$, $L_w=50$ m, $\sigma_u=\sigma_v=1.06$, $\sigma_w=0.7$ m/s and altitude = $50$ m, where $\sigma_u$, $\sigma_v$, and $\sigma_w$ are the intensities of the turbulence along the vehicle frame axes, and $L_u$, $L_v$, and $L_w$ are spatial wavelengths. 


We assume the UAV is also affected by a nonlinear dynamic drag force  given by
\begin{IEEEeqnarray}{c}
F_\mathrm{d}(\dot{\bfxi}) =  \rho C_\mathrm{d} A \,\norm{\dot{\bfxi}}\,\dot{\bfxi}/2=0.00715\,\norm{\dot{\bfxi}}\,\dot{\bfxi}
\end{IEEEeqnarray}
where $C_d=0.8$ is the non-dimensional aerodynamic drag coefficient, $A=0.01425$ m$^2$ is the planform area of the aircraft and $\rho=1.255$ kg/m$^3$ is the density of air.
The numerical simulations were performed using the SPIRAL algorithm in the \Julia programming language \cite{del2024spiral, bezanson2017julia}. SPIRAL is a third-order integration algorithm for the rotational motion of extended bodies which requires only one force calculation per time step, but it does not require rotor normalization at each time step.

\subsection[First Scenario: Recovering from an Initial 180 Flip]{First Scenario: Recovering from an Initial 180$\!\:{}^{\circ}$ Flip} \label{sub: flip}

\begin{figure}[!htb]
    \centering
    \includegraphics[width=.9\linewidth]{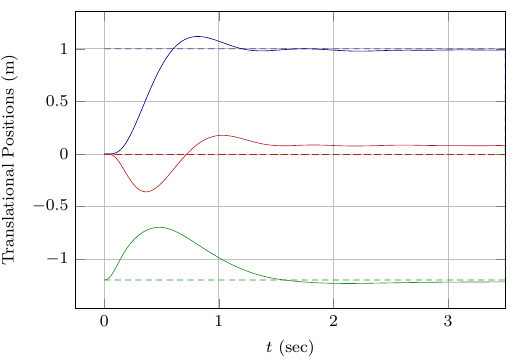}
    \caption{Position $\bfxi$ with respect to $\mathbf{e}_1$ (\textcolor{myNavy}{\solidSrule[3mm]}), $\mathbf{e}_2$ (\textcolor{myFireBrick}{\solidSrule[3mm]}), and $\mathbf{e}_3$ (\textcolor{myForestGreen}{\solidSrule[3mm]}) and desired position $\bfxi_\mathrm{d}$ with respect to $\mathbf{e}_1$ (\textcolor{myNavy}{\dashedrule}), $\mathbf{e}_2$ (\textcolor{myFireBrick}{\dashedrule}), and  $\mathbf{e}_3$ (\textcolor{myForestGreen}{\dashedrule}). The tracking is effectively achieved on axis $\mathbf{e}_1$ and $\mathbf{e}_3$. However, there is an offset on $\mathbf{e}_2$ due to the effect of the wind, which cannot be corrected due to the lack of integrators in the controller.}
    \label{fig:firstxpos}
\end{figure}

The first scenario consists of the UAV recovering from an initial configuration with $180^{\circ}$ rotation in $\phi$, i.e., upside down, and $180^{\circ}$ rotation in $\psi$, i.e., nose pointing to the back. From this configuration, the vehicle is tasked to hover with a desired orientation aligned with the NED inertial frame.
The initial translational position and velocity are set to 
\begin{IEEEeqnarray}{c}\IEEEnonumber
    \bfxi(0)=-1.2\mathbf{e}_3,~~\dot{\bfxi}(0) = 0
\end{IEEEeqnarray}
The translational reference is defined by $\bfxi_{\mathrm{d}} =  -1.2\mathbf{e}_3$. The initial rotation is given by the rotor $R_0 = \e_2\e_3$, a preferred rotation rotor is set to $R_\mathrm{p}=1$. Fig. \ref{fig:firstxpos} illustrates the UAV’s position throughout the simulation, showing convergence toward the translational position reference on axis $\mathbf{e}_1$ and $\mathbf{e}_3$. The offset observed in $\mathbf{e}_2$ is a result of wind disturbance, which the controller cannot fully compensate for due to the absence of integral action.

\begin{figure}[!htb]
    \centering
    \includegraphics[width=.9\linewidth]{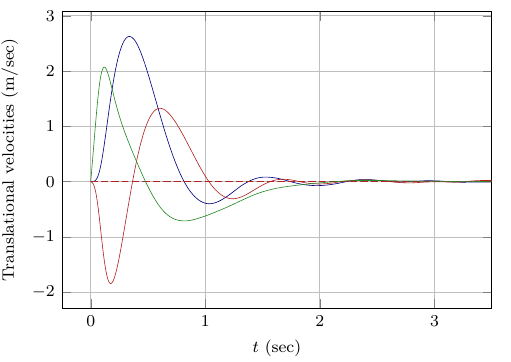}
    \caption{Velocity $\dot{\bfxi}$ with respect to $\mathbf{e}_1$ (\textcolor{myNavy}{\solidSrule[3mm]}), $\mathbf{e}_2$ (\textcolor{myFireBrick}{\solidSrule[3mm]}), and $\mathbf{e}_3$ (\textcolor{myForestGreen}{\solidSrule[3mm]}) and desired  velocity $\dot{\bfxi}_{\mathrm{d}}$ with respect to  $\mathbf{e}_1$ (\textcolor{myNavy}{\dashedrule}), $\mathbf{e}_2$ (\textcolor{myFireBrick}{\dashedrule}), and  $\mathbf{e}_3$ (\textcolor{myForestGreen}{\dashedrule}). Velocities tend to zero showing the stabilization of the UAV. }
    \label{fig:firstvpos}
\end{figure}

\begin{figure}[!htb]
    \centering   
    \includegraphics[width=.9\linewidth]{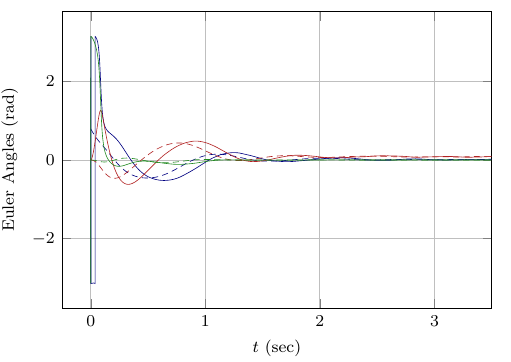}
    \caption{Current $R$ and desired $R_\mathrm{d}$ rotations transformed into Euler angles using \eqref{eq:RtoEuler}: $\roll$  (\textcolor{myNavy}{\solidSrule[3mm]}), $\pitch$  (\textcolor{myFireBrick}{\solidSrule[3mm]}),  $\yaw$  (\textcolor{myForestGreen}{\solidSrule[3mm]}),  $\roll_\mathrm{d}$  (\textcolor{myNavy}{\dashedrule}), $\pitch_\mathrm{d}$  (\textcolor{myFireBrick}{\dashedrule}), and $\yaw_\mathrm{d}$  (\textcolor{myForestGreen}{\dashedrule}). The initial rotation $R=\e_2\e_3$ is equivalent to an initial rotation $\roll=-\pi$ (roll) and  $\yaw=-\pi$ (yaw) \color{black}. Note that Euler angles cannot distinguish between $\pm \pi$, leading to ambiguity -- an issue that does not arise when using GA.}
    \label{fig:firsteulerangles}
\end{figure}

The UAV's translational velocities v.s.~the translational velocity references are shown in Fig. \ref{fig:firstvpos}.
For visualization purposes, Fig. \ref{fig:firsteulerangles} shows the Euler angles v.s.~the rotational reference, which are extracted using \eqref{eq:RtoEuler}. In the initial moments of the simulation, it becomes evident that Euler angles cannot distinguish between $-\pi$ and $\pi$, as shown by the abrupt jumps of the vertical lines from $-3.14$ to $+3.14$. This ambiguity is inherently avoided when using GA rotors, which provide a continuous and unambiguous representation of rotation.

\begin{figure}[!htb]
    \centering
    \includegraphics[width=.99\linewidth]{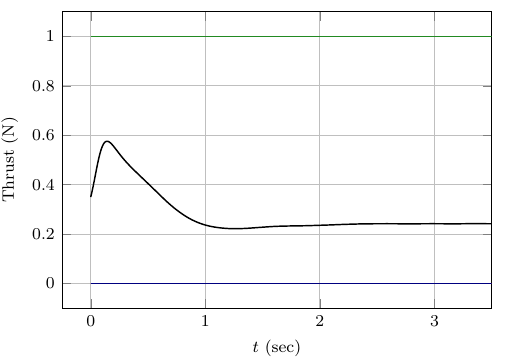}
    \caption{Thrust magnitude $f$ (\textcolor{black}{\solidSrule[3mm]}) and direction $\bft$ with respect to the body frame  $\mathbf{b}_1$ (\textcolor{myNavy}{\solidSrule[3mm]}), $\mathbf{b}_2$ (\textcolor{myFireBrick}{\solidSrule[3mm]}),  $\mathbf{b}_3$ (\textcolor{myForestGreen}{\solidSrule[3mm]}). Thrust is in the $\mathbf{b}_3$ direction and the magnitude tending to a constant to compensate the gravity while hovering.}
    \label{fig:firstturnt}
\end{figure}

\begin{figure}[!htb]
    \centering
    \includegraphics[width=.9\linewidth]{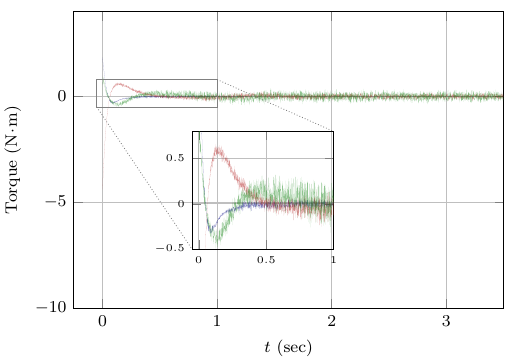}
    \caption{Torque \scalebox{1.5}{$\uptau$} 
    with respect to the body frame $\mathbf{b}_1\mathbf{b}_2$ (\textcolor{myNavy}{\solidSrule[3mm]}), $\mathbf{b}_2\mathbf{b}_3$ (\textcolor{myFireBrick}{\solidSrule[3mm]}), and $\mathbf{b}_3\mathbf{b}_1$ (\textcolor{myForestGreen}{\solidSrule[3mm]}). We observe a large energy injected a the begin of the simulation and then a ripple in torques predominates, which is principally due to the turbulent component of the wind affecting the UAV.}
    \label{fig:firsttaus}
\end{figure}

\begin{figure}[!htb]
    \centering
    \includegraphics[width=\linewidth]{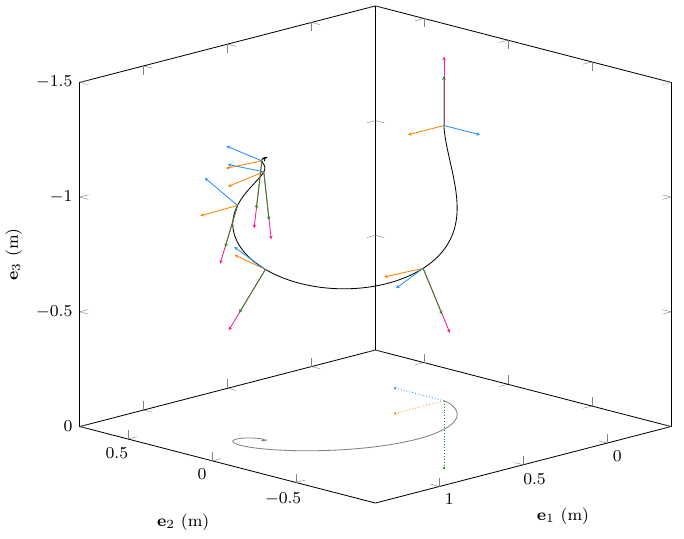}
    \caption{The flip recovery trajectory illustrated in $3$D. The body frame of the UAV $\mathbf{b}_1$ (\textcolor{orange}{\solidSrule[3mm]}), $\mathbf{b}_2$ (\textcolor{myDodgerBlue}{\solidSrule[3mm]}), and $\mathbf{b}_3$ (\textcolor{myForestGreen}{\solidSrule[3mm]}), starts in an inverted position in $\phi$ and $\psi$ dynamics, i.e., upside down and pointing backwards. The inertial reference frame $\mathbf{e}_1$ (\textcolor{orange}{\dashedrule}), $\mathbf{e}_2$ (\textcolor{myDodgerBlue}{\dashedrule}), and $\mathbf{e}_3$ (\textcolor{myForestGreen}{\dashedrule}) is situated using NED convention at the origin. The image of the UAV's position $\bfxi$ (\textcolor{black}{\solidSrule[3mm]}) and the corresponding $\mathbf{e}_1\mathbf{e}_2$ projection (\textcolor{myGray}{\solidSrule[3mm]}) are shown. Additionally, the thrust direction $\bft$ (\textcolor{myDeepPink}{\solidSrule[3mm]}) is shown at instants with the body frame. Units are meters, but unit vectors representing the body frame and thrust direction have been scaled for illustrative clarity.}
    \label{fig:firstplot3d}
\end{figure}

Fig. \ref{fig:firstturnt} shows the thrust direction and magnitude; we can see that the thrust is only in the direction of $\mathbf{b}_3$ and the magnitude stabilizes to a value that compensates for gravity.
%
%
%
Fig.~\ref{fig:firsttaus} shows the time evolution of the torque $\Tau$, which is stronger at the beginning and up to time $0.4$ sec to quickly recover from the inverted configuration.
%
Fig.~\ref{fig:firstplot3d} shows the $3$D trajectory executed by the UAV during the flip recovery. We can see that the control objective is successfully achieved, as the UAV recovers correctly from the inverted position.

These results support the hypothesis that by adopting a GA-based model, a low-computational-cost control strategy can be designed, enabling the execution of complex maneuvers—e.g. recovering from an inverted position—with efficiency and reliability.

\subsection{Second Scenario: Tracking a $3$D Rhodonea Curve }
The second scenario consists of trajectory tracking of a $3$D rhodonea curve (rose path) in a $3$D space, with a preferred UAV heading orientation.
The initial translational position and velocity are set to 
\begin{IEEEeqnarray}{c}\IEEEnonumber
\bfxi(0)=50\mathbf{e}_1+35\mathbf{e}_3,~~\dot{\bfxi}(0) = 0
\end{IEEEeqnarray}
and a time-dependent position reference is set to 

\begin{IEEEeqnarray}{rCl}\IEEEnonumber
    \bfxi_{\mathrm{d}} &=& 50\cos(0.376 t) \cos(0.752 t )\mathbf{e}_1\\
    &~&\quad +~ 50 \sin(0.376 t) \cos(0.752 t)\mathbf{e}_2 +(35-3.75t)\mathbf{e}_3.\IEEEnonumber
\end{IEEEeqnarray}
%
%
The quad tilt-rotor aircraft allows for a time-variant thrust direction, therefore, we set the thrust direction to start along $\mathbf{b}_3$. Next, from approximately time $t=10$ to $t=11$, the thrust direction gradually transitions to a combination of $\mathbf{b}_1$ and $\mathbf{b}_3$, where it remains for the remainder of the trajectory tracking mission. Mathematically, this can be represented as
\begin{IEEEeqnarray}{c}
    \turnt = R_\turnt \mathbf{b}_3  R_\turnt^\dagger   \label{eq:thrustvarianta}   
\end{IEEEeqnarray} 
with $R_\turnt = \exp\!\big({-\mathbf{b}_1\mathbf{b}_3\varphi(t)/2}\big)$  parameterized by the angle 
\begin{IEEEeqnarray}{c} \label{eq:thrustvariantb}
 \varphi(t) = \frac{\pi/4}{1+e^{-10(t-10.5)}}
\end{IEEEeqnarray} 

which is measured w.r.t $\mathbf{b}_3$ in the $\mathbf{b}_1\mathbf{b}_3$ plane. 
The initial rotation is given by the rotor $R_0 = 1$ and the preferred rotation reference is set to 
\begin{IEEEeqnarray}{c}\IEEEnonumber
    R_\mathrm{p} = \exp\big(-\!\mathbf{e}_1\mathbf{e}_2 0.45\sin(0.1\pi t)\big)
\end{IEEEeqnarray} 
which corresponds to a time variation in the angle $\uppsi$.

\begin{figure}[!htb]
    \centering
    \includegraphics[width=.87\linewidth]{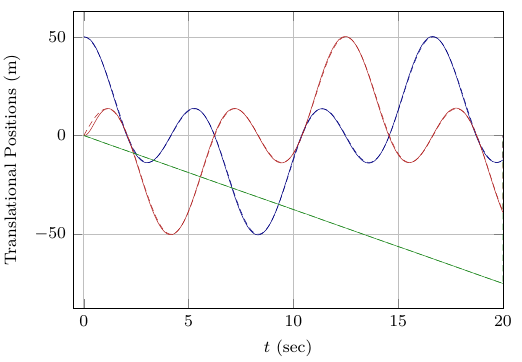}
    \caption{Position $\bfxi$ with respect to $\mathbf{e}_1$ (\textcolor{myNavy}{\solidSrule[3mm]}), $\mathbf{e}_2$ (\textcolor{myFireBrick}{\solidSrule[3mm]}), and $\mathbf{e}_3$ (\textcolor{myForestGreen}{\solidSrule[3mm]}) and desired position $\bfxi_\mathrm{d}$ with respect to $\mathbf{e}_1$ (\textcolor{myNavy}{\dashedrule}), $\mathbf{e}_2$ (\textcolor{myFireBrick}{\dashedrule}), and  $\mathbf{e}_3$ (\textcolor{myForestGreen}{\dashedrule}).}
    \label{fig:secxpos}
\end{figure}

\begin{figure}[htb]
    \centering
    \includegraphics[width=.87\linewidth]{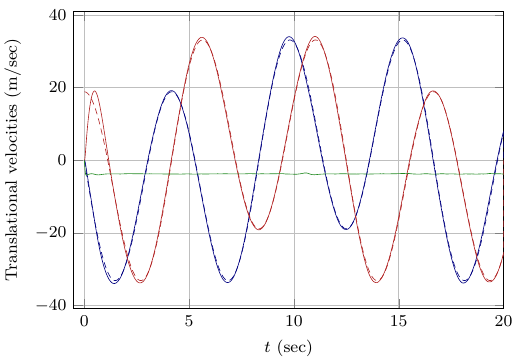}
    \caption{Velocity $\dot{\bfxi}$  with respect to $\mathbf{e}_1$ (\textcolor{myNavy}{\solidSrule[3mm]}), $\mathbf{e}_2$ (\textcolor{myFireBrick}{\solidSrule[3mm]}), and $\mathbf{e}_3$ (\textcolor{myForestGreen}{\solidSrule[3mm]}) and desired  velocity $\dot{\bfxi}_{\mathrm{d}}$ with respect to  $\mathbf{e}_1$ (\textcolor{myNavy}{\dashedrule}), $\mathbf{e}_2$ (\textcolor{myFireBrick}{\dashedrule}), and  $\mathbf{e}_3$ (\textcolor{myForestGreen}{\dashedrule}). }
    \label{fig:secvpos}
\end{figure}

\begin{figure}[!htb]
    \centering   
    \includegraphics[width=.87\linewidth]{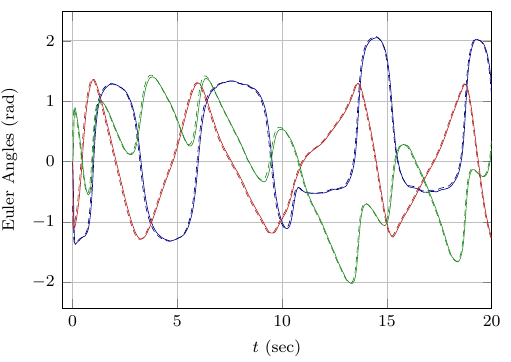}
    \caption{Current and desired rotations $R$ and $R_\mathrm{d}$ transformed into Euler angles using \eqref{eq:RtoEuler}: $\roll$  (\textcolor{myNavy}{\solidSrule[3mm]}), $\pitch$  (\textcolor{myFireBrick}{\solidSrule[3mm]}),  $\yaw$  (\textcolor{myForestGreen}{\solidSrule[3mm]}),  $\roll_\mathrm{d}$  (\textcolor{myNavy}{\dashedrule}), $\pitch_\mathrm{d}$  (\textcolor{myFireBrick}{\dashedrule}), and $\yaw_\mathrm{d}$  (\textcolor{myForestGreen}{\dashedrule}). The rotation correction at around $t=11$ sec. is executed to adapt to the variation of the thrust direction.}
    \label{fig:seceulerangles}
\end{figure}

\begin{figure}[!htb]
    \centering
    \includegraphics[width=\linewidth]{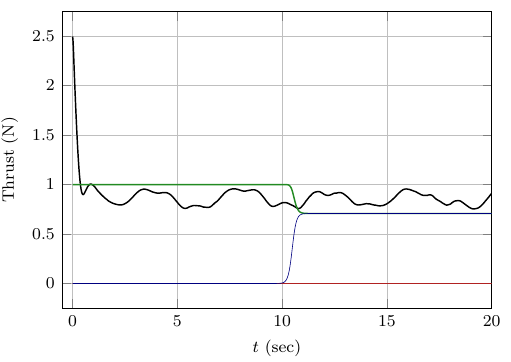}
    \caption{Thrust magnitude $f$ (\textcolor{black}{\solidSrule[3mm]}) and direction $\bft$ with respect to the body frame  $\mathbf{b}_1$ (\textcolor{myNavy}{\solidSrule[3mm]}), $\mathbf{b}_2$ (\textcolor{myFireBrick}{\solidSrule[3mm]}),  $\mathbf{b}_3$ (\textcolor{myForestGreen}{\solidSrule[3mm]}). From $0$ sec to $10$ sec the system operates like a quadrotor with thrust directed along $\mathbf{b}_3$. Approximately, between $10$ sec and $11$ sec the thrust direction undergoes a modification, ultimately stabilizing to a configuration suitable for horizontal flight with a thrust magnitude of $\sqrt{2}/2(\mathbf{b}_1+\mathbf{b}_3)$}
    \label{fig:secturnt}
\end{figure}

\begin{figure}[!htb]
    \centering
    \includegraphics[width=\linewidth]{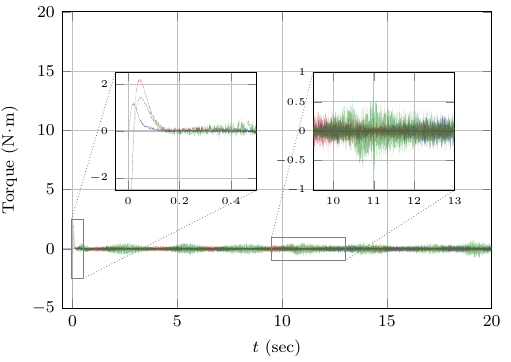}
    \caption{Torque \scalebox{1.5}{$\uptau$} 
    with respect to the body frame $\mathbf{b}_1\mathbf{b}_2$ (\textcolor{myNavy}{\solidSrule[3mm]}), $\mathbf{b}_2\mathbf{b}_3$ (\textcolor{myFireBrick}{\solidSrule[3mm]}), and $\mathbf{b}_3\mathbf{b}_1$ (\textcolor{myForestGreen}{\solidSrule[3mm]}).}
    \label{fig:sectaus}
\end{figure}

Fig. \ref{fig:secxpos} shows the UAV position v.s.~the translational position reference while following the rhodonea curve, exhibiting appropriate convergence. 
The corresponding translational velocities v.s.~the translational velocity reference are shown in Fig. \ref{fig:secvpos}. 
For visualization purposes, Fig. \ref{fig:seceulerangles} shows the Euler angles v.s.~the rotational reference, obtained using the transformation in~\eqref{eq:RtoEuler}. Notice a rotation correction executed at around $t=10$ sec, which is performed in order to adapt to the variation of the thrust direction.
Fig. \ref{fig:secturnt} shows the thrust magnitude and direction relative to the body frame. During the first 10 seconds, the system behaves like a conventional quad rotorcraft, with thrust aligned along the $\mathbf{b}_3$ axis. Between $t\approx10$ and $t\approx11$ sec., the thrust direction is gradually adjusted, ultimately transitioning to a configuration suitable for sustained horizontal flight.
Fig. \ref{fig:sectaus} shows the time evolution of torque $\Tau$, which is stronger at the beginning and up to $t=2$ sec., and also between time $10$ and $11$ sec., to compensate for and correct the UAV rotation.

\begin{figure*}[!htb]
    \centering
    \includegraphics[width=.9\linewidth]{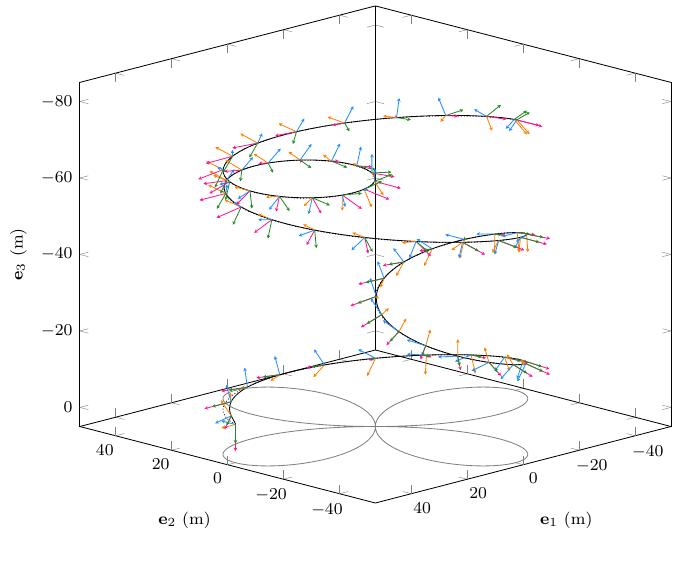}
    \caption{The flight trajectory illustrated in $3$D. The image of the UAV's position $\bfxi$ (\textcolor{black}{\solidSrule[3mm]}), the corresponding $\mathbf{e}_1\mathbf{e}_2$ projection (\textcolor{myGray}{\solidSrule[3mm]}), and desired position $\bfxi_\mathrm{d}$ (\textcolor{black}{\dashedrule}) are shown. The body frame of the UAV $\mathbf{b}_1$ (\textcolor{orange}{\solidSrule[3mm]}), $\mathbf{b}_2$ (\textcolor{myDodgerBlue}{\solidSrule[3mm]}), and $\mathbf{b}_3$ (\textcolor{myForestGreen}{\solidSrule[3mm]}), starts at the origin and then tracks a $3$D trajectory. Additionally, the thrust direction $\bft$ (\textcolor{myDeepPink}{\solidSrule[3mm]}) is shown at instants with the body frame and its direction is time-variant with the law from \eqref{eq:thrustvarianta} and \eqref{eq:thrustvariantb}. Units are meters, but unit vectors representing the body frame and thrust direction have been scaled for illustrative clarity.}
    \label{fig:secplot3d}
\end{figure*}

Fig. \ref{fig:secplot3d} shows the $3$D trajectory executed by the UAV, from which we can conclude that the trajectory tracking control objective is successfully achieved.

These results also align with our hypothesis that, by adopting a GA-based model, it is possible to design a low-computational-cost control strategy capable of achieving high-accuracy trajectory tracking. 


\section{COMPARATIVE ANALYSIS}\label{Discussion}
A comparative analysis between the control strategy proposed in this work and six representative, state-of-the-art control algorithms is presented in Table~\ref{tab:comparison}. The selected references share key properties with our approach, including trajectory tracking of a quadrotor, formal stability guarantees, and the avoidance of singularities. Given these shared characteristics, the controllers in \cite{Leok2010, lee2010geometric, Lee2013, Lee2010Complex,Sharma2020Left, Jia2022} were chosen as conceptual references.

In the algorithms proposed in \cite{Leok2010, lee2010geometric, Lee2013, Lee2010Complex}, stability is only guaranteed when the initial attitude error corresponds to an angle less than $180^\circ$. By following this restriction, almost global asymptotic tracking is established for the quadrotor’s attitude, position, and velocity. Similarly, \cite{Sharma2020Left} restricts exponential stability to initial conditions where a Lyapunov-like function involving attitude and angular velocity errors is limited by a positive constant that depends on the controller gains. In \cite{Jia2022}, the desired reference quaternion presents a singularity when the quadrotor is facing downward, and its control scheme establishes that the  rotational and translational closed-loop systems are shown to be input-to-state stable, with bounded disturbance-observer error and convergence of the attitude tracking errors to an arbitrarily small residual set. While such numerical or stability-related restrictions do not occur in a nominal quadrotor operation, they limit the quadrotor to perform extreme maneuvers such as the recovery from an inverted orientation demonstrated in Subsection~\ref{simulation}.\ref{sub: flip}. Our approach does not exhibit such actuation-related limitations, while achieving globally asymptotically tracking in the presence of perturbations with unbounded amplitude and bounded energy.

A key difference w.r.t. our method lies in the fact that the algorithms in \cite{Leok2010, lee2010geometric, Lee2013, Lee2010Complex, Sharma2020Left} rely on the calculation of the time derivatives of the desired angular velocity. Furthermore, \cite{Leok2010, lee2010geometric, Lee2013, Lee2010Complex} require the computation of the time derivatives of intermediate controller steps, while \cite{Jia2022} indicates the first and second time derivatives of the translational controller are obtained through a numerical differentiation. Conversely, reference trajectories generated by our proposed controller are obtained through algebraic expressions, eliminating the need of differentiation algorithms in both the references and the control strategy. 




The algorithms in \cite{Leok2010, lee2010geometric, Lee2013, Lee2010Complex, Sharma2020Left} require measurement of the first derivatives of the position and orientation, while \cite{Jia2022} additionally depends on the second derivative of the position. Our control strategy requires measurements of the second and third derivatives of the position and the first derivative of the orientation.


While \cite{Leok2010, lee2010geometric, Lee2010Complex, Sharma2020Left} do not consider any disturbances in their design, both \cite{Lee2013, Jia2022} along with the proposed algorithm explicitly account for them. Specifically, our approach and \cite{Jia2022} consider the effect of aerodynamic drag and wind disturbances on the quadrotor's dynamics, whereas \cite{Jia2022} additionally evaluates robustness against dynamic center of gravity (CoG) shifts and external input torques.


\begin{table*}[!t]
\small
\renewcommand{\arraystretch}{1.3}
\setlength{\tabcolsep}{3pt}
\centering
\caption{Comparative Analysis of Closely Related Control Algorithms}
\label{tab:comparison}
\scalebox{0.8}{

\begin{tabular}{
  c 
  >{\centering\arraybackslash}p{4.75cm} 
  >{\centering\arraybackslash}p{5.25cm} 
  >{\centering\arraybackslash}p{2.25cm}
  >{\centering\arraybackslash}p{3.5cm} 
  >{\centering\arraybackslash}p{3cm} 
}

\toprule
\textbf{Reference} & \textbf{Stability} & \textbf{Restrictions} & \makecell[c]{\textbf{Memoryless}\\  \textbf{Controller}} & \makecell[c]{\textbf{Disturbances}} & \textbf{Uncertainties}\\
\midrule
\cite{Leok2010, lee2010geometric} \cite{Lee2010Complex} & Almost Globally Asymptotic & Initial attitude error $<2$ & No & No & No\\[0.8ex]
\cite{Lee2013} & Almost Globally Asymptotic & Initial attitude error $<2$ & No & No & Unstructured additive bounded uncertainties \\[0.8ex]
\cite{Sharma2020Left} & Almost Globally Exponential & Initial conditions limited by controller gains & No & No & No \\[0.8ex]
\cite{Jia2022} & ISS, Disturbance observer error bounded & Singularity when facing downward & No & \makecell[c]{Sinusoidal wind, \\Step torque, Drag} & Dynamic CoG\\
Proposed  & Globally Asymptotic & Unrestricted Initial Conditions & Yes & \makecell[c]{Dynamic Drag, \\Dryden Turbulence} & No\\
\bottomrule
\end{tabular}
}
\end{table*}

\color{black}
\section{CONCLUSIONS}\label{conclusions}

GA was utilized to model the translational and rotational dynamics of objects in $3$D space. This work demonstrates how such a modeling approach can simplify the control design process, in particular for trajectory tracking in underactuated aerial systems. The proposed control strategy features a cascaded structure, where a linear rotational subsystem drives a translational subsystem through a nonlinear interconnection. This architecture enables the development of an initial control strategy that is both simple and effective, with closed-loop stability rigorously established using input-to-state stability (ISS) methods. A key advantage of the approach is the analytical derivation of rotational references directly from translational ones, thereby eliminating the need for online differentiation algorithms. The resulting control law is computationally efficient, requiring only basic operations, no memory storage, and no iterative search loops.

The controller’s performance is validated through two numerical simulation scenarios using a variant of a quad tilt-rotor UAV. The first scenario demonstrates recovery from an inverted initial position to achieve stable hovering at a desired location. The second scenario involves tracking a $3$D rhodonea curve trajectory while maintaining a specified heading orientation. In both cases the controller was able to generate the control signals required to successfully accomplish the mission objectives. 

\subsection{Future work}

Future work will build on the current results to develop more advanced control strategies aimed at enhancing disturbance rejection also in the rotational motion and enabling coordinated control of both thrust magnitude and direction in underactuated hybrid aircraft.

Additionally, future work will include a systematic sensitivity analysis to quantify how controller gains affect performance metrics and to evaluate the robustness of the proposed approach under varying parameters.

\section*{ACKNOWLEDGMENT}
This work was supported in part by the U.S. National Science Foundation (NSF) under Grant 2318288 and by Agencia I+D+i PICT 2021-I-A-0730, Argentina.

\bibliographystyle{IEEEtran}
\bibliography{main}

\FloatBarrier 

\begin{IEEEbiography}[{\includegraphics[width=1in,height=1.25in,clip,keepaspectratio]{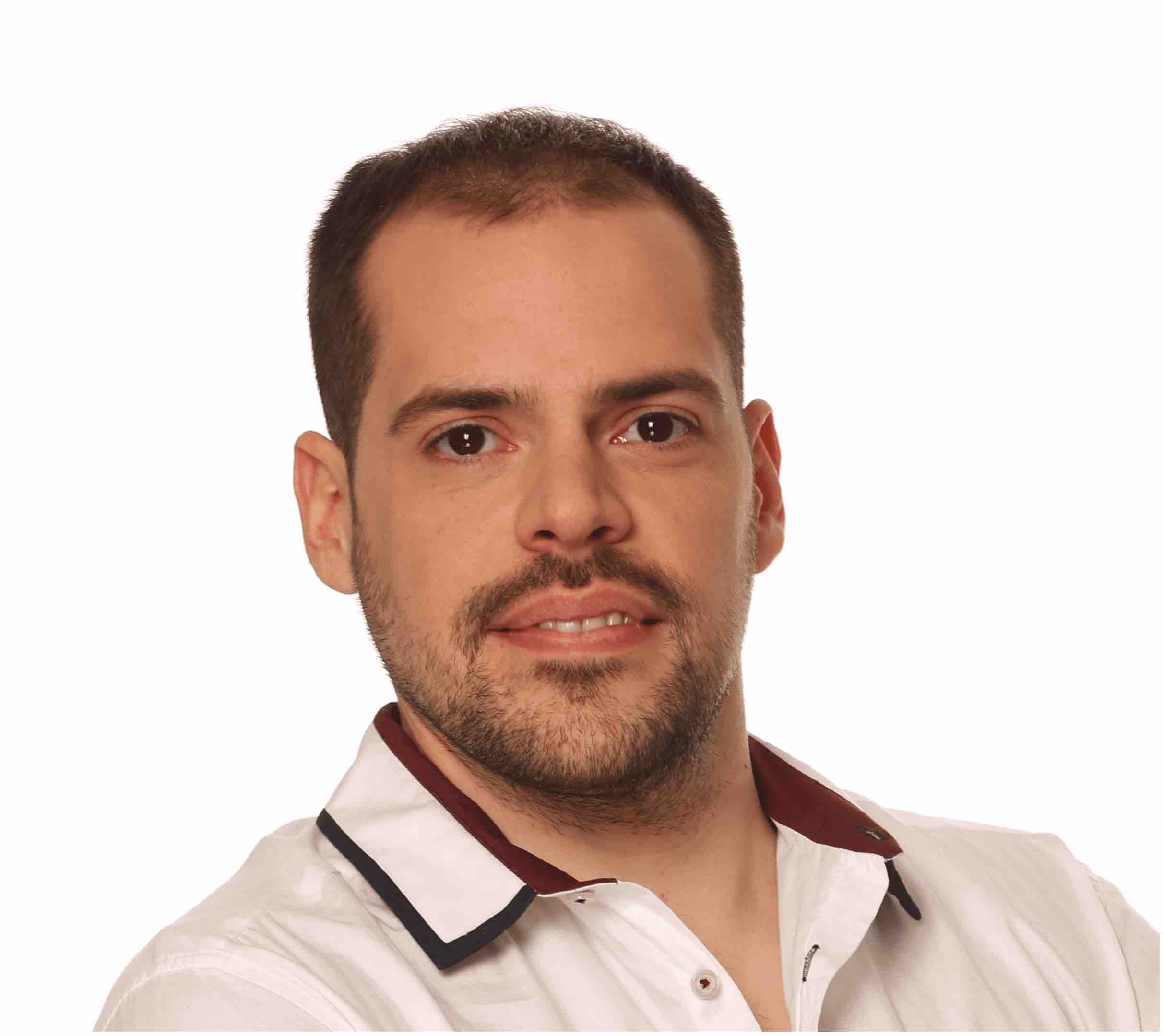}}]{Ignacio Rubio Scola} received the Ph.D. degree in Control from Université Grenoble Alpes, France in 2015. 
From 2018 he was an Assistant Professor at UNR Argentina, and from 2022 he is a full researcher at Department of Industrial Products Engineering of the INTI and at CONICET Argentina. He is interested in nonlinear, hybrid and unmanned systems, particularly in low-complexity observers-based controllers.
\end{IEEEbiography}

\begin{IEEEbiography}[{\includegraphics[width=1in,height=1.25in,clip,keepaspectratio]{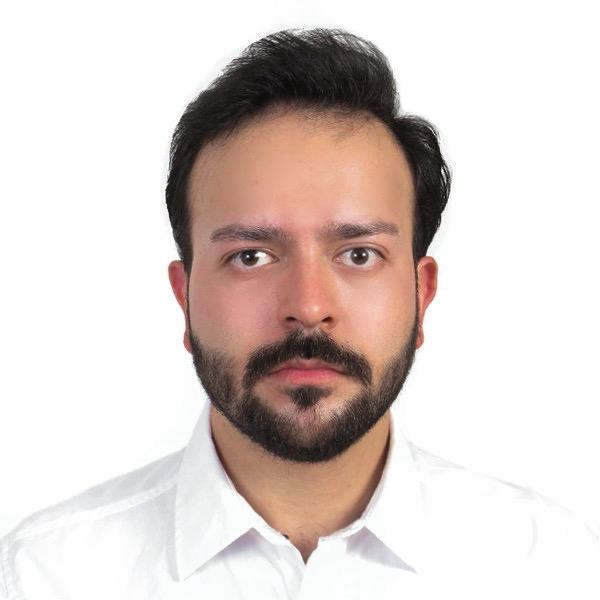}}]{Omar Alejandro Garcia Alcantara} received the M.Sc. degree in Autonomous 
Systems from the Center for Research and Advanced Studies of the National Polytechnic Institute, Mexico in 2023, and his B.S. degree in Aeronautical Engineering from the National Polytechnic Institute, México in 2018. At present, he's pursuing his PhD. in Electrical Engineering at New Mexico State University. His research interests include Neuromorphic Computing, and design and control of UAVs. 
\end{IEEEbiography}

\begin{IEEEbiography}[{\includegraphics[width=1in,height=1.25in,clip,keepaspectratio]{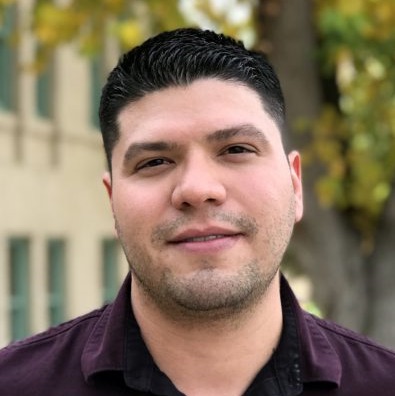}}]{Steven Sandoval} received the B.S.~Electrical Engineering and M.S.~Electrical Engineering from New Mexico State University in 2007 and 2010 respectively, and the Ph.D.~degree in Electrical Engineering from Arizona State University in 2016. Currently, he is an Associate Professor at the Klipsch School of Electrical and Computer Engineering at New Mexico State University. Research interests include time-frequency analysis and geometric algebra.\end{IEEEbiography}

\begin{IEEEbiography}[{\includegraphics[width=1in,height=1.25in,clip,keepaspectratio]{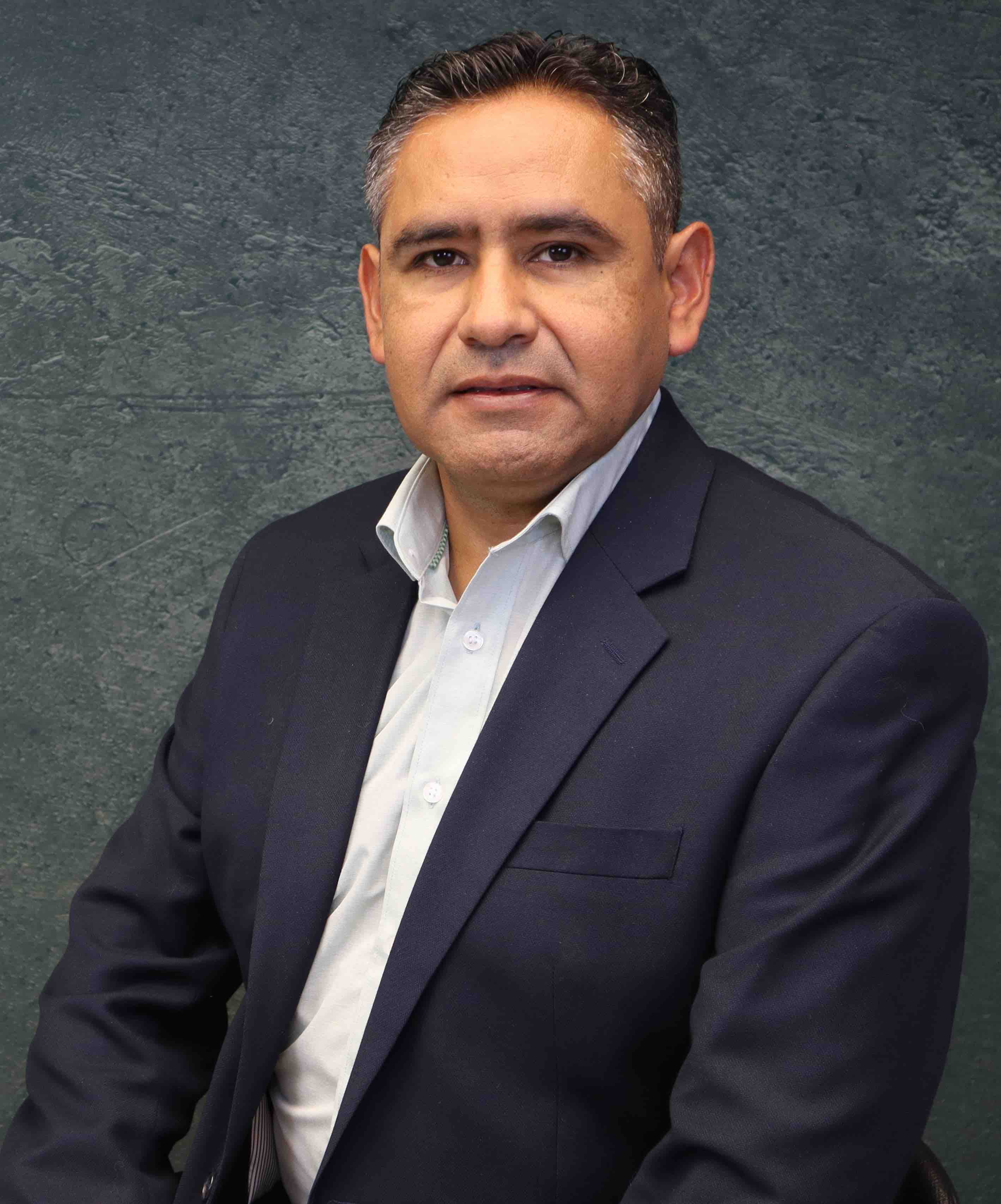}}]{Eduardo Steed Espinoza Quesada} (Senior Member, IEEE)
received the Ph.D. degree in Automatic Control from the Center for Research and Advanced Studies of the National Polytechnic Institute, Mexico in 2013. He was a Postdoctoral Researcher at the Nevada Advance Autonomous Systems Innovation Center at University of Nevada, USA in 2016. At present, he holds a Professor position at CINVESTAV, Mexico. His research interests include Neuromorphic Computing, UASs applications, design and control of eVTOL aircraft.
\end{IEEEbiography}

\begin{IEEEbiography}[{\includegraphics[width=1in,height=1.25in,clip,keepaspectratio]{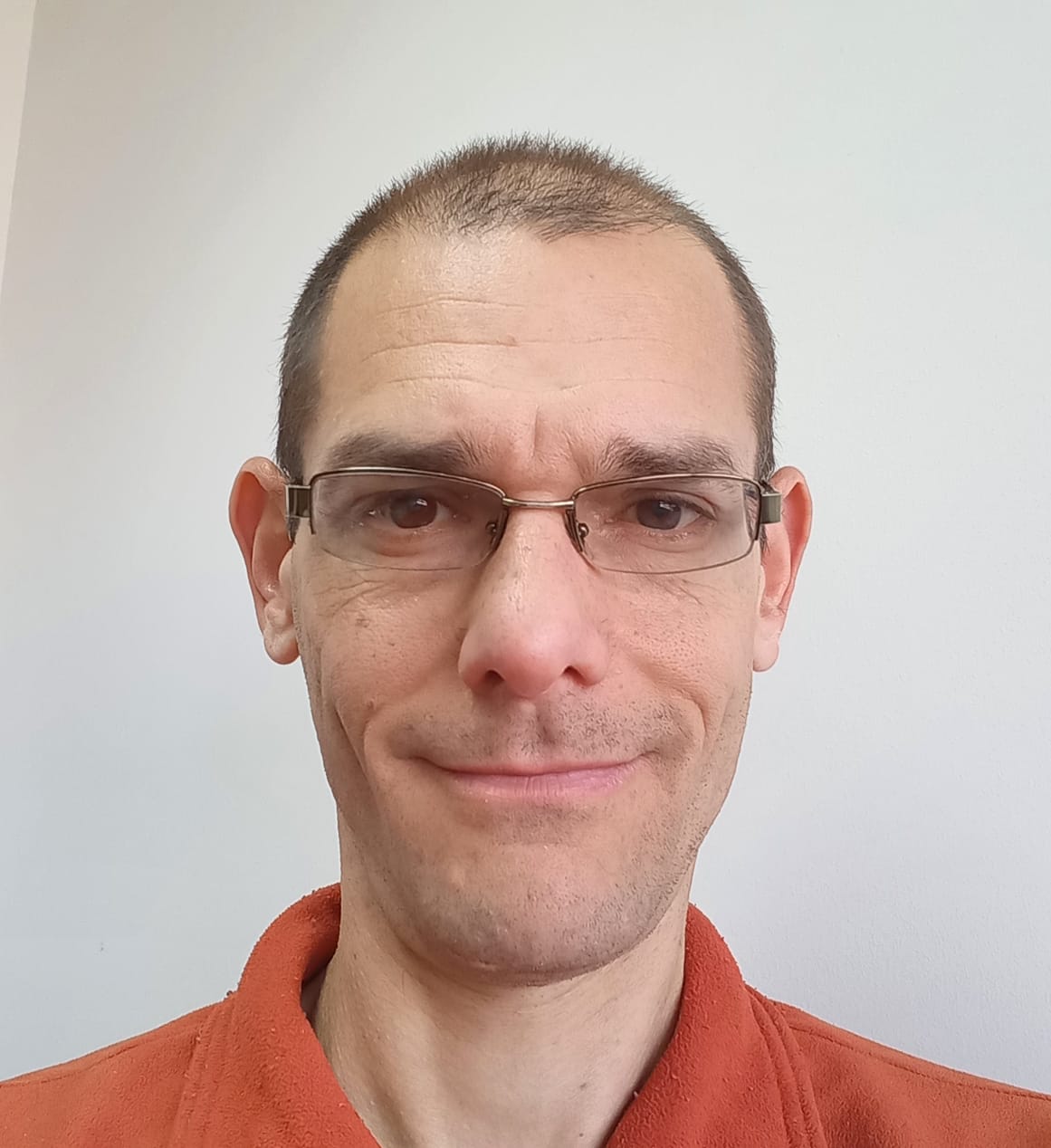}}]{Hernan Haimovich} (Senior Member, IEEE) received the Electronics Engineering degree with highest honours in 2001 from the Universidad Nacional de Rosario (UNR), Argentina, and the Ph.D. degree from The University of Newcastle, Australia, in 2006. After a brief period as postdoctoral researcher, Dr. Haimovich became a permanent Investigator of the Argentine Research Council CONICET, currently at the Principal Investigator level. Dr. Haimovich is also Professor at the School of Electronics Engineering, UNR, and serves as Associate Editor for the journal Automatica. His research interests include control theory for nonlinear, switched, time-varying, impulsive, hybrid and/or discontinuous systems, and related applications.
\end{IEEEbiography}

\begin{IEEEbiography}[{\includegraphics[width=1in,height=1.25in,clip,keepaspectratio]{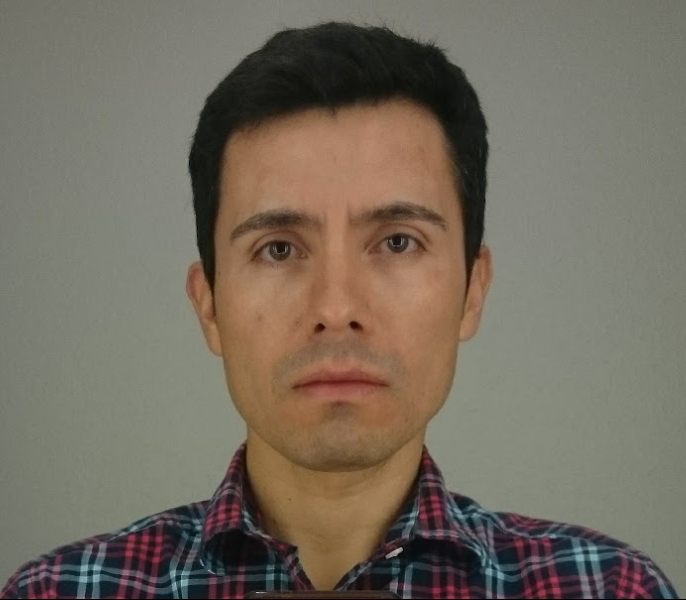}}]{Luis Rodolfo Garcia Carrillo} (Senior Member, IEEE) received his 
Ph.D. in Control Systems from the University of Technology of Compiegne, (Sorbonne Universités) France, in 2011. From 2012 to 2013, he was a postdoctoral researcher at the Center for Control, Dynamical Systems and Computation at the University of California, Santa Barbara. Dr. Garcia Carrillo currently holds an Associate Professor position with the Klipsch School of Electrical and Computer Engineering at New Mexico State University, in Las Cruces, New Mexico USA. He also serves as Autonomous Systems Associate Editor for IEEE Transactions on Aerospace and Electronic Systems (TAES). His research interests include control systems, robotics, autonomous systems, and the use of computer vision in feedback control.
\end{IEEEbiography}

\end{document}